\documentclass[superscriptaddress,nofootinbib,tightenlines,preprintnumbers]{revtex4-2}
\pdfoutput=1
\usepackage[utf8]{inputenc}

\usepackage[pdftex,breaklinks,colorlinks,linktocpage,
citecolor=blue,
urlcolor=blue]{hyperref}
\usepackage{subcaption}
\usepackage{bm}
\usepackage{amssymb}
\usepackage{amsfonts}
\usepackage{mathrsfs}
\usepackage{amsmath}
\usepackage{amsthm}
\usepackage{xcolor}
\usepackage{bbold}
\usepackage{braket}
\usepackage{physics}
\usepackage{graphicx}
\usepackage{tikz}
\usepackage[normalem]{ulem}
\usepackage{cancel}
\usepackage{comment}
\usetikzlibrary{decorations.pathmorphing}

\usepackage{caption}
\newcommand{\be}{\begin{equation}}
\newcommand{\ee}{\end{equation}}
\newcommand{\bea}{\begin{eqnarray}}
\newcommand{\eea}{\end{eqnarray}}
\def\bml{\begin{subequations}}
\def\blea{\bml\begin{eqnarray}}
\def\eml{\end{subequations}}
\def\elea{\end{eqnarray}\eml}

\newcommand{\beq}{\begin{equation}}

\newcommand{\eeq}{\end{equation}}

\theoremstyle{definition}

\makeatletter
\def\l@subsubsection#1#2{}
\makeatother

\begin{document}

\title{Mass and entropy of asymptotically flat eternal quantum black holes in 2D}

\author{Jean Alexandre}
\email{jean.alexandre@kcl.ac.uk}
\affiliation{Department of Physics, King’s College London,
Strand, London, WC2R 2LS, United Kingdom}

\author{Eleni-Alexandra
Kontou}
\email{eleni.kontou@kcl.ac.uk}
\affiliation{Department of Mathematics, King’s College London,
Strand, London WC2R 2LS, United Kingdom}

\author{Diego Pardo Santos}
\email{diego.pardo@kcl.ac.uk}
\affiliation{Department of Mathematics, King’s College London,
Strand, London WC2R 2LS, United Kingdom}

\author{Silvia Pla}
\email{silvia.pla-garcia@tum.de}
\affiliation{Physik-Department, Technische Universit\"at M\"unchen, James-Franck-Str., 85748 Garching,
Germany}

\author{Andrew Svesko}
\email{andrew.svesko@kcl.ac.uk}
\affiliation{Department of Mathematics, King’s College London,
Strand, London WC2R 2LS, United Kingdom}

\begin{abstract}

\noindent Semi-classical dilaton gravity in (1+1)-dimensions remains one of the only arenas where quantum black holes
can be exactly constructed, fully accounting for backreaction due to quantum matter. Here we provide a comprehensive analysis of the mass and thermodynamic properties of static asymptotically flat quantum black holes both analytically and numerically. First, we analytically investigate eternal quantum black hole solutions to a one-parameter family of analytically solvable models interpolating between Russo-Susskind-Thorlacius and Bose, Parker, and Peleg gravities. Examining these models in a semi-classically allowed parameter space, we find naked singularities may exist for quantum fields in the Boulware state.  Using a quasi-local formalism, where we confine the black hole to a finite sized cavity, we derive the conserved energy and analyze the system's thermal behavior. Specifically, we show the semi-classical Wald entropy precisely equals the generalized entropy, accounting for both gravitational and fine grained matter entropies, and we find a range where the quantum black holes are thermally stable. Finally, we numerically construct eternal black hole solutions to semi-classical Callan-Giddings-Harvey-Strominger gravity and find their thermal behavior is qualitatively different from their analytic counterparts. In the process, we develop an analytic expansion of the solutions and find it accurately approximates the full numerical solutions in the semi-classical limit.

\end{abstract}

\maketitle

\newpage

\tableofcontents

\newpage

\section{Introduction} \label{sec:intro}

\noindent Black holes provide a window into the nature of quantum gravity. Indeed, some of our best insights into quantum aspects of gravity follow from studying quantum black holes, i.e., solutions to the semi-classical Einstein equations, 
\beq G_{\mu\nu}+\Lambda g_{\mu\nu}=8\pi G_{d}\langle T_{\mu\nu}^{\text{ren}}\rangle\;.\label{eq:scEineom}\eeq
Here, $\langle T^{\text{ren}}_{\mu\nu}\rangle$ refers to the expectation value of the renormalized stress-tensor due to quantized matter fields permeating the classical background spacetime $g_{\mu\nu}$. A first step in solving (\ref{eq:scEineom}) requires an analysis of quantum fields in a \emph{fixed} curved background. Such a treatment indicates black holes thermally radiate \cite{Hawking:1975vcx} at a temperature identified with their surface gravity. Then the classical laws of black hole mechanics \cite{Bardeen:1973gs} may be identified with laws of thermodynamics \cite{Hawking:1976de}, where the thermal entropy is proportional to the co-dimension-2 horizon area \cite{Bekenstein:1972tm,Bekenstein:1973ur}.
Importantly, such identification between geometry and thermodynamics are only justified upon coupling quantum matter to gravity via semi-classical backreaction. Indeed, that black holes evaporate is a consequence of backreaction. 

Uncovering exact black hole solutions to semi-classical gravity is an open challenge. In $(3+1)$-dimensional gravity and higher, no black hole solutions to (\ref{eq:scEineom}) exactly incorporating backreaction are known.\footnote{York \cite{York:1983zb,York:1984wp} and collaborators \cite{Hochberg:1992rd,Hochberg:1992xt,Anderson:1994hh} previously examined the first linearized correction to the Einstein tensor on the perturbed Schwarzschild  metric due to (originally) a conformally coupled scalar in the Hartle-Hawking vacuum using an approximate form of $\langle T_{\mu\nu}\rangle$ suggested by Page \cite{Page:1982fm}.} Indeed, solutions to \eqref{eq:scEineom} in $3+1$ dimensions have only been shown in cosmological settings due to their high levels of symmetry (see \cite{Wald:1977up,Fischetti:1979ue,Mukhanov:1996ak,Eltzner:2010nx, Meda:2020smb, Gottschalk:2018kqt}).  Progress can be made by descending to unworldly lower dimensions, where the problem of quantum backreaction becomes technically tractable. In particular, the semi-classical Einstein equation in $(2+1)$-dimensions can be solved to first order in a perturbative expansion in backreaction for, e.g., a conformally coupled quantum field \cite{Souradeep:1992ia,Steif:1993zv,Matschull:1998rv,Casals:2016ioo,Casals:2019jfo,Emparan:2022ijy}. Backreaction, however, modifies the geometry such that corrections enter on the order of the Planck scale, where quantum gravitational effects cannot be ignored and the semi-classical approximation is inconsistent.

In this article, we descend even lower to $(1+1)$-dimensional semi-classical dilaton gravity. The reason is three-fold: (i) the semi-classical approximation is self-consistent; (ii) the semi-classical field equations are exactly solvable, and (iii) some 2D semi-classical models effectively characterize aspects of higher-dimensional black holes. Combined, the lower-dimensional framework, albeit simplified, serves as a proving ground to precisely examine problems in semi-classical gravity.\footnote{Assuming a spherical-wave approximation such that the quantum matter stress-tensor can be evaluated as done in two-dimensions, four-dimensional quantum black holes and their dynamics can be numerically analyzed, cf. \cite{Parentani:1994ij,Boyanov:2025otp}.} In particular, we provide a comprehensive treatment of the geometry and horizon thermodynamics of eternal asymptotically flat 2D quantum black holes. 

The models of interest we explore are semi-classical Callan-Giddings-Harvey-Strominger (CGHS) gravity \cite{Callan:1992rs} or modifications thereof that render the system analytically solvable \cite{Russo:1992ax,Bose:1995pz,Fabbri:1995bz,Cruz:1995zt}. Importantly, such models are in a regime where the semi-classical approximation is valid, in the sense that metric fluctuations can be suppressed and gravitational degrees of freedom may be treated as effectively classical. In the higher-dimensional context, it is difficult to work self-consistently in a semi-classical approximation: metric fluctuations (graviton loops) enter at the same order as matter fluctuations, such that it is inconsistent to treat the background geometry classically \cite{Ford:1982wu}. More precisely, suppose gravity with coupling $G_{d}$ interacts with $N$ quantum matter fields (the matter fields need not interact themselves). Only in a large-$N$ expansion with $G_{d}N$  held fixed are graviton loops subdominant to matter loops in perturbation theory, such that the gravitational field may be effectively treated as classical, cf. \cite{Hartle:1981zt,Hu:2004gf}.\footnote{Equivalently, this semi-classical limit has $G_{d}\to0$ and $NG_{d}\equiv\ell^{d-2}$ held fixed, for length $\ell$, controlling the strength of backreaction, and spacetime dimension $d$. For $d=2$, one fixes $NG_{2}=\ell$.} In practice it is difficult to have such control in gravity (3+1)-dimensions and higher. In 2D semi-classical dilaton models, however, where dilaton gravity is coupled to a large number of conformal fields, the semi-classical field equations are solvable.\footnote{Appealing to holography, exact quantum black holes can be constructed in three-dimensions \cite{Emparan:2002px} and (in principle) higher dimensions, a consequence of working in the large-$N$ planar limit of AdS/CFT duality. See \cite{Panella:2024sor} for a review.}

There is a long history of exploring the semi-classical extensions of the CGHS model (cf. \cite{Giddings:1992fp,Strominger:1994tn,Thorlacius:1994ip,Fabbri:2005mw} for reviews), where quantum conformal matter is modelled by the 1-loop Polyakov-Liouville action. CGHS coupled to Polyakov alone breaks an important symmetry enjoyed by the classical model, rendering this semi-classical extension only solvable numerically. Exactly, analytically, solvable models, namely, Russo-Susskind-Thorlacius (RST) \cite{Russo:1992ax}  or Bose-Parker-Peleg (BPP) \cite{Bose:1995pz}, arise from restoring this gauge symmetry at the quantum level. These models, particularly RST, have been well-studied, especially how they characterize evaporating black holes and whether the information puzzle has a resolution in semi-classical gravity \cite{Fiola:1994ir}. After a decades long dormant period, semi-classical 2D dilaton gravity had a revival in large part due to the advent of the `island rule’ \cite{Penington:2019npb,Almheiri:2019psf,Almheiri:2019hni}, a prescription that explicitly shows the entropy of Hawking radiation follows a unitary Page curve \cite{Gautason:2020tmk,Hartman:2020swn}.

Less explored, is a one-parameter family of semi-classical extensions of CGHS interpolating between the RST and BPP models \cite{Fabbri:1995bz,Cruz:1995zt}. In this article we fill this gap, uncovering a number of features which we now briefly highlight: 
\begin{itemize}
\item \emph{Quasi-local energy and thermodynamics:} A peculiarity of the classical CGHS eternal black hole is that the Gibbons-Hawking prescription \cite{Gibbons:1976ue} needs to be appended to attain sensible thermodynamics. More specifically, the on-shell Euclidean action (including a local counterterm to remove infrared divergences) for the eternal black hole vanishes identically. To remedy this, one may introduce an auxiliary finite boundary a la York \cite{York:1986it} and derive quasi-local thermodynamics from the tree-level partition function, e.g., \cite{Davis:2004xi,Bergamin:2007sm,Grumiller:2007ju}. We adapt this formalism to exact quantum black hole solutions of the semi-classical interpolating model -- deriving novel local counterterms in the process -- and derive their quasi-local Brown-York energy, and microcanonical entropy. We show the Iyer-Wald entropy is exactly equal to the sum of a gravi-dilaton contribution plus the von Neumann matter entropy. That is, the semi-classical Wald entropy is equal to the generalized entropy 
\beq
\label{eq:intro3}S_{\text{Wald}}=S_{\text{grav}}+S_{\text{vN}}=S_{\text{gen}}\;,\eeq
as anticipated in \cite{Pedraza:2021cvx}, but only shown explicitly for semi-classical Jackiw-Teitelboim (JT) gravity. Interestingly, for the eternal black hole solution to the whole family of analytically solvable models, the entropy evaluates to
\beq S_{\text{Wald}}=S_{\text{classical}}+S_{\text{thermal gas}}\;,\label{eq:swaldintro}\eeq
where $S_{\text{classical}}$ is exactly equal to the entropy of the classical CGHS black hole. 
Finally, evaluating the heat capacity, we show the quantum black hole, over a certain range of parameter space, is thermally stable. 
\item \emph{Numerical quantum black holes:} 
When we consider quantum backreaction on the classical CGHS background, the gauge symmetry responsible for analytic solvability breaks. We are left with two coupled second-order ordinary differential equations and a first-order constraint for the dilaton and the metric, which can only be solved numerically. We focus on static thermal states, which are fixed by imposing regularity at the horizon. A new feature of the resulting black hole solutions is that the Hawking temperature depends in a non-trivial way on the initial condition chosen for the dilaton value at the horizon \(\phi_{H}\). Another relevant aspect of these equations is that for large \(|\phi_{H}|\), the classical solution is recovered. Exploiting this fact, we have constructed an approximate analytic solution based on an expansion in the small parameter $\frac{N}{12}e^{2\phi_H}$. The resulting approximation agrees well with numerical solutions and captures the leading semi-classical corrections. We have extended the quasi-local Hamiltonian formalism to evaluate the ADM mass of these solutions and evaluated the entropy via the Iyer-Wald prescription. The breaking of the classical gauge symmetry directly affects the black hole entropy in this model. For semi-classical CGHS, we find that the black hole entropy does not evaluate to \eqref{eq:intro3}. Instead, we find corrections that, at first order in a large-\(|\phi_{H}|\) expansion, are proportional to $1+2\phi_{H}$.

\item \emph{Naked singularities:} We analyze the singularity structure of the quantum black holes when the quantum matter is in the Hartle-Hawking and the Boulware vacua. Here, singularities are defined as spacetime points where the Ricci scalar diverges. Naively, in either vacuum, the semi-classical models admit solutions with naked singularities, singularities not hidden behind a horizon. We show that the existence of such naked singularities are primarily incompatible with the semi-classical regime of validity, $N\ll M/\lambda$ where $M$ is a parameter associated with the ADM mass of the black hole and $\lambda$ the length scale of the problem. Notably, while no naked singularities in the Hartle-Hawking vacuum are semi-classically valid, we find a large mass regime consistent with the semi-classical approximation where naked singularities exist in the Boulware vacuum. This constitutes a violation of weak cosmic censorship in (1+1) spacetime dimensions.  
\end{itemize}

\noindent The remainder of this article is as follows. In Section \ref{sec:CGHSbhs} we review the classical CGHS model of dilaton gravity, introducing the quasi-local Brown-York tensor to compute the ADM mass and entropy of static black holes. We then move to exactly solvable models of semi-classical gravity in Section \ref{sec:BPPbhs} focusing on a model that interpolates between the well-known RST and BPP theories of dilaton-gravity. After detailing the structure of the exact quantum black hole solutions, we examine the null and averaged null energy conditions of this family of theories, as well as explore the singularity structure of the quantum black holes. The horizon thermodynamics of these exact black hole solutions are analyzed in Section \ref{sec:massentqbhs}, using, for the first time, quasi-local methods. In Section \ref{sec:numericalqBHs} we numerically construct quantum black hole solutions to the semi-classical CGHS model, also numerically computing the thermodynamic quantities. We conclude in Section \ref{sec:discussion} including a discussion of future work. To keep the article self-contained, we have also included several appendices.

\vspace{5mm}

\noindent \textbf{Conventions.} When analyzing the Lorentzian solutions we use the metric signature $(-,+)$. We work in units where $c=\hbar=1$. The gravitational constant in $d$ dimensions is denoted as $G_d$.

\section{CGHS gravity and classical black holes} \label{sec:CGHSbhs}

\noindent Here we summarize the essentials of the Callan, Giddings, Harvey, Strominger (CGHS) model of two-dimensional dilaton gravity \cite{Callan:1992rs}, including its solubility and black hole solutions. We then analyze the thermal description of CGHS black holes, where, for consistency, we find it necessary to introduce a timelike boundary, which we choose to obey Dirichlet boundary conditions.

\subsection{CGHS gravity}

\noindent The CGHS model of two-dimensional dilaton gravity \cite{Callan:1992rs} arises from a spherical dimensional reduction of four-dimensional extremal magnetically charged black holes with a dilaton. A dilaton is a scalar field, in four-dimensions coupled in a specific way with other fields, that represents a low--energy limit of string theory. The main idea is that the area of the transverse codimension-2 sphere goes like $e^{-2\phi}$ so the dilaton in two-dimensions incorporates elements of the four-dimensional theory. For more details see \cite{Witten:1991yr,Giddings:1992kn,Callan:1992rs} and the review \cite{Harvey:1992xk}.

The CGHS model is described by the Lorentzian action
\beq I_{\text{CGHS}}=\frac{1}{2\pi}\int_{\mathcal{M}} d^{2}x\sqrt{-g}e^{-2\phi}(R+4(\nabla\phi)^{2}+4\lambda^{2})
+I_{\text{mat}}\;,\label{eq:CGHSLor}\eeq
with 
\beq I_{\text{mat}}=-\frac{1}{4\pi}\int_{\mathcal{M}} d^{2}x\sqrt{-g}\sum_{i=1}^{N}(\nabla f_{i})^{2}\;.\label{eq:matactCGHS}\eeq
Here $\phi$ is the dilaton, $\{f_{i}\}$ denote $N$ minimally-coupled matter (massless scalar)  fields, and $\lambda^{2}$ serves as a cosmological constant, with $\lambda^{-1}$ a length scale.

The effective two-dimensional Newton's constant is identified as the coefficient of the Ricci scalar, 
\beq
\frac{1}{16\pi G_{2}}=\frac{1}{2\pi}e^{-2\phi} \,.
\eeq
To have a well-posed variational problem for spacetimes $\mathcal{M}$ with boundary $\partial M$ obeying Dirichlet boundary conditions (where the fields at the boundary are fixed), the action (\ref{eq:CGHSLor}) is supplemented by a Gibbons-Hawking-York boundary term, 
\beq I_{\text{GHY}}=\frac{1}{\pi}\int_{\partial M} dt\sqrt{-\gamma}e^{-2\phi}K\;,\label{eq:GHYtermCGHS}\eeq
where $\gamma_{\mu\nu}$ is the induced metric endowed on the (spatial) boundary $\partial M$ and $K$ is the trace of the extrinsic curvature.  

The metric, dilaton, and matter field equations are, respectively,
\beq 
\begin{split}
T_{\mu\nu}^{\text{mat}}&=4e^{-2\phi}\left(\nabla_{\mu}\nabla_{\nu}\phi-g_{\mu\nu}\Box\phi+g_{\mu\nu}(\nabla\phi)^{2}-g_{\mu\nu}\lambda^{2}\right)\;,\\
0&=e^{-2\phi}[4\lambda^{2}+4\Box\phi-4(\nabla\phi)^{2}+R]\;,\\
0&=\Box f_{i}\;,
\end{split}
\label{eq:CGHSeoms}\eeq
where the classical matter stress-tensor is 
\beq T_{\mu\nu}^{\text{mat}}\equiv-\frac{4\pi}{\sqrt{-g}}\frac{\delta I_{\text{mat}}}{\delta g^{\mu\nu}}=\sum_{i=1}^{N}\left[(\nabla_{\mu}f_{i})(\nabla_{\nu}f_{i})-\frac{1}{2}g_{\mu\nu}(\nabla f_{i})^{2}\right]\;,
\label{eq:classstresstens}\eeq
with matter action $I_{\text{mat}}$ (\ref{eq:matactCGHS}). Observe that $g^{\mu\nu}T_{\mu\nu}^{\text{mat}}=0$, and so
\beq 0=4e^{-2\phi}(-\Box\phi+2(\nabla\phi)^{2}-2\lambda^{2})\;.\label{eq:vanstresstenstrace}\eeq
Combined with the dilaton equation of motion (\ref{eq:CGHSeoms}) yields $R=-2\Box\phi$.

Notably, the CGHS model is exactly solvable. To wit, move to conformal gauge
\beq ds^{2}=-e^{2\rho}dw^{+}dw^{-}\;,\label{eq:confgauge}\eeq
with lightcone coordinates $w^{\pm}$ for a set of Minkowski coordinates $(w^{0},w^{1})$, and $\rho=\rho(w^{+},w^{-})$. In this gauge, the action (\ref{eq:CGHSLor}) becomes (reviewed in Appendix \ref{app:eoms}) 
\beq 
I_{\text{CGHS}}=\frac{1}{\pi}\int dw^{+}dw^{-}\biggr[2\partial_{-}(\phi-\rho)(\partial_{+}e^{-2\phi})+\lambda^{2}e^{2(\rho-\phi)}+\frac{1}{2}\sum_{i=1}^{N}(\partial_{+}f_{i})(\partial_{-}f_{i})\biggr]\;.
\label{eq:CGHSrewriteST}\eeq
The dilaton and matter equations of motion  (\ref{eq:CGHSeoms})  become
\beq
\begin{split}
0&=e^{-2(\rho+\phi)}\left[2\partial_{+}\partial_{-}\rho+\lambda^{2}e^{2\rho}-4\partial_{+}\partial_{-}\phi+4(\partial_{+}\phi)(\partial_{-}\phi)\right]\;,\\
0&=e^{-2\rho}\partial_{+}\partial_{-}f_{i}\;.
\end{split}
\label{eq:dilamatEOMconf}\eeq
The metric equations, meanwhile, are
\beq 
\begin{split}
 &T^{\text{mat}}_{\pm\pm}= 4e^{-2\phi}\left[\partial_{\pm}^{2}\phi-2(\partial_{\pm}\rho)(\partial_{\pm}\phi)\right]\;,\\
 & T^{\text{mat}}_{\pm\mp}=8e^{-2\phi}(\partial_{+}\phi)(\partial_{-}\phi)+2\lambda^{2}e^{2(\rho-\phi)}-4e^{-2\phi}\partial_{+}\partial_{-}\phi\;,
\end{split}
\label{eq:meteomCGHSconfgauge}\eeq
with stress-tensor components (\ref{eq:classstresstens})
\beq T_{\pm\pm}^{\text{mat}}=\sum_{i=1}^{N}(\partial_{\pm}f_{i})(\partial_{\pm}f_{i})\;,\quad T_{\pm\mp}^{\text{mat}}=0\;.\eeq
From here, a general solution to the field equations (\ref{eq:dilamatEOMconf}) and (\ref{eq:meteomCGHSconfgauge}) may be expressed as integrals over two free-field equations \cite{Callan:1992rs}. 

Solutions of such generality are cumbersome to work with. More convenient solutions can be obtained using that the conformal gauge (\ref{eq:confgauge}) does not fix the conformal subgroup of diffeomorphisms. Indeed, varying (\ref{eq:CGHSrewriteST}) with respect to $\rho$ gives 
\beq 0=e^{-2\phi}\left[4(\partial_{-}\phi)(\partial_{+}\phi)-2\partial_{+}\partial_{-}\phi+\lambda^{2}e^{2\rho}\right]\;.\label{eq:rhoeomCGHS}\eeq
Subtracting the $\rho$ equation from the $\phi$ equation of motion (\ref{eq:dilamatEOMconf}) yields
\beq 
0=2e^{-2(\rho+\phi)}\partial_{+}\partial_{-}(\rho-\phi)\;.\label{eq:freefieldeq}\eeq
This implies $\Box(\rho-\phi)=0$ and such that the combination $(\rho-\phi)$ is a free field, with general solution $(\rho-\phi)=\upsilon_{+}(w^{+})+\upsilon_{-}(w^{-})$ for arbitrary gauge functions $\upsilon_{\pm}(w^{\pm})$. There is thus a residual gauge symmetry, with conserved current $j^{\mu}=\partial^{\mu}(\phi-\rho)$. The \emph{Kruskal gauge} sets $\upsilon_{\pm}=0$, i.e., $\rho=\phi$. 

Note that the gauge choice $\rho=\phi$ does not imply that $\rho$ transforms as a scalar field; it transforms as a component of the metric. Working in Kruskal gauge results in simple closed form eternal and dynamical black holes solutions.

\subsection{Black hole solutions}\label{subsec:bhclas}

\noindent In Kruskal gauge, $\rho=\phi$, the metric equations (\ref{eq:meteomCGHSconfgauge}) simplify to 
\beq
\begin{split}
&T^{\text{mat}}_{\pm\pm}=-2\partial_{\pm}^{2}(e^{-2\phi})\;,\\
&T_{\pm\mp}^{\text{mat}}=2\partial_{+}\partial_{-}(e^{-2\phi})+2\lambda^{2}\;.
\end{split}
\label{eq:metCGHSeomKruskal}\eeq
Solving this system of equations, along with the matter equations, leads to eternal and dynamical black hole geometries, as we now briefly review (see \cite{Harvey:1992xk,Strominger:1994tn,Thorlacius:1994ip,Fabbri:2005mw} for additional details). 

\vspace{2mm}

In vacuum, where $f_{i}=0$, the metric equations (\ref{eq:metCGHSeomKruskal}) are easily solved, yielding,
\beq  ds^{2}=-e^{2\phi}dx^{+}dx^{-}\;,\quad e^{-2\phi}=\frac{M}{\lambda}-\lambda^{2}x^{+}x^{-}\;,\label{eq:eternalBH}\eeq
where we designate the coordinate frame in Kruskal gauge with null coordinates $(x^{\pm})$, and $M$ is an integration constant related to the mass of the system (see below). With matter, $f_{i}\neq0$, one can construct exact solutions of \emph{dynamical} black hole, i.e., those formed under gravitational collapse. This is a notable advantage of two-dimensional models of dilaton gravity, however, we will not explore such solutions in this article. 

 $\bullet$ \emph{Linear dilaton vacuum.} When $M=0$, the geometry (\ref{eq:eternalBH}) is dubbed the linear dilaton vacuum. It is easy to show the Ricci scalar is $R=-2\Box\phi=0$ everywhere, such that the vacuum geometry is flat. This is made manifest via the coordinate transformation $\lambda x^{\pm} =\pm e^{\pm\lambda\sigma^{\pm}}$ for coordinates $(\sigma^{+},\sigma^{-})$ such that $e^{-2\phi}=e^{\lambda(\sigma^{+}-\sigma^{-})}$ and $ds^{2}=d\sigma^{+}d\sigma^{-}$. Introducing temporal and spatial coordinates $\sigma^{0}=\frac{1}{2}(\sigma^{+}+\sigma^{-})$ and $\sigma^{1}=\frac{1}{2}(\sigma^{+}-\sigma^{-})$, it is apparent that
\beq
\phi=-\lambda \sigma^{1} \,. 
\eeq
Combined with treating $e^{2\phi}$ as an effective gravitational coupling constant, regimes of strong and weak coupling coincide with $\sigma^{1}\to-\infty$ and $\sigma^{1}\to+\infty$, respectively.

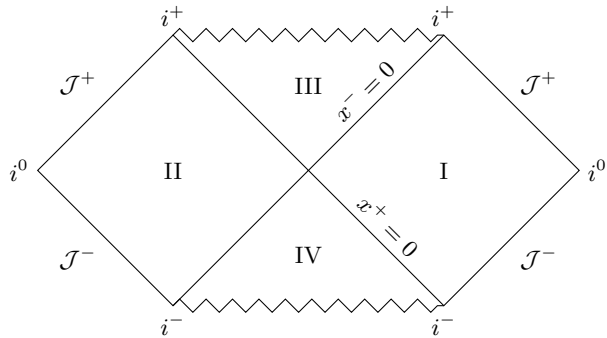
\begin{figure}[t!]
	\centering
	\begin{tikzpicture}[scale=0.45]
		\node (I)    at ( 4,0)   {I};
		\node (II)   at (-4,0)   {II};
		\node (III)  at (0, 2.5) {III};
		\node (IV)   at (0,-2.5) {IV};
		
		\path  
		(II) +(90:4)  coordinate[label=90:$i^+$]  (IItop)
		+(-90:4) coordinate[label=-90:$i^-$] (IIbot)
		+(0:4)   coordinate                  (IIright)
		+(180:4) coordinate[label=180:$i^0$] (IIleft)
		;
		\draw (IIleft) -- 
		node[midway, above left]    {$\cal{J}^+$}
		node[midway, below, sloped] {}
		(IItop) --
		node[midway, above, sloped] {}
		(IIright) -- 
		node[midway, below, sloped] {}
		(IIbot) --
		node[midway, above, sloped] {}
		node[midway, below left]    {$\cal{J}^-$}    
		(IIleft) -- cycle;
		
		\path 
		(I) +(90:4)  coordinate[label=90:$i^+$] (Itop)
		+(-90:4) coordinate[label=-90:$i^-$] (Ibot)
		+(180:4) coordinate (Ileft)
		+(0:4)   coordinate[label=0:$i^0$] (Iright)
		;
		\draw  (Ileft) --node[midway, above, sloped] {$x^-=0$} (Itop) -- node[midway, above right]    {$\cal{J}^+$}
		node[midway, below, sloped] {}(Iright) -- node[midway, above, sloped] {}
		node[midway, below right]    {$\cal{J}^-$}(Ibot) --node[midway, above, sloped] {$x^+=0$} (Ileft) -- cycle;
		
		\draw[decorate,decoration=zigzag] (IItop) -- (Itop)
		node[midway, above, inner sep=2mm] {};
		
		\draw[decorate,decoration=zigzag] (IIbot) -- (Ibot)
		node[midway, below, inner sep=2mm] {};
	\end{tikzpicture}
	\caption{Penrose diagram for a static two-dimensional dilatonic black hole.}
	\label{fig:BHdia}
\end{figure}

$\bullet$ \emph{Eternal black hole.} For $M>0$, the geometry (\ref{eq:eternalBH}) describes an eternal black hole, see Fig. \ref{fig:BHdia}. The global event horizon consists of two null segments $\{x^{+}=0,x^{-}>0\}$ and $\{x^{-}=0,x^{+}>0\}$ and bifurcation `2-sphere' at $x^{+}x^{-}=0$, shrouding a (spacelike) curvature singularity where the Ricci curvature, 
\beq R=8e^{-2\phi}\partial_{+}\partial_{-}\phi=\frac{4M\lambda}{\frac{M}{\lambda}-\lambda^{2}x^{+}x^{-}}\;,\eeq
diverges at $x^{+}x^{-}=\frac{M}{\lambda^{3}}$. For $M<0$, the horizon disappears, leaving behind an unphysical naked (timelike) singularity. 
Notice the horizon value of the dilaton $\phi_{H}$ goes like $e^{2\phi_{H}}=\lambda/M$ such that large $M$ makes gravity arbitrarily weak at the horizon. Further, along the horizon $R=4\lambda^{2}$, independent of $M$, while in the asymptotic region $x^{+}x^{-}\to-\infty$, $R\to0$.  

In the weak-coupling regime ($\sigma^{1}\to\infty$) the black hole approaches the linear dilaton vacuum, $\phi\to-\lambda\sigma^{1}$ and $\rho\to0$, up to exponentially small corrections. 

Finally, a comment on nomenclature. The vacuum geometry (\ref{eq:eternalBH})  is the eternal CGHS black hole in \emph{Kruskal coordinates} $(x^{+},x^{-})$. Meanwhile, the coordinate transformation $\lambda x^{\pm}=\pm  e^{\pm\lambda \sigma^{\pm}}$ puts the static geometry (\ref{eq:eternalBH}) into the \emph{Eddington-Finkelstein} form
\beq 
ds^{2}=-\frac{d\sigma^{+}d\sigma^{-}}{\left(1+\frac{M}{\lambda}e^{-\lambda(\sigma^{+}-\sigma^{-})}\right)}\;, \quad e^{-2\phi}=\frac{M}{\lambda}+e^{\lambda(\sigma^{+}-\sigma^{-})}\;.
\label{eqn:EFcoord}\eeq
We will further introduce $\sigma^{\pm}=(\sigma^{0}\pm\sigma^{1})$ for time and spatial coordinates $\sigma^{0}$ and $\sigma^{1}$, respectively. When $M>0$, the Eddington-Finkelstein geometry asymptotically approaches the (flat) linear dilaton vacuum. In these coordinates the bifurcation point is located at $(\sigma^{+}-\sigma^{-})\propto\sigma^{1}\to-\infty$ such that these coordinates only cover the exterior of the black hole. For additional details about coordinate systems, see Appendix \ref{app:coordsys}.

\subsection{Asymptotic and quasi-local mass} 

\subsubsection{Notions of mass}

\noindent The notion of mass in general relativity is not local as in (non-gravitational) classical mechanics; it cannot be cast as a volume integral over some local mass density along a hypersurface, a consequence of the equivalence principle. In fact, for stationary and asymptotically flat spacetimes, we discuss two notions of mass: 
\begin{itemize}
    \item Arnowitt--Deser--Misner (ADM) mass \cite{Arnowitt:1962hi}: Originally defined using an ADM split of the spacetime in the Hamiltonian formulation of general relativity, the ADM mass $M_{\text{ADM}}$ is expressed as a proper flux integral over a codimension-2 sphere at asymptotic spatial infinity. Specifically, in four-dimensional general relativity, in an asymptotically Euclidean coordinate system on a spacelike hypersurface $\Sigma$ (cf. Eq. (11.2.14) of \cite{Wald:1984rg}),\footnote{Another common expression for the ADM mass is (cf. Eq. (4.81) of \cite{Poisson:2009pwt}) $M_{\text{ADM}}=-\frac{1}{8\pi G}\int_{S_{\infty}}(k-k_{0})\sqrt{\sigma}d^{2}\theta$, where $\sigma_{AB}$ is the metric on the codimension-2 sphere at spatial infinity, $k=\sigma^{AB}k_{AB}$ is the trace of the extrinsic curvature of $S_{\infty}$ embedded in a codimension-1 spacelike slice, while $k_{0}$ is the extrinsic curvature of $S_{\infty}$ embedded in flat space. This form of the ADM mass directly follows from the gravitational Hamiltonian upon imposing the Hamiltonian and momentum constraints.} 
    \beq M_{\text{ADM}}=\frac{1}{16\pi G}\oint_{S_{\infty}}(\partial^{\beta}h_{\alpha\beta}-\delta^{\gamma \delta}\partial_{\alpha}h_{\gamma\delta})n^{\alpha}dS\;,\label{eq:defnADMGR}\eeq
    for outward pointing unit normal $n^{\mu}$ to the codimension-2 sphere $S_{\infty}$ at spatial infinity, and $h_{\mu\nu}=g_{\mu\nu}-\eta_{\mu\nu}$ represents asymptotic deviations of the full metric $g_{\mu\nu}$ from the flat metric $\eta_{\mu\nu}$. If the asymptotic geometry is time-translation invariant, the ADM mass is a globally conserved `charge', defined with respect to the asymptotic symmetries at spatial infinity \cite{Regge:1974zd}. 
    \item Komar mass \cite{Komar:1958wp}: For stationary spacetimes, such that there exists a time-like Killing vector $\zeta^{\mu}_{(t)}$ (with normalization $\zeta^{2}=-1$ at spatial infinity) it is written as a surface integral over a codimension-2 sphere at spatial infinity (in $d$-dimensional spacetime)
    \beq M_{\text{Komar}}=-\frac{1}{8\pi G_{d}}\oint_{S_{\infty}}dS_{\alpha\beta}\nabla^{\alpha}\zeta^{\beta}_{(t)}=\frac{1}{8\pi G_{d}}\oint_{S_{\infty}}d^{d-2}x\sqrt{\Theta}u_{\alpha}n_{\beta}\nabla^{\alpha}\zeta^{\beta}_{(t)}\;,\label{eq:komarmassGR}\eeq
    for binormal area-element $dS_{\alpha\beta}=-\frac{1}{2}(u_{\alpha}n_{\beta}-u_{\beta}n_{\alpha})d^{d-2}x\sqrt{\Theta}$ with (future pointing) timelike and spacelike unit normals $u^{\alpha}$ and $n^{\alpha}$, respectively, and $\Theta_{AB}$ is the induced metric on the codimension-2 sphere. For stationary spacetimes, the ADM mass is equivalent to the Komar mass. 
\end{itemize}

In addition to these global definitions of mass, defined as surface integrals at asymptotic spatial or null infinity, there also exist quasi-local notions of mass, i.e., mass associated to gravitating systems of finite extent. Originally listed as an open problem in classical general relativity \cite{Penrose1982SOMEUP}, by now there are a host of quasi-local definitions of mass  (see \cite{Szabados:2009eka} for a review). We will focus on the quasi-local definition proposed by Brown and York \cite{BY1992og,Brown:1992br} (see also \cite{Hawking:1995fd,Hawking:1996ww}). In particular, for a finite timelike boundary $B$ obeying Dirichlet boundary conditions,
with induced metric $\gamma_{\mu\nu}$ and extrinsic curvature $K_{\mu\nu}$, that intersects a spatial hypersurface $\Sigma$, the Brown-York Hamiltonian in general relativity can be cast as
\beq H_{\zeta}=\oint_{\partial\Sigma}\zeta^{\mu}u^{\nu}\tau_{\mu\nu}\;,\qquad \tau_{\mu\nu}\equiv \frac{2}{\sqrt{-\gamma}}\frac{\delta I_{\text{cl}}}{\delta \gamma^{\mu\nu}}=-\frac{1}{8\pi G}(K_{\mu\nu}-K\gamma_{\mu\nu})\;.\eeq
Here $\zeta^{\mu}$ is a diffeomorphism generating vector
field and $\epsilon_{\partial\Sigma}$ is the volume element of the codimension-2 boundary $\partial\Sigma$ of $\Sigma$. Further, $\tau_{\mu\nu}$ is the quasi-local Brown-York stress-tensor \cite{Brown:1992br}, defined as the variation of the on-shell action $I_{\text{cl}}$ of the finite subregion (a functional of the induced metric on the boundary). Notably, $H_{\zeta}$ has the form of a generator of a boundary isometry with Killing vector $\zeta$; indeed, $\zeta$ approaches a Killing vector of the spatial boundary metric. By appropriately choosing the vector field $\zeta$, one obtains the Brown-York quasi-local energy and angular momentum. Note that it can be shown the Brown-York Hamiltonian coincides with the ADM Hamiltonian in the limit where the finite boundary $B$ is sent to spatial infinity in an asymptotically flat spacetime. The notion of quasi-local Brown-York charges has also been extended to the case when $B$ is a null hypersurface, cf. \cite{Brown:1996bw,Booth:2001gx,Donnay:2019jiz,Jafari:2019bpw,Chandrasekaran:2021hxc,Odak:2023pga}.

Since the CGHS model characterizes two-dimensional asymptotically flat spacetimes, it is possible to define analogues of the ADM mass. In particular, for ADM mass, historically, the first approach was to construct a pseudotensor from the linearization of the metric equations of motion due to deviations of the metric $g_{\mu\nu}$ and the dilaton $\phi$ at asymptotic spatial infinity \cite{Witten:1991yr} (see also \cite{Callan:1992rs,Park:1992sd}). We review this method in Appendix \ref{app:Waldformalism}. A Noether charge method may be employed to attain a different form of the stress-energy pseudotensor \cite{Navarro-Salas:1994cyz}. Another approach is to construct the ADM Hamiltonian, first accomplished in \cite{Bilal:1993wm}, working directly in conformal gauge.

Below we will instead use covariant phase space methods \cite{Crnkovic:1986ex,Lee:1990nz,Wald:1993nt,Iyer:1994ys} to construct the analogue of the quasi-local Brown-York Hamiltonian \cite{Brown:1992br}, which recovers the ADM energy at spatial infinity. We summarize this formalism and provide additional details in Appendix \ref{app:Waldformalism}.

\subsubsection{Brown-York and ADM energy}

\noindent The quasi-local Brown-York stress-tensor$\tau_{\mu\nu}$ was computed for arbitrary two-dimensional dilaton theories of gravity in \cite{Pedraza:2021cvx} (see Appendix \ref{app:Waldformalism}). Adapting \eqref{eq:BYstresstensandP} to the case of the classical CGHS model, we find 
\beq \tau^{\mu\nu}_{\text{CGHS}}=-\frac{2}{\pi}e^{-2\phi}\gamma^{\mu\nu}n^{\alpha}\partial_{\alpha}\phi\;,\quad \varepsilon_{\text{CGHS}}=\frac{2}{\pi}e^{-2\phi}n^{\alpha}\partial_{\alpha}\phi\;,\label{eq:BYstressCGHS}\eeq
where $\varepsilon_{\text{CGHS}}\equiv u_{\mu}u_{\nu}\tau^{\mu\nu}_{\text{CGHS}}$ is the quasi-local energy density, cf. (\ref{eq:quasilocalen}). Thus, the quasi-local Hamiltonian is 
\beq H_{\zeta}=-\frac{2}{\pi}\oint_{\partial\Sigma}\epsilon_{\partial\Sigma}e^{-2\phi}g_{\mu\nu}\zeta^{\mu}u^{\nu}(n^{\alpha}\partial_{\alpha}\phi)=\frac{2}{\pi}\oint_{\partial\Sigma}\epsilon_{\partial\Sigma}\mathcal{N}e^{-2\phi}n^{\alpha}\partial_{\alpha}\phi\;,\label{eq:BYHamilCGHS}\eeq
where $\mathcal{N}\equiv-g_{\mu\nu}\zeta^{\mu}u^{\nu}$ is the lapse, with $\zeta^{\mu}$ is normalized such that $\mathcal{N}=1$ at asymptotic spatial infinity. Since the bulk spacetime is two-dimensional, the integral over the codimension-2 surface $\partial\Sigma$ in $H_{\zeta}$ amounts to evaluating the integrand at the location where the finite timelike boundary intersects the spacelike hypersurface $\Sigma$. 

As in standard general relativity, the ADM mass follows from sending $\partial\Sigma\to S_{\infty}$ and $\mathcal{N}\to1$. Indeed, in this limit $H_{\zeta}$ precisely matches the on-shell Hamiltonian constructed in \cite{Carrasco:2023fcj} following a standard Hamiltonian formulation.\footnote{In particular, consider the boundary contribution to the on-shell ADM Hamiltonian in Eq. (C.22) of \cite{Carrasco:2023fcj}, with unit lapse and vanishing shift and $f=e^{-2\phi}/2\pi$.} As written, however, the ADM mass is divergent. Indeed, consider the eternal CGHS black hole 
in Eddington-Finkelstein coordinates, where $\zeta^{\mu}=\delta^{\mu}_{0}=e^{\rho}u^{\mu}$, $n^{\mu}=e^{-\rho}\delta^{\mu}_{1}$, such that the lapse $\mathcal{N}=e^{\rho}$. The Hamiltonian (\ref{eq:BYHamilCGHS}) is
\beq
\begin{split} 
H_{\zeta}&=\frac{2}{\pi}\oint_{\partial\Sigma}\epsilon_{\partial\Sigma}e^{-2\phi}\partial_{1}\phi=-\frac{2}{\pi}\oint_{\partial\Sigma}\epsilon_{\partial\Sigma}e^{2\lambda\sigma^{1}}\lambda\;,
\end{split}
\label{eq:ADMHamiltonianWOsub}\eeq
which diverges (to $-\infty$) at spatial infinity, $\sigma^{1}\to+\infty$. Further, the integrand coincides with that for the linear dilaton vacuum. Hence, using a background subtraction method to regulate the divergence by subtracting the linear dilaton vacuum would, peculiarly, lead to a vanishing ADM energy.

To ameliorate the situation, we thus introduce a local boundary  counterterm to the Gibbons-Hawking-York boundary action (\ref{eq:GHYtermCGHS}), 
\beq I_{\text{ct}}=-\frac{1}{\pi}\int_{B}dt\sqrt{-\gamma}e^{-2\phi}(2\lambda)\;.\label{eq:localctCGHS}\eeq
 For higher-dimensional asymptotically flat backgrounds, ordinarily background subtraction is sufficient to deal with long range infrared divergences, while there is not a generally accepted diffeomorphism invariant local boundary counterterm method (in contrast with asymptotically anti de Sitter backgrounds). We will return to this notable difference for two-dimensional dilaton black holes when we analyze the thermodynamics. For now, we recall that, at the level of the CGHS action, there is an effective cosmological constant, and in that sense it is not so surprising a counterterm method works for two-dimensional dilaton black holes. Indeed, the local counterterm (\ref{eq:localctCGHS}) has the same form as the local counterterm used in the context of AdS$_{2}$ Jackiw-Teteilboim gravity (see, e.g., Eq. (2.2) of \cite{Pedraza:2021cvx}). The counterterm can be derived using Hamilton-Jacobi methods \cite{Davis:2004xi,Bergamin:2007sm}, which we review in Appendix \ref{app:Waldformalism}. 

 With the counterterm (\ref{eq:localctCGHS}), the Brown-York stress-tensor and energy density are modified to
\beq \tau^{\mu\nu}_{\text{CGHS}}=-\frac{2}{\pi}e^{-2\phi}\gamma^{\mu\nu}\left(n^{\alpha}\partial_{\alpha}\phi+\lambda\right)\;,\quad \varepsilon_{\text{CGHS}}=\frac{2}{\pi}e^{-2\phi}\left(n^{\alpha}\partial_{\alpha}\phi+\lambda\right)\;,
\label{eq:BYtensCGHS}\eeq
while the Hamiltonian becomes 
\beq H_{\zeta}=\frac{2}{\pi}\oint_{\partial\Sigma}\epsilon_{\partial\Sigma}\mathcal{N}e^{-2\phi}(n^{\alpha}\partial_{\alpha}\phi+\lambda)\;.\label{eq:quasiHamCGHS}\eeq
We now notice the energy density $\varepsilon_{\text{CGHS}}=0$ in the linear dilaton vacuum. Meanwhile, working in Eddington-Finkelstein gauge, asymptotically, 
\beq 
\begin{split}
\lim_{\sigma^{1}\to+\infty}\varepsilon_{\text{CGHS}}
&=\lim_{\sigma^{1}\to+\infty}\frac{2}{\pi}e^{2\lambda\sigma^{1}}\left(\lambda+Me^{-2\lambda\sigma^{1}}-\lambda e^{-\rho}\right)\\
&\approx \lim_{\sigma^{1}\to+\infty} \frac{2}{\pi}e^{2\lambda\sigma^{1}}\left(\lambda+ Me^{-2\lambda\sigma^{1}}-\lambda\left[1+\frac{1}{2}\frac{M}{\lambda}e^{-2\lambda\sigma^{1}}\right]\right)\\
&=\frac{M}{\pi}\;.
\end{split}
\eeq
Thus, the ADM energy for the classical CGHS black hole is
\beq M_{\text{ADM}}=\lim_{\sigma^{1}\to+\infty}H_{\zeta}\;,\eeq
such that $M=\pi M_{\text{ADM}}$.\footnote{The factor of $\pi$ difference between $M$ and $M_{\text{ADM}}$ can resolved by defining the Brown-York stress-tensor to be $\tau^{\mu\nu}=\frac{2\pi}{\sqrt{-\gamma}}\frac{\delta I}{\delta\gamma^{\mu\nu}}$, resulting in $M_{\text{ADM}}=M$.} 

Further, notice that expanding $\phi=-\lambda\sigma^{1}+\varphi$, for $\varphi$ an asymptotic deviation of the linear dilaton vacuum, it follows
\beq H_{\zeta}=\frac{2}{\pi}\oint_{\partial\Sigma}\epsilon_{\partial\Sigma}e^{2\lambda\sigma^{1}}(-\lambda+\partial_{1}\varphi+\lambda e^{\rho})\;,\eeq
at leading order in $\varphi$. Subsequently taking the limit $\sigma^{1}\to+\infty$ recovers the familiar form of the ADM energy \cite{Witten:1991yr} computed via a pseudotensor method, i.e., 
\beq M_{\text{ADM}}=\frac{2}{\pi}\oint_{\partial\Sigma}\epsilon_{\partial\Sigma}e^{2\lambda\sigma^{1}}(\partial_{1}\varphi+\lambda\rho)\biggr|_{\sigma^{1}\to+\infty}\;,\eeq
where we treated $\rho$ as small in this limit. 

Away from spatial infinity, the quasi-local energy associated with an observer at the finite timelike boundary $B$ is given by the Hamiltonian (\ref{eq:quasiHamCGHS}). Specifically, for the eternal black hole in Schwarzschild coordinates (\ref{eq:schwargaugeapp})
\beq
\begin{split} 
&ds^{2}=-f(r)(d\sigma^{0})^{2}+f^{-1}(r)dr^{2}\;,\qquad f(r)=\left(1-\frac{M}{\lambda}e^{-2\lambda r}\right)\;,\\
&e^{-2\phi}=e^{2\lambda r}\;,\label{eq:schwargauge} 
\end{split}
\eeq
the Brown-York quasi-local energy (density) is 
\beq E_{\text{BY}}\equiv \varepsilon_{\text{CGHS}}=\frac{2\lambda e^{2\lambda r_{B}}}{\pi}\left(1-\mathcal{N}(r_{B})\right)\;,\quad \mathcal{N}(r_{B})=\sqrt{1-\frac{M}{\lambda}e^{-2r_{B}\lambda}}\;,\label{eq:BYeneternalCGHS}\eeq
where $r=r_{B}$ denotes the location of the finite Dirichlet boundary. In Schwarzschild coordinates the horizon is located at $e^{2\lambda r_{H}}=M/\lambda$ (the positive root of $f(r_{H})=0$), such that the lapse is $\mathcal{N}(r_{B})=\sqrt{1-e^{2\lambda(r_{H}-r_{B})}}$. Thus, when the Dirichlet boundary is coincident with the horizon, the quasi-local energy of the eternal CGHS black hole vanishes. From the viewpoint of imposing Dirichlet boundary conditions on fields, we can opt to rewrite the quasi-local Brown-York energy in terms of the value of the dilaton at the boundary, $\phi_{B}$, 
\beq E_{\text{BY}}=\frac{2\lambda}{\pi}e^{-2\phi_{B}}(1-\mathcal{N}(\phi_{B}))\;,\label{eq:BYeneternalCGHSv2}\eeq
with $\mathcal{N}(\phi_{B})=(1-\frac{M}{\lambda}e^{2\phi_{B}})^{1/2}$. It is easy to verify that in the limit the finite boundary is pushed to asymptotic spatial infinity, $\phi_{B}\to-\infty$, we recover $E_{\text{BY}}\to M_{\text{ADM}}$.

\subsection{Classical thermodynamics}

\noindent Stationary black holes are known to obey a set of four laws that serve as mechanical analogues to the four laws of thermodynamics \cite{Bardeen:1973gs}. In particular, the first law relates the variation of the ADM mass induced by a perturbation of the metric (and matter fields on the background should there be any), to the variation of the black hole event horizon area, $\delta M_{\text{ADM}}=\frac{\kappa}{2\pi}\delta(A_{H}/4G_{\text{N}})$, where the proportionality constant $\kappa$ is the surface gravity evaluated on the horizon.\footnote{The surface gravity $\kappa$ is defined via $\nabla_{\mu}\xi^{2}=-2\kappa\xi_{\mu}$, for Killing horizon generator $\xi^{\mu}$ (with normalization chosen such that $\kappa$ is positive on the future Killing horizon).} It is easy to verify that for the eternal CGHS black hole (\ref{eq:schwargauge}) the Killing vector $\zeta^{\mu}=\partial^{\mu}_{\sigma^{0}}$ generates the Killing horizon with surface gravity $\kappa=\frac{1}{2}|f'(r_{H})|=2Me^{-2\lambda r_{H}}=\lambda$. Notably, the surface gravity is independent of the mass parameter $M$. Consequently, the first law of CGHS black hole mechanics trivially reads
\beq \delta M_{\text{ADM}}=\frac{1}{\pi}\delta M=\frac{\kappa}{2\pi}\delta\left(\frac{2M}{\lambda}\right)\;.\label{eq:classfirstlawatinft}\eeq
Note that $2M/\lambda$ is equal to the horizon ``area''. Indeed, using $\frac{1}{4G_{2}}=2e^{-2\phi}$, and $A=1$ for the area of a point, then $(\frac{A}{4G_{2}})|_{H}=2e^{-2\phi_{H}}=\frac{2M}{\lambda}$. 
Meanwhile, notice the first order variation of the Brown-York energy (\ref{eq:BYeneternalCGHSv2}) with respect to $M$ and $\phi_{B}$ gives the following quasi-local first law
\beq 
\begin{split} 
\delta E_{\text{BY}}
&=\frac{\lambda}{2\pi \mathcal{N}}\delta\left(\frac{2M}{\lambda}\right)-\frac{\lambda}{\pi\mathcal{N}}(\mathcal{N}-1)^{2}\delta (e^{-2\phi_{B}})\;,
\end{split}
\label{eq:quasilocfirst}\eeq
where we used $Me^{2\phi_{B}}=\lambda(1-\mathcal{N}^{2})$.

Comparing the mechanical first law (\ref{eq:classfirstlawatinft}) to the ordinary first law of thermodynamics, it is suggestive to interpret the eternal CGHS black hole as a thermodynamic system with internal energy  $M_{\text{ADM}}$, temperature $T=\kappa/2\pi$, and entropy $S=2M/\lambda$. This interpretation can be realized by appealing to the Gibbons-Hawking prescription for constructing the canonical gravitational partition function $\mathcal{Z}$ using the Euclidean gravitational path integral \cite{Gibbons:1976ue}, 
\beq \mathcal{Z}(\beta)=\text{tr}(e^{-\beta H})=\int\mathcal{D}\psi e^{-I_{E}[\psi]}\;.\label{eq:canonpartfunc}\eeq
Here $\beta$ denotes the inverse temperature, here held fixed, and $\mathcal{D}\psi$ denotes the functional integration measure over field variables $\{\psi\}=\{g_{\mu\nu},\phi\}$, subject to boundary conditions defining the canonical thermal ensemble (fixed $\beta$), and $I_{E}$ is the Euclidean gravity action. Presently, the total Euclidean  action for the CGHS model is, including the Gibbons-Hawking-York boundary and local counterterm, 
\beq I_{E}=-\frac{1}{2\pi}\int_{\mathcal{M}}d^{2}x\sqrt{g}e^{-2\phi}(R+4(\nabla\phi)^{2}+4\lambda^{2})-\frac{1}{\pi}\int_{\partial\mathcal{M}}dt_{E}\sqrt{\gamma}e^{-2\phi}(K-2\lambda)\;,\label{eq:EucactionCGHS}\eeq
where $t_{E}$ is a Euclidean time coordinate with periodicity $t_{E}\sim t_{E}+\beta$, such that the boundary of the Euclidean manifold $\partial\mathcal{M}=S^{1}$. We could append this action with the matter fields $f_{i}$, however, we will refrain doing so for the time being.

The periodicity $\beta$ is fixed so as to remove the conical singularity located at the horizon of the Euclidean black hole. Indeed, the eternal CGHS black hole in Euclideanized Rindler coordinates (\ref{eq:rindlercoordapp}), 
\beq ds^{2}_{E}=e^{2\rho}[(\lambda X)^{2}dT_{E}^{2}+dX^{2}]\;,\quad e^{-2\rho}=e^{-2\phi}=\frac{M}{\lambda}+\lambda^{2} X^{2}\;, \label{eq:EucRindlercoord}\eeq
has a conical singularity at the horizon $X=0$. The singularity is removed via the identification $T_{E}\sim T_{E}+\beta$, with
\beq \beta=\frac{2\pi}{\lambda}\;.\label{eq:betaH}\eeq
Equivalently, $\beta=\frac{2\pi}{\kappa}$ for surface gravity $\kappa$. Thus, the periodicity of the Euclidean time circle is fixed to be the inverse Hawking temperature $T^{-1}_{\text{H}}=\beta_{\text{H}}=\frac{2\pi}{\kappa}$ \cite{Hawking:1975vcx}. Unlike higher-dimensional black holes and the AdS$_{2}$ black hole in Jackiw-Teitelboim gravity, notice the temperature $T_{\text{H}}$ is independent of the black hole mass, and is simply a constant. 

The goal now is to derive the thermodynamic energy and entropy for the CGHS black hole using the gravitational partition function (\ref{eq:canonpartfunc}). In practice, this is accomplished using a saddle-point approximation, expanding the Euclidean action about its classical solutions. At leading order in a stationary phase approximation (``tree-level''), the partition function is given by the on-shell Euclidean action
\beq \mathcal{Z}(\beta)\approx e^{-I_{E}^{\text{on-shell}}}\;.\eeq
To this end, consider the two-dimensional bulk contribution to the Euclidean action (\ref{eq:EucactionCGHS}). Implementing the dilaton equation of motion (\ref{eq:CGHSeoms}) to remove the Ricci scalar in the two-dimensional bulk contribution to the Euclidean action (\ref{eq:EucactionCGHS}) yields
\beq
\begin{split} 
I_{E}^{\text{on-shell}}&=-\frac{1}{2\pi}\int_{\mathcal{M}}d^{2}x\sqrt{g}e^{-2\phi}[8(\nabla\phi)^{2}-4\Box\phi]+I_{\partial\mathcal{M}}\\
&=\frac{1}{\pi}\int_{\mathcal{M}}d^{2}x\sqrt{g}\nabla^{\mu}[2e^{-2\phi}(\nabla_{\mu}\phi)]+I_{\partial\mathcal{M}}\;,
\end{split}
\eeq
for boundary action $I_{\partial\mathcal{M}}$, and to arrive to the second line we used $e^{-2\phi}(-4\Box\phi)=\nabla^{\mu}(-4e^{-2\phi}\nabla_{\mu}\phi)-8e^{-2\phi}(\nabla\phi)^{2}$. Performing the integral leaves us with a boundary action
\beq I_{E}^{\text{on-shell}}=\frac{1}{\pi}\int_{\partial\mathcal{M}}dt_{E}\sqrt{\gamma}\left[2e^{-2\phi}n^{\mu}\nabla_{\mu}\phi-e^{-2\phi}(K-2\lambda)\right]\;.\label{eq:partonIECGHS}\eeq
Here, it is understood that the integration is over the Euclidean time circle evaluated at the boundary at asymptotic spatial infinity. Further, note that had we included conformal matter fields $f_{i}$ via the action (\ref{eq:matactCGHS}), then the matter does not contribute to the on-shell action (\ref{eq:partonIECGHS}) provided one assumes $f_{i}n^{\mu}\nabla_{\mu}f_{i}|_{\partial\mathcal{M}}=0$ (using that $\Box f_{i}=0$ on-shell). 

We can proceed by evaluating the action (\ref{eq:partonIECGHS}) in the eternal black hole background. Specifically, working in Rindler coordinates (\ref{eq:EucRindlercoord}), we find\footnote{Here it is useful to know $\sqrt{\gamma}=\sqrt{\gamma_{T_{E}T_{E}}}=\lambda Xe^{\rho}$, $n^{\mu}=e^{-\rho}\partial^{\mu}_{X}$, while $n^{\mu}\nabla_{\mu}\phi=-X\lambda^{2}e^{\rho}$ and $K=\nabla_{\mu}n^{\mu}=\frac{M}{X\lambda}e^{\rho}$. Further, the domain of integration in the on-shell action is $T_{E}\in[0,\beta_{\text{H}}]$.}
\beq I_{E}^{\text{on-shell}}=\frac{1}{\pi}\lim_{X\to\infty}\frac{2\pi}{\lambda}\left(-2X^{2}\lambda^{3}-M+2X^{2}\lambda^{3}\sqrt{1+\frac{M}{X^{2}\lambda^{3}}}\right)=0\;.\eeq
Thus, curiously, the on-shell action vanishes for any $\beta$.
This naively suggests the eternal CGHS black hole has vanishing free energy, $F=\frac{1}{\beta_{\text{H}}}I_{E}^{\text{on-shell}}$. To better understand this deficiency  in applying the Gibbons-Hawking prescription, below we will instead employ the quasi-local method of York \cite{York:1986it}.

Before moving on, let us briefly comment on our introduction of the local counterterm (\ref{eq:localctCGHS}). First, note that without the counterterm, the on-shell action diverges in the infrared. We could opt for the background subtraction method, i.e., subtracting the reference linear dilaton vacuum solution, as often done in the context of asymptotically flat spacetimes in higher-dimensions \cite{Gibbons:1976ue}. Replacing $K$ with $[K]=K-K_{0}$, where $K_{0}$ denotes the trace of the extrinsic curvature in the linear dilaton vacuum ($M=0$), is insufficient as the action still diverges. Instead, we could subtract the entire on-shell action evaluated in the linear dilaton vacuum from the on-shell action evaluated in the black hole background. That is, 
\beq
\begin{split} 
I^{\text{on-shell}}_{E,\text{reg}}\equiv I^{\text{on-shell}}_{E,\text{BH}}-I^{\text{on-shell}}_{E,\text{LDV}}=\frac{\beta_{\text{H}}}{\pi}\left(-2X^{2}\lambda^{3}-M\right)-\frac{\beta_{\text{H}}}{\pi}(-2X^{2}\lambda^{3})=-M_{\text{ADM}}\beta_{\text{H}}\;.
\end{split}
\eeq
According to standard thermodynamics, we would erroneously conclude the thermal energy $E$ and entropy $S$ are 
\beq E=\left(\frac{\partial I^{\text{on-shell}}_{E,\text{reg}}}{\partial \beta_{\text{H}}}\right)=-M_{\text{ADM}}\;,\quad S=\beta_{\text{H}}E-I_{E,\text{reg}}^{\text{on-shell}}=0\;.\eeq
Thus, as when we computed the ADM mass of the eternal black hole (see around (\ref{eq:ADMHamiltonianWOsub})), subtracting the linear dilaton vacuum is insufficient. 

The issue with background subtraction, in general, is that it is ambiguous which reference background should be subtracted. Typically this is not a problem in asymptotically flat backgrounds, however, it is known to be problematic for asymptotically anti-de Sitter backgrounds, see, e.g., \cite{Vanzo:1997gw,Emparan:1999pm}. Thus, often for AdS spacetimes, one instead employs a local counterterm method for regulating infrared divergences \cite{Henningson:1998gx,Kraus:1999di,Emparan:1999pm}, where no such ambiguities arise.\footnote{Note, however, in \cite{Gibbons:1992rh} background subtraction method was utilized to study the thermodynamics of Witten's black hole and gave sensible results. It was confirmed in \cite{Davis:2004xi} the background subtraction and local counterterms methods for Witten's black hole were consistent.}

\subsection{Quasi-local thermodynamics}

\noindent Following York \cite{York:1986it}, we consider an observer located at a finite timelike boundary $B$ outside of the eternal black hole.\footnote{Previously, York's program was adapted for the CGHS black hole \cite{Davis:2004xi}, and (A)dS$_{2}$ black holes in Jackiw-Teteilboim gravity in \cite{Lemos:1996bq,Svesko:2022txo,Aguilar-Gutierrez:2024nst}.} The periodicity of the Euclidean time circle remains $\beta=\beta_{\text{H}}$ (\ref{eq:betaH}). The proper length of the circle at the boundary $B$, however, is 
\beq \beta_{\text{T}}=\int_{0}^{\beta_{\text{H}}}dt_{E}\mathcal{N}(\phi_{B})
=\beta_{\text{H}}\mathcal{N}(\phi_{B})\;.
\label{eq:tolmanT}\eeq
This is simply the (inverse) Tolman temperature $\beta_{\text{T}}^{-1}=T$, the redshifted temperature measured locally at $B$. In the limit the boundary $B$ is sent to asymptotic spatial infinity, the Tolman temperature coincides with the Hawking temperature.

Assuming the metric and dilaton obey Dirichlet boundary conditions on $B$, the thermal canonical ensemble is defined as the ensemble with fixed Tolman temperature and $\phi_{B}$, the value of the dilaton at $B$,
\beq \text{Canonical ensemble:}\quad (\beta_{\text{T}},\phi_{B})\quad \text{fixed}\;.\eeq
As in the standard Gibbons-Hawking prescription, the canonical partition function $\mathcal{Z}(\beta_{T},\phi_{B})$, is expressed as a Euclidean gravitational path integral, which we evaluate using a saddle-point approximation. At leading order, we only need the on-shell Euclidean action (\ref{eq:partonIECGHS}) evaluated at the finite boundary $B$. Specifically,  we find (see also \cite{Davis:2004xi})\footnote{It is useful to know $A_{B}=\frac{M}{\lambda}\frac{\beta_{\text{H}}^{2}}{\beta_{H}^{2}-\beta_{\text{T}}^{2}}$ (which diverges as the boundary is pushed to spatial infinity) 
$$\frac{(\beta_{\text{T}}-\beta_{\text{H}})^{2}}{\beta_{\text{H}}^{2}}=e^{2\phi_{B}}\left[2X_{B}^{2}\lambda^{2}+\frac{M}{\lambda}-2X_{B}\lambda\sqrt{\frac{M}{\lambda}+X_{B}^{2}\lambda^{2}}\right]\;,$$
where $X_{B}$ denotes the Dirichlet cavity surrounding the black hole.}   
\beq 
\begin{split}
I^{\text{on-shell}}_{E}
&=-2A_{B}\frac{(\beta_{\text{T}}-\beta_{\text{H}})^{2}}{\beta_{\text{H}}^{2}}\;.
\end{split}
\label{eq:EucCGHSYork}\eeq
where we introduced the ``area'' $A_{B}\equiv e^{-2\phi_{B}}$. Evidently, fixed $\phi_{B}$ implies fixed $A_{B}$, and thus we can characterize the canonical ensemble as one where $\beta_{\text{T}}$ and $A_{B}$ are both fixed. 
Using $Z(\beta_{\text{T}},\phi_{B})\approx e^{-I_{E}^{\text{on-shell}}(\beta_{\text{T}},A_{B})}$, standard thermodynamics allows us to compute the energy $E$, entropy $S$, and free energy $F$
\beq 
\begin{split}
&E=\left(\frac{\partial I_{E}^{\text{on-shell}}}{\partial\beta_{\text{T}}}\right)_{\hspace{-1mm}A_{B}}=-4A_{B}\frac{(\beta_{\text{T}}-\beta_{\text{H}})}{\beta_{\text{H}}^{2}}=\frac{2M_{\text{ADM}}\beta_{\text{H}}}{\beta_{\text{T}}+\beta_{\text{H}}}\;,\\
&S=\beta_{\text{T}}E-I_{E}^{\text{on-shell}}=M_{\text{ADM}}\beta_{\text{H}}=\frac{2M}{\lambda}\;,\\
& F=E-\frac{1}{\beta_{\text{T}}}S=\frac{1}{\beta_{\text{T}}}I_{E}\;.
\end{split}
\label{eq:thermovarCGHS}\eeq

Comparing to (\ref{eq:BYeneternalCGHSv2}), it is easy to verify the thermal energy $E$ is equivalent to the quasi-local energy, $E=E_{\text{BY}}$. Finally, it is natural to define a ``surface pressure'' $\sigma$,
\beq \sigma\equiv-\left(\frac{\partial E}{\partial A_{B}}\right)_{\hspace{-1mm}S}=2\frac{(\beta_{\text{T}}-\beta_{\text{H}})^{2}}{\beta_{\text{T}}\beta^{2}_{\text{H}}}=\frac{\lambda}{\pi\mathcal{N}}(\mathcal{N}-1)^{2}\;.\label{eq:surfpress}\eeq
Consequently, we can interpret the mechanical first law (\ref{eq:quasilocfirst}) as a genuine first law of (quasi-local) thermodynamics, 
\beq \delta E=T\delta S-\sigma \delta A_{B}\;,\label{eq:firstlawthermo}\eeq
for temperature $T=T_{\text{H}}/\mathcal{N}$. 
Additionally, the thermodynamic variables $E,T,S,\sigma$ and $A_{B}$ obey the Euler relation, 
\beq E=TS-\sigma A_{B}\;.\eeq
In the limit the boundary $B$ is sent to spatial infinity, where $\beta_{\text{T}}\to\beta_{\text{H}}$, it follows, $E\to T_{\text{H}}S$, $\sigma\to0$, and $F\to0$.

The heat capacity at fixed $A_{B}$, meanwhile, 
\beq C_{A_{B}}=-\beta_{\text{T}}^{2}\left(\frac{\partial E}{\partial\beta_{\text{T}}}\right)_{\hspace{-1mm} A_{B}}=4A_{B}\frac{\beta_{\text{T}}^{2}}{\beta_{\text{H}}^{2}}\;,\label{eq:heatcapCGHS}\eeq
is positive for all $\beta_{\text{T}}>0$ (vanishing only for $\beta_{\text{T}}=0$). Thus, the eternal CGHS black hole with a finite Dirichlet wall is thermally stable. Notice in the limit the wall approaches asymptotic infinity (where $A_{B}\to\infty$) the heat capacity diverges to positive infinity, as remarked in \cite{Fiola:1994ir}. This is consistent with the definition of heat capacity for thermal systems with constant temperature: an infinite change in heat produces a negligible change in temperature. 

It is straightforward to invert the thermodynamic formulae (\ref{eq:thermovarCGHS}) to express the entropy $S$ solely as a function of the Tolman temperature $T$, and variable conjugate to the surface pressure, $A_{B}$, 
\beq S(T,A_{B})=2A_{B}\left(1-\frac{\lambda^{2}}{4\pi^{2}T^{2}}\right)\;.\eeq
Evidently, the canonical entropy is non-negative for Tolman temperatures $T\geq T_{\text{H}}$. Meanwhile, the entropy in the microcanonical ensemble (a fixed $E$ ensemble) for constant $A_{B}$ is 
\beq S(E,A_{B})=\frac{\pi E}{2A_{B}\lambda^{2}}(4A_{B}\lambda-\pi E)\;.\eeq
The entropy is easily shown to be concave in $E$, a characteristic trait for thermally stable systems.

\section{Exact semi-classical gravity and quantum black holes} \label{sec:BPPbhs}

\noindent In this section, we turn to the study of quantum black holes, showcasing the utility of two-dimensional dilaton gravity. After briefly reviewing the construction of the semi-classical CGHS model, where the classical CGHS action is appended by the Polyakov term, we turn to exactly solvable models of two-dimensional semi-classical gravity. We review the geometric construction of eternal quantum black holes, where we analyze the null and averaged null energy conditions. We also interrogate the singularity structure, where we find naked singularities can occur for quantum matter in the Boulware state in a large mass regime.

\subsection{Quantum effects on a fixed background} 

\noindent Quantum matter fields on a fixed classical four-dimensional black hole background formed under collapse reveal the black hole emits thermal radiation at a temperature proportional to the surface gravity of the black hole \cite{Hawking:1975vcx}. The same effect occurs for two-dimensional black holes \cite{Callan:1992rs,Giddings:1992ff}. Consider two asymptotically flat regions near asymptotic infinity, the ``in'' region $I^{-}_{L}$ and ``out'' region $I^{+}_{R}$, and treat $f_{i}$ as quantum fields. Observers stationed in these regions will make measurements on fields $f_{i}$, where formally their observations are related via Bogoliubov transformations, which encode the thermal nature of the radiation.  For two-dimensional conformal matter fields, there is a simpler way to characterize black hole radiation, tied to the existence of the conformal (trace) anomaly \cite{Christensen:1977jc}. 

Classically, the stress-tensor (\ref{eq:classstresstens}) has vanishing trace. When leading 1-loop quantum effects are incorporated, however, the trace of the expectation value of the quantum stress-tensor, for any state, is 
\beq \langle T^{\mu}_{\;\mu}\rangle=\frac{c}{24} R\;,\label{eq:traceanom}\eeq
for central charge $c$ of a two-dimensional conformal field theory. Presently, $c=N$, the number of massless scalar fields. Working in conformal gauge, where $g^{\mu\nu}T_{\mu\nu}=-4e^{-2\rho}T_{\pm \mp}$ and $R=8e^{-2\rho}\partial_{+}\partial_{-}\rho$, it follows
\beq \langle T_{\pm\mp}\rangle=-\frac{N}{12}\partial_{+}\partial_{-}\rho\;.\label{eq:Tpmanom}\eeq
Further, by imposing covariant conservation $\nabla^{\mu}\langle T_{\mu\nu}\rangle$ and using (\ref{eq:Tpmanom}), the other components of the quantum stress-tensor are worked out to be
\beq
\langle T_{++}\rangle=-\frac{N}{12}[(\partial_{+}\rho)^{2}-\partial^{2}_{+}\rho+t_{+}(x^{+})]\;,\label{eq:Tppappv2}\eeq
\beq \langle T_{--}\rangle=-\frac{N}{12}[(\partial_{-}\rho)^{2}-\partial^{2}_{-}\rho+t_{-}(x^{-})]\;.\label{eq:Tmmappv2}\eeq
Here the functions $t_{\pm}$ arise as functions of integration, and additional input is needed to fix their form. For example, for a black hole formed under collapsing matter, functions $t_{\pm}$ are fixed by imposing boundary conditions such that there is no incoming radiation along past null infinity.

Note that for a collapsing black hole, as one approaches future null infinity it follows (see, e.g., \cite{Thorlacius:1994ip})
\beq \langle T_{--}\rangle_{\text{collapse}} \to\frac{N\lambda^{2}}{48}\left[1-\frac{1}{(1+P_{\infty}e^{\sigma_{-}})^{2}}\right]\;,\label{eq:enfluxpreback}\eeq
for pressure at null infinity $P_{\infty}$,
while the remaining stress-energy components vanish. Thus, there is a non-vanishing outgoing flux along $I_{R}^{+}$. In the far past, $\sigma_{-}\to-\infty$, the flux exponentially vanishes, while approaching the horizon the flux approaches the constant value, understood to be the energy flux due to Hawking radiation. Integrating the total energy along all of future null infinity, however, results in infinity. This is nonsensical as the black hole cannot radiate more energy than it has. To remedy the situation requires accounting for backreaction due to quantum matter fields characterizing Hawking radiation on the classical background geometry.

\subsection{Incorporating backreaction}

\noindent Including backreaction amounts to letting the quantum stress-tensor $\langle T_{\mu\nu}\rangle$ source the classical metric equations of motion. In particular, combining the CGHS metric equations of motion (\ref{eq:meteomCGHSconfgauge}), whilst replacing the classical stress tensor with the $`\pm\mp'$ components of the quantum stress-tensor (\ref{eq:Tpmanom}) give\footnote{Here we follow the convention that $\langle T_{\mu\nu}\rangle\equiv -\frac{2\pi}{\sqrt{-g}}\frac{\delta I_{\text{Poly}}}{\delta g^{\mu\nu}}$ for quantum effective action $I_{\text{Poly}}$. Note the factor of two difference from the classical stress-energy tensor defined in (\ref{eq:classstresstens}).}
\beq e^{-2\phi}[2\partial_{+}\partial_{-}\phi-4(\partial_{+}\phi)(\partial_{-}\phi)-\lambda^{2}e^{2\rho}]=\frac{N}{12}\partial_{+}\partial_{-}\rho\;.\label{eq:rhoeomwoutRST}\eeq
The other metric equations will be modified by the quantum-stress tensor (\ref{eq:Tppappv2}) and (\ref{eq:Tmmappv2}), as we will show momentarily. 

These semi-classical quantum matter contributions have an effective description in terms of the non-local 1-loop Polyakov action 
\beq I_{\text{Poly}}=-\frac{N}{96\pi}\int_{\mathcal{M}}d^{2}x\sqrt{-g}R\Box^{-1}R\;,\label{eq:Polyactv1}\eeq
where $\Box^{-1}$ is the Green's function of the D'Alembertian operator $\Box$. Indeed, working in conformal gauge $ds^{2}=e^{2\rho}dw^{+}dw^{-}$ for arbitrary lightcone coordinates $(w^{+},w^{-})$, the Polyakov action can be put into a local form, i.e., locally depends on coordinates\footnote{To see this, recall in conformal gauge $R=-2\Box\rho=8e^{-2\rho}\partial_{+}\partial_{-}\rho$ such that $R\Box^{-1}R=(8e^{-2\rho}\partial_{+}\partial_{-}\rho)(-2\rho)$. Substituting this into the non-local action (\ref{eq:Polyactv1}) with $\sqrt{-g}=e^{2\rho}/2$ and performing integration by parts (dropping boundary terms) yields the local action (\ref{eq:PolylocactCG}).} 
\beq I_{\text{Poly}}=-\frac{N}{12\pi}\int_{\mathcal{M}} dw^{+}dw^{-}(\partial_{+}\rho)(\partial_{-}\rho)\;.\label{eq:PolylocactCG}\eeq
In fact, by introducing  an auxiliary scalar field $Z$, the Polyakov action can be placed in the local form\footnote{For comparison, our $Z=-\varphi$ of Eq. (5.56) of \cite{Fabbri:2005mw}, and $Z=-2\chi$ in Eq. (2.11) of \cite{Pedraza:2021cvx}.}
\beq I_{\text{Poly}}=-\frac{N}{48\pi}\int_{\mathcal{M}} d^{2}x\sqrt{-g}\left[\frac{1}{2}(\nabla Z)^{2}-RZ\right]\;.
\label{eq:Polyactv2}\eeq
The equation of motion for the auxiliary field $Z$ is
\beq \Box Z+R=0\;.\label{eq:eomZ}\eeq
The formal solution of $Z$ may be written as $Z(x)=-\int_{\mathcal{M}} d^{2}y\sqrt{-g(y)}G(x,y)R(y)$ (for Green's function $G(x,y)\equiv\Box^{-1}$ obeying $\Box_{x}G(x,y)=(-g(x))^{-1/2}\delta^{2}(x-y)$) such that upon substitution into (\ref{eq:Polyactv1}) one recovers the nonlocal action.\footnote{To see this, perform an integration by parts to rewrite $(\nabla Z)^{2}=-Z\Box Z$, dropping a total derivative. Implementing the formal solution to the auxiliary equation of motion, $Z=-\Box^{-1}R$, the non-local action (\ref{eq:Polyactv1}) follows, which also commonly written as $I_{\text{Poly}}=-\frac{N}{96\pi}\int d^{2}x\sqrt{-g(x)}\int d^{2}y\sqrt{-g(y)}R(x)G(x,y)R(y)$, for Green's function $G(x,y)$.}

Thus, semi-classical backreaction is incorporated by appending the classical CGHS model by the Polyakov action (\ref{eq:CGHSLor}) \cite{Callan:1992rs}
\beq I=I_{\text{CGHS}}+I_{\text{Poly}}\;,\label{eq:semiclassCGHS}\eeq
where the quantum matter is encoded by the Polyakov term (for dynamical black holes, one can additionally include fields $\{f_{i}\}$). Using the local action (\ref{eq:Polyactv2}), the semi-classical theory (\ref{eq:semiclassCGHS}) has metric equations
\beq 2e^{-2\phi}\left[\nabla_{\mu}\nabla_{\nu}\phi-g_{\mu\nu}\Box\phi+g_{\mu\nu}(\nabla\phi)^{2}-g_{\mu\nu}\lambda^{2}\right]=\langle T_{\mu\nu}\rangle\;,\label{eq:semiCGHSmetEOM}\eeq
for quantum stress-tensor
\beq \langle T_{\mu\nu}\rangle\equiv-\frac{2\pi}{\sqrt{-g}}\frac{\delta I_{\text{Poly}}}{\delta g^{\mu\nu}}=\frac{N}{24}\left[\nabla_{\mu}\nabla_{\nu}Z-g_{\mu\nu}\Box Z+\frac{1}{2}\left((\nabla_{\mu}Z)(\nabla_{\nu}Z)-\frac{1}{2}(\nabla Z)^{2}g_{\mu\nu}\right)\right]\;.\label{eq:TmunuPoly}\eeq
Easily, $g^{\mu\nu}\langle T_{\mu\nu}\rangle$ recovers the 1-loop conformal anomaly (\ref{eq:traceanom}).

Working in conformal gauge, the solution to the auxiliary equation of motion (\ref{eq:eomZ}) is easily attained to be 
\beq Z=2\rho-2\xi\;,\label{eq:solnZconfg}\eeq
where $\xi$ is some solution to the wave equation $\Box\xi=0=\partial_{+}\partial_{-}\xi$, such that $\xi(w^{+},w^{-})=\xi_{+}(w^{+})+\xi_{-}(w^{-})$. Further, the components of the quantum stress-tensor (\ref{eq:TmunuPoly}) in conformal gauge are
\beq 
\begin{split}
&\langle T_{\pm\mp}\rangle=-\frac{N}{24}\partial_{\pm}\partial_{\mp}Z\;,\\
&\langle T_{\pm\pm}\rangle=\frac{N}{24}\left[\partial^{2}_{\pm}Z-2(\partial_{\pm}\rho)(\partial_{\pm}Z)+\frac{1}{2}(\partial_{\pm}Z)^{2}\right]\;.
\end{split}
\eeq
With the solution (\ref{eq:solnZconfg}), the components (\ref{eq:Tpmanom}) --- (\ref{eq:Tmmappv2}) follow, for functions $t_{\pm}(w^{\pm})$ 
\beq t_{+}(w^{+})\equiv \partial^{2}_{+}\xi-(\partial_{+}\xi)^{2}\;,\quad t_{-}(w^{-})\equiv \partial^{2}_{-}\xi-(\partial_{-}\xi)^{2}\;.\label{eq:tpmcongen}\eeq
The functions $t_{\pm}$ are in fact related to how one characterizes the definition of vacuum states,  as we will review momentarily.

Technically, the semi-classical $\rho$-equation of motion (\ref{eq:rhoeomwoutRST}) is not a consistent equation in $\hbar$. To wit, temporarily restoring factors of $\hbar$ (\emph{viz.} $N\to N\hbar$), the left-hand side is of the order $\mathcal{O}(\hbar^{0})$, while the right-hand side is of the order $\mathcal{O}(\hbar)$. Solutions to this equation would include all powers of $\hbar$, however, order $\mathcal{O}(\hbar^{2})$ contributions to the solution would be modified by $\mathcal{O}(\hbar^{2})$ corrections to (\ref{eq:rhoeomwoutRST}). This signals a breakdown in a perturbative expansion in small $\hbar$. Instead, the ``semi-classical limit'' is understood to be a perturbative expansion in $1/N$ for large $N$ (setting $\hbar=1$) with $Ne^{2\phi}$ fixed \cite{Callan:1992rs}. Indeed, both sides of  (\ref{eq:rhoeomwoutRST}) are of the order $\mathcal{O}(N^{1})$, while corrections will be of order $\mathcal{O}(N^{0})$ and thus safely neglected in a large-$N$ expansion. Interpreting $e^{2\phi}\sim G_{2}$ as the two-dimensional Newton's constant, this is the familiar semi-classical limit for which the semi-classical Einstein equations follow, cf. \cite{Hartle:1981zt}. 

At this stage, it is worth commenting on the hierarchy of relevant scales, and, correspondingly, the validity of our semi-classical approximation. Recall that the classical CGHS model arises from a spherical dimensional reduction of near-extremal stringy black holes. To keep curvature corrections small from this higher-dimensional perspective, one works with `large' black holes such that $r_{H}/(\lambda^{-1})\gg1$, and, subsequently, $S_{\text{BH}}^{\text{cl}}\sim e^{-2\phi_{H}}=e^{2\lambda r_{H}}=M/\lambda\gg1$. Now, the semi-classical approximation is meant to be understood  quantizing $N$-massless scalar fields whilst leaving the background geometry classical. This is is possible provided $N\gg1$, such that we can safely ignore stringy-like ghost corrections to the dilaton or metric. At the same time, for $N$ to amount to a correction to the classical set-up, we impose $M/\lambda\gg N$.\footnote{This requirement is consistent with requiring backreaction modify the geometry such that the energy flux (\ref{eq:enfluxpreback}) does not approach a non-zero constant, e.g., \cite{Banks:1992ba}.} Combined, the semi-classical regime of validity is 
\beq
e^{-2\phi_{H}}=M/\lambda\gg N\gg1\;.\label{eq:scvalidity}
\eeq
The analogous regime of validity for the semi-classical JT model was established in \cite{Pedraza:2021cvx}. 

Further, since the Polyakov action does not modify the dilaton equation of motion (\ref{eq:dilamatEOMconf}), it can be combined with the $\rho$ equation of motion (\ref{eq:rhoeomwoutRST}) to eliminate $2\partial_{+}\partial_{-}\phi$ such that (\ref{eq:rhoeomwoutRST}) may be cast as
\beq 2\left(1-\frac{N}{12}e^{2\phi}\right)\partial_{+}\partial_{-}\rho=4(\partial_{+}\phi)(\partial_{-}\phi)+\lambda^{2}e^{2\rho}\;.\eeq
Thus, the left-hand side will vanish in the event the dilaton reaches the critical value 
\be
\phi_{c}=\frac{1}{2}\ln(12/N)\,.
\ee
Unless the right-hand side similarly vanishes at this critical value, the second-derivatives of $\rho$ and $\phi$ will diverge, leading to curvature singularities of the quantum black hole \cite{Russo:1992ht}. In this region, however, where the fields grow as order $\mathcal{O}(N)$, the semi-classical large-$N$ approximation breaks down, and it is therefore premature to establish whether there is in fact such a singularity.

Altogether, the semi-classical theory (\ref{eq:semiclassCGHS}) describes dominant semi-classical effects, including both Hawking radiation and  backreaction in black hole backgrounds. In terms of perturbation theory, the large-$N$ action (CGHS plus Polyakov) describes gravi-dilaton tree diagrams plus quantum matter at 1-loop; gravi-dilaton loops are suppressed relative to the matter loop. To proceed and solve the semi-classical gravitational equations of motion, one must choose the state of the quantum matter. Let us therefore briefly relate the properties of $t_{\pm}$ to (normal-ordered) stress-tensors and choice of vacuum for quantum matter.

\subsubsection{Normal-ordered stress tensors and vacuum states}
\label{subsubsec:normalorderedSET}

\noindent First note that $\rho$ transforms as a tensor. Specifically, under a conformal transformation from $(w^{\pm})$ to another set of (null) conformal coordinates $(z^{\pm})$, (see, \emph{e.g.}, Eq. (5.68) of \cite{Fabbri:2005mw})
\beq \rho(w^{+},w^{-})\to\rho(z^{+},z^{-})=\rho(w^{+},w^{-})+\frac{1}{2}\ln\left(\frac{dw^{+}}{dz^{+}}\frac{dw^{-}}{dz^{-}}\right)\;.\label{eq:rhotranslaw}\eeq
Meanwhile, the auxiliary field $Z=2\rho-2\xi$ transforms as a scalar. Thus, functions $\xi_{\pm}$ must compensate for (\ref{eq:rhotranslaw}) and transform under a conformal transformation as 
\beq \xi_{\pm}(w^{\pm})\to\xi_{\pm}(z^{\pm})=\xi_{\pm}(z^{\pm}(w^{\pm}))+\frac{1}{2}\ln\frac{dw^{\pm}}{dz^{\pm}}\;.\label{eq:transformxi}\eeq
 Consequently, functions $t_{\pm}(w^{\pm})$ transform as
\beq t_{\pm}(z^{\pm})=\left(\frac{dw^{\pm}}{dz^{\pm}}\right)^{2}t_{\pm}(w^{\pm})+\frac{1}{2}\{w^{\pm},z^{\pm}\}\;,\label{eq:tpmtrans}\eeq
where $\{w^{\pm},z^{\pm}\}$ is the Schwarzian derivative of $w^{\pm}$ with respect to $z^{\pm}$,
\beq \{w^{\pm},z^{\pm}\}\equiv\frac{(w^{\pm})'''}{(w^{\pm})'}-\frac{3}{2}\left(\frac{(w^{\pm})''}{(w^{\pm})'}\right)^{2}\;, \eeq
for $(w^{\pm})'\equiv\frac{dw^{\pm}}{dz^{\pm}}$.

 It turns out $t_{\pm}$ are identified with the expectation value of the \emph{normal-ordered} stress tensor \cite{Fabbri:2005mw}  
\beq \langle \Psi | :T_{\pm\pm}(w^{\pm}):| \Psi \rangle \equiv -\frac{N}{12}t_{\pm}(w^{\pm})\;,\label{eq:normordgen}\eeq
where $|\Psi\rangle$ is some unspecified quantum state.
By normal ordering one means
\beq 
\label{eqn:normord}
:T_{\pm\pm} (w^\pm):\;\equiv T_{\pm\pm}(w^\pm)-\langle 0_{w}|T_{\pm\pm}(w^\pm)|0_{w}\rangle \mathbb{1}\;.\eeq
Here $|0_{w}\rangle$  denotes the vacuum state with respect to the $(w^{\pm})$ coordinate system, i.e., the vacuum state defined with respect to the positive frequency modes in coordinates $w^{\pm}$, ${a}_{w}|0_{w}\rangle=0$. Clearly then, $\langle 0_{w}|:T_{\pm\pm}(w^{\pm}):|0_{w}\rangle=0$.

The identification (\ref{eq:normordgen}) follows from the observation that, in conformal gauge, about an arbitrary point in an arbitrary curved two-dimensional spacetime, the quantum stress-tensor is related to its normal-ordering via 
\beq \langle\Psi|T_{\pm\pm}(w^{\pm})|\Psi\rangle=-\frac{N}{12}[(\partial_{\pm}\rho)^{2}-\partial^{2}_{\pm}\rho]+\langle \Psi | :T_{\pm\pm}(w^{\pm}):| \Psi \rangle\;.\label{eq:qstressrelnorm}\eeq
Comparing to components (\ref{eq:Tmmappv2}), the identification (\ref{eq:normordgen}) follows. Moreover, from transformation \eqref{eq:tpmtrans} we see the  normal-ordered stress tensor obeys an  anomalous transformation law under a conformal transformation  $w^{\pm}\to z^{\pm}(w^{\pm})$,
\beq :T_{\pm\pm}(z^{\pm}):\;=\left(\frac{dw^{\pm}}{dz^{\pm}}\right)^{2}\,:T_{\pm\pm}(w^{\pm}):-\frac{N}{24}\{w^{\pm},z^{\pm}\} \mathbb{1} \;.\label{eq:transnormord}\eeq
Notice that taking the expectation value with respect to $|0_w \rangle$ on both sides yields
\beq \langle 0_{w}|:T_{\pm\pm}(z^{\pm}):|0_{w}\rangle=-\frac{N}{24}\{w^{\pm},z^{\pm}\}\;.\label{eq:changeinvacstnov2}\eeq

Aside from the above transformation rules, the main take away is the following observation: the vacuum state $|0_{w}\rangle$ is the state such that the functions $t_{\pm}(w^{\pm})$ vanish. Indeed, from the identification (\ref{eq:normordgen})
\beq\langle 0_w| :T_{\pm\pm}(w^{\pm}):|0_w \rangle=0\quad \Longleftrightarrow \quad t_{\pm}(w^{\pm})=0\;.\label{eq:gendefvacstate}\eeq
Thence, the functions $t_{\pm}$ are state-dependent. In particular, when in the vacuum state $|0_{w}\rangle$, the expectation value of the quantum stress-tensor is 
\beq \langle 0_{w}|T_{\pm\pm}(w^{\pm})|0_{w}\rangle=-\frac{N}{12}[(\partial_{\pm}\rho)^{2}-\partial^{2}_{\pm}\rho]\;.\eeq
If the background was two-dimensional flat Minkowski space, where $|0_{w}\rangle$ is the Minkowski vacuum, the quantum stress-tensor vanishes, as expected.
We emphasize, moreover, that while $t_{\pm}(w^{\pm})=0$ when in the vacuum state $|0_{w}\rangle$, generally $t_{\pm}(z^{\pm})\neq0$ in the same state. Finally, under a change of vacuum, the expectation value of the quantum stress-tensor obeys an anomalous transformation (combining normal ordering (\ref{eqn:normord}) and \eqref{eq:changeinvacstnov2})
\beq \langle 0_{w}|T_{\pm\pm}(z^{\pm})|0_{w}\rangle=\langle 0_{z}|T_{\pm\pm}(z^{\pm})|0_{z}\rangle-\frac{N}{24}\{w^{\pm},z^{\pm}\}\;.\label{eq:changeinvacst}\eeq
This transformation will be particularly useful when specifying vacuum states of the quantum matter in a particular background geometry.

Finally, the stress-energy tensor itself, being a rank-2 tensor, follows a standard coordinate transformation law
\beq \langle \Psi | T_{\pm\pm}(z^{\pm})|\Psi \rangle=\left(\frac{dw^{\pm}}{dz^{\pm}}\right)^{2}\langle \Psi | T_{\pm\pm}(w^{\pm})|\Psi\rangle\;.\label{eq:coordtranstmunu}\eeq
This transformation is consistent with the anomalous transformation law for the normal-ordered stress tensor, since combining \eqref{eq:normordgen} with   \eqref{eq:transnormord} and (\ref{eq:changeinvacst}) yields \eqref{eq:coordtranstmunu}.

\subsubsection{Choice of vacuum}
\label{subsubsec:vacuum}

\noindent In the context of semi-classical black hole physics, there are typically two distinct choices of vacuum for the quantum matter fields in an eternal black hole background. Namely, the Boulware and Hartle-Hawking vacuum states.  For a black hole formed under dynamical collapse, there is also the ``in'' vacuum state. Finally, there is also the Unruh vacuum, which can be viewed as a particular limit of the in-vacuum (and sometimes viewed as a vacuum state for the eternal black hole \cite{fabbri-navarro}). Here we summarize the essentials of the Boulware and Hartle-Hawking quantum states in the context of the semi-classical CGHS black hole. Our presentation here follows Section 6.2.4 of \cite{Fabbri:2005mw} and \cite{Pedraza:2021cvx} (with additional details relegated to Appendix \ref{app:coordsys}). 

Before we characterize each of the vacuum states, let us first apply the above formulae for the relation between (normal-ordered) quantum-stress tensors for conformal quantum fields in  the black hole backgrounds of interest. Specifically, for the eternal CGHS black hole, recall the metric in Eddington-Finkelstein coordinates $(\sigma^{\pm})$ (\ref{eqn:EFcoord}), 
\beq ds^{2}=-e^{2\rho(\sigma^{+},\sigma^{-})}d\sigma^{+}d\sigma^{-}\;,\quad \rho(\sigma^{+},\sigma^{-})=-\frac{1}{2}\ln\left(1+\frac{M}{\lambda}e^{-\lambda(\sigma^{+}-\sigma^{-})}\right)\;,\eeq
and Kruskal coordinates $(x^{\pm})$ (\ref{eq:eternalBH}), 
\beq ds^{2}=-e^{2\rho(x^{+},x^{-})}dx^{+}dx^{-}\;,\quad \rho(x^{+},x^{-})=-\frac{1}{2}\ln\left(\frac{M}{\lambda}-\lambda^{2}x^{+}x^{-}\right)\;,\eeq
where recall $x^{\pm}=\pm\lambda^{-1}e^{\pm\lambda \sigma^{\pm}}$.
The conformal factor $\rho$ is related in these two coordinates via the transformation (\ref{eq:rhotranslaw}) using $\{z^{\pm}\}=\{\sigma^{\pm}\}$ and $\{w^{\pm}\}=\{x^{\pm}\}$, 
\beq \label{eq:rhosigma-rhox}
\rho(\sigma^{+},\sigma^{-})=\rho(x^{+},x^{-})+\frac{\lambda}{2}(\sigma^{+}-\sigma^{-})=\rho(x^{+},x^{-})+\frac{1}{2}\ln(-\lambda^{2}x^{+}x^{-})\;.\eeq
Moreover, the functions $\xi_{\pm}$ transform as (\ref{eq:transformxi})
\beq
\begin{split}
&\xi_{\sigma^{+}}=\xi_{x^{+}}+\frac{\lambda\sigma^{+}}{2}=\xi_{x^{+}}+\frac{1}{2}\ln(\lambda x^{+})\;,\\
&\xi_{\sigma^{-}}=\xi_{x^{-}}-\frac{\lambda\sigma^{-}}{2}=\xi_{x^{-}}+\frac{1}{2}\ln(-\lambda x^{-})\;,
\end{split}
\eeq
from which it follows $\xi(\sigma^{+},\sigma^{-})=\xi_{\sigma^{+}}+\xi_{\sigma^{-}}$ is 
\beq \xi(\sigma^{+},\sigma^{-})=\xi(x^{+},x^{-})+\frac{\lambda}{2}(\sigma^{+}-\sigma^{-})=\xi(x^{+},x^{-})+\frac{1}{2}\ln(-\lambda^{2}x^{+}x^{-})\;,\label{eq:xitransEFK}\eeq
with $\xi(x^{+},x^{-})=\xi_{x^{+}}+\xi_{x^{-}}$.
Lastly, from the transformation (\ref{eq:tpmtrans}) the integration functions $t_{\pm}$ in these coordinate systems are related via\footnote{Here note the Schwarzian derivative $\{x^{\pm},\sigma^{\pm}\}=-\frac{\lambda^{2}}{2}$.}
\beq
\begin{split}
&t_{\sigma^{+}}=e^{2\lambda\sigma^{+}}t_{x^{+}}-\frac{\lambda^{2}}{4}=(\lambda x^{+})^{2}t_{x^{+}}-\frac{\lambda^{2}}{4}\;,\\
 &t_{\sigma^{-}}=e^{-2\lambda\sigma^{-}}t_{x^{-}}-\frac{\lambda^{2}}{4}=(\lambda x^{-})^{2}t_{x^{-}}-\frac{\lambda^{2}}{4}\;.
\end{split}
\label{eq:tpmtransEFK}\eeq
With these formulae in hand, the normal-ordered stress-tensors in the two coordinate systems for the eternal black hole are related via the anomalous transformation (\ref{eq:transnormord})
\beq 
\begin{split}
:T_{\sigma^{\pm}\sigma^{\pm}}(\sigma^{\pm}):\,=\, (\lambda x^{\pm})^{2}:T_{x^{\pm}x^{\pm}}(x^{\pm}):+\frac{N\lambda^{2}}{48} \mathbb{1} \;.   
\end{split}
\label{eq:normstresstransEFK}\eeq

Finally, from the quantum stress-tensor components (\ref{eq:Tpmanom}) and (\ref{eq:qstressrelnorm}), we have, in Eddington-Finkelstein coordinates
\beq 
\begin{split}
&\langle \Psi|T_{\sigma^{\pm}\sigma^{\pm}}(\sigma^{\pm})|\Psi\rangle=-\frac{N\lambda^{2}}{48}\left[1-\left(1+\frac{M}{\lambda}e^{-\lambda(\sigma^{+}-\sigma^{-})}\right)^{-2}\right]-\frac{N}{12}t_{\sigma^{\pm}}\;,\\
 &\langle \Psi|T_{\sigma^{+}\sigma^{-}}(\sigma^{\pm})|\Psi\rangle=-\frac{N M\lambda}{24}\frac{e^{-\lambda(\sigma^{+}-\sigma^{-})}}{\left(1+\frac{M}{\lambda}e^{-\lambda(\sigma^{+}-\sigma^{-})}\right)^{2}}\;,
\end{split}
\label{eq:qscompsEF}\eeq
where we used the identification (\ref{eq:normordgen}). In Kruskal coordinates, 
\beq 
\begin{split}
&\langle \Psi|T_{x^{+}x^{+}}(x^{\pm})|\Psi\rangle=\frac{N\lambda^{2}}{48}\frac{(\lambda x^{-})^{2}}{\left(\frac{M}{\lambda}-\lambda^{2}x^{+}x^{-}\right)^{2}}-\frac{N}{12}t_{x^{+}}\;,\quad  \langle \Psi|T_{x^{-}x^{-}}(x^{\pm})|\Psi\rangle=\frac{N\lambda^{2}}{48}\frac{(\lambda x^{+})^{2}}{\left(\frac{M}{\lambda}-\lambda^{2}x^{+}x^{-}\right)^{2}}-\frac{N}{12}t_{x^{-}}\;,\\
&\langle \Psi|T_{x^{+}x^{-}}(x^{\pm})|\Psi\rangle=-\frac{NM\lambda}{12}\frac{1}{\left(\frac{M}{\lambda}-\lambda^{2}x^{+}x^{-}\right)^{2}}\;.
\end{split}
\label{eq:qscompsKrusk}\eeq
At this point the integration functions $t_{\pm}$ remain undetermined. Let us now review the characteristic traits of each of the aforementioned vacua. 

\vspace{4mm}

\noindent $\bullet$ \emph{Boulware vacuum.} The Boulware vacuum state $|\text{B}\rangle$ \cite{Boulware:1974dm} is defined by  positive frequency modes in (static) Eddington-Finkelstein coordinates $(\sigma^{\pm})$. By definition of normal-ordering, 
\beq \langle \text{B}|:T_{\sigma^{\pm}\sigma^{\pm}}(\sigma^{\pm}):|\text{B}\rangle=0\;,\eeq
such that $t_{\sigma^{\pm}}(\sigma^{\pm})=0$ from the identification (\ref{eq:normordgen}). Hence, using (\ref{eq:qscompsEF}), the stress-tensor has components 
\beq 
\begin{split}
 &\langle \text{B}|T_{\sigma^{\pm}\sigma^{\pm}}(\sigma^{\pm})|\text{B}\rangle=-\frac{N\lambda^{2}}{48}\left[1-\left(1+\frac{M}{\lambda}e^{-\lambda(\sigma^{+}-\sigma^{-})}\right)^{-2}\right]\;,
\end{split}
\label{eq:qscompsEFBoul}\eeq
while $\langle T_{\sigma^{+}\sigma^{-}}\rangle$ has the same form for any state since it is state independent. In the limit near spatial infinity $\sigma^{1}\equiv \frac{1}{2}(\sigma^{+}-\sigma^{-})\to+\infty$, the components vanish,
\beq \langle \text{B}|T_{\sigma^{\pm}\sigma^{\pm}}(\sigma^{\pm})|\text{B}\rangle\to 0\;.\eeq
Likewise, the stress-tensor components $\langle \text{B}|T_{\sigma^{+}\sigma^{-}}(\sigma^{\pm})|\text{B}\rangle\to0$ at spatial infinity. Thus, all of the stress-tensor components at asymptotic spatial infinity reduce to the case for the linear dilaton (``Minkowski'') vacuum. 

Meanwhile, at the horizon, $\sigma^{1}\to-\infty$, 
\beq \langle \text{B}|T_{\sigma^{\pm}\sigma^{\pm}}(\sigma^{\pm})|\text{B}\rangle\to -\frac{N\lambda^{2}}{48}\;,\quad \langle \text{B}|T_{\sigma^{\pm}\sigma^{\mp}}(\sigma^{\pm})|\text{B}\rangle\to0\;.\eeq
The expectation value of the stress tensor in static Eddington-Finkelstein coordinates gives rise to a negative energy, which is interpreted as a Casimir energy in the presence of a boundary at the black hole horizon. 

To better understand the physics at the horizon, it is useful to instead turn to Kruskal coordinates $(x^{\pm})$. The stress-tensor components (\ref{eq:qscompsKrusk}) are
\beq 
\begin{split}
&\langle \text{B}|T_{x^{+}x^{+}}(x^{\pm})|\text{B}\rangle=\frac{N\lambda^{2}}{48}\frac{(\lambda x^{-})^{2}}{\left(\frac{M}{\lambda}-\lambda^{2}x^{+}x^{-}\right)^{2}}-\frac{N}{48 (x^{+})^{2}}\;,\quad  \langle \text{B}|T_{x^{-}x^{-}}(x^{\pm})|\text{B}\rangle=\frac{N\lambda^{2}}{48}\frac{(\lambda x^{+})^{2}}{\left(\frac{M}{\lambda}-\lambda^{2}x^{+}x^{-}\right)^{2}}-\frac{N}{48 (x^{-})^{2}}\;,\\
&\langle \text{B}|T_{x^{\pm}x^{\mp}}(x^{\pm})|\text{B}\rangle=-\frac{NM\lambda}{12}\frac{1}{\left(\frac{M}{\lambda}-\lambda^{2}x^{+}x^{-}\right)^{2}}\;,
\end{split}
\label{eq:qscompsKruskboul}\eeq
where from transformation (\ref{eq:tpmtransEFK}) we used that in the Boulware vacuum $t_{x^{\pm}}=\frac{1}{4(x^{\pm})^{2}}$. Thus, at the bifurcation point of the horizon ($x^{+}x^{-}=0$), the components of the stress-tensor is 
\beq \langle \text{B}|T_{x^{\pm}x^{\pm}}(x^{\pm})|\text{B}\rangle\to-\frac{N}{48(x^{\pm})^{2}}\;,\quad \langle \text{B}|T_{x^{\pm}x^{\mp}}(x^{\pm})|\text{B}\rangle=-\frac{N\lambda^{3}}{12M}\;.\label{eq:regularityBoul}\eeq
Thus, in Kruskal coordinates, the expectation value of the stress-tensor in the Boulware vacuum diverges along the past ($x^{+}=0$) and future ($x^{-}=0$) horizons. Likewise, the expectation value of the normal-ordered stress-tensor diverges at the horizon, as can be seen using the anomalous transformation (\ref{eq:normstresstransEFK}).

\vspace{4mm}

\noindent $\bullet$ \emph{Hartle-Hawking vacuum.} The Hartle-Hawking vacuum $|\text{HH}\rangle$ \cite{Hartle:1976tp} is defined by positive frequency modes in Kruskal coordinates ($x^{\pm})$. By definition, the normal-ordered stress-tensor vanishes,
\beq \langle \text{HH}|:T_{x^{\pm}x^{\pm}}(x^{\pm}):|\text{HH}\rangle=0\;,\eeq
as do the state-dependent functions $t_{x^{\pm}}(x^{\pm})=0$. Consequently, using tensor components (\ref{eq:qscompsKrusk}), it follows, 
\beq 
\begin{split}
&\langle \text{HH}|T_{x^{\pm}x^{\pm}}(x^{\pm})|\text{HH}\rangle=\frac{N\lambda^{2}}{48}\frac{(-\lambda^2 x^{+}x^{-})^{2}}{\left(\frac{M}{\lambda}-\lambda^{2}x^{+}x^{-}\right)^{2}}\frac{1}{(\pm \lambda x^{\pm})^{2}}\;.
\end{split}
\label{eq:qscompsKruskHH}\eeq
Unlike for the Boulware state (\ref{eq:regularityBoul}), the expectation value of the stress-tensor in the Hartle-Hawking vacuum is regular along the horizon $x^{\pm}\to0$.

To see how the expectation value of the stress-tensor in the Hartle-Hawking vacuum behaves at asymptotic infinity, it is useful to transform to Eddington-Finkelstein coordinates ($\sigma^{\pm}$). Via (\ref{eq:normstresstransEFK}) it immediately follows
\beq \langle \text{HH}|:T_{\sigma^{\pm}\sigma^{\pm}}(\sigma^{\pm}):|\text{HH}\rangle=\frac{N\lambda^{2}}{48}\;,\eeq
along with $t_{\sigma^{\pm}}=-\frac{\lambda^{2}}{4}$. Subsequently, using the tensor components (\ref{eq:qscompsEF})
\beq \langle \text{HH}|T_{\sigma^{\pm}\sigma^{\pm}}(\sigma^{\pm})|\text{HH}\rangle=\frac{N\lambda^{2}}{48}\frac{1}{\left(1+\frac{M}{\lambda}e^{-\lambda(\sigma^{+}-\sigma^{-})}\right)^{2}}\;.\eeq
In the asymptotic region at spatial infinity $(\sigma^{+}-\sigma^{-})\to\infty$, it follows
\beq \langle \text{HH}|T_{\sigma^{\pm}\sigma^{\pm}}(\sigma^{\pm})|\text{HH}\rangle\to \frac{N\lambda^{2}}{48}=\frac{N\pi^{2}}{12}T_{H}^{2}\;,\eeq
for Hawking temperature $T_{H}=\lambda/2\pi$. Thus, with respect to a static observer in ($\sigma^{\pm}$) coordinates, at asymptotic infinity, the Hartle-Hawking vacuum state is interpreted as the state with the energy density of a thermal bath of particles at Hawking temperature $T_{H}$. Thence, the Hartle-
Hawking state allows one to identify the black hole as a thermal system with a temperature, thermodynamic energy and entropy, unlike the Boulware vacuum.

\vspace{5mm}
In semi-classical gravity, the next step would be to have an appropriate quantum stress-tensor (for some specific choice of vacuum) source the metric equations of motion to incorporate quantum matter backreaction on the classical background. This cannot be achieved analytically for the semi-classical CGHS model (\ref{eq:semiclassCGHS}), but can be solved numerically (as we further describe in Section \ref{sec:numericalqBHs}). Let us first move on to exactly solvable models.

\subsection{Exactly solvable semi-classical gravity}

\noindent The reason the semi-classical CGHS model cannot be solved analytically is because the Polyakov term breaks the residual Kruskal gauge symmetry (\ref{eq:freefieldeq}), $\rho=\phi$, the classical CGHS model (with or without matter) featured. The same type of residual gauge symmetry can be restored  by adding appropriate covariant counterterms to the semi-classical action (\ref{eq:semiclassCGHS}). Specifically, building off of \cite{deAlwis:1992emy,deAlwis:1992hv,Bilal:1992kv,Giddings:1992ae}, Russo-Susskind-Thorlacius (RST) proposed adding to the semi-classical action the following \cite{Russo:1992ax}
\beq I_{\text{RST}}=-\frac{N}{96\pi}\int_{\mathcal{M}}d^{2}x\sqrt{-g}\;2\phi R\;,\label{eq:RSTterm}\eeq
yielding an analytically solvable model. Similarly, Bose-Parker-Peleg (BPP) \cite{Bose:1995pz} found adding 
\beq I_{\text{BPP}}=\frac{N}{24\pi}\int_{\mathcal{M}}d^{2}x\sqrt{-g}[-\phi R+(\nabla\phi)^{2}]\;\label{eq:BPPterm}\eeq
to the semi-classical CGHS action yields a soluble model with analytic black hole solutions.\footnote{Unlike the classical CGHS model, the semi-classical RST and BPP models cannot be obtained as the dimensional reduction of higher-dimensional theories of gravity; the RST and BPP terms are added in by hand at the two-dimensional level. These semi-classical models, however, have a higher-dimensional, doubly-holographic pedigree, where the dilaton $\phi$ captures fluctuating modes of an end-of-the-world brane in three-dimensional anti-de Sitter spacetime \cite{Neuenfeld:2024gta}.}

Here we focus on an exactly solvable one-parameter interpolation of the RST and BPP models \cite{Fabbri:1995bz,Cruz:1995zt}. In particular, to the semi-classical CGHS model (\ref{eq:semiclassCGHS}) we add 
\beq I_{a}=\frac{N}{24\pi}\int d^{2}x\sqrt{-g}\left[(a-1)\phi R+(1-2a)(\nabla\phi)^{2}\right]\;.\label{eq:IbRSTBPPact}\eeq
In principle $a$ is any real parameter, where for $a=1/2$ the action $I_{a}$ reduces to the RST term (\ref{eq:RSTterm}), while $a=0$ returns the BPP term (\ref{eq:BPPterm}). Note, however, that for $a>1/2$, the sign of the kinetic term changes sign. We will therefore primarily focus on values of $0\leq a\leq 1/2$. Thus, the semi-classical theory of interest is 
\beq I=I_{\text{CGHS}}+I_{\text{Poly}}+I_{a}\;,\label{eq:fullactRSTBPP}\eeq
which we may further couple to conformal matter $\{f_{i}\}$, with classical action (\ref{eq:matactCGHS}).  Additionally, one should include Gibbons-Hawking-York surface terms to ensure a well-posed variational problem, as well as local counterterms. We will return to these terms in Section \ref{sec:massentqbhs}.

Restoration of Kruskal gauge symmetry via the $I_{a}$ term becomes clear when working in conformal coordinates. 
To see this, note that the total action (\ref{eq:fullactRSTBPP}) in conformal gauge can be written as
\beq 
\begin{split}
I&=\frac{1}{\pi}\int_{\mathcal{M}}dw^{+}dw^{-}\biggr[2\partial_{-}(\phi-\rho)\partial_{+}\left(e^{-2\phi}-\frac{N}{24}(\phi-\rho)+\frac{Na}{12}\phi\right)+\lambda^{2}e^{2(\rho-\phi)}+\frac{1}{2}\sum_{i=1}^{N}(\partial_{+}f_{i})(\partial_{-}f_{i})\biggr]\;,
\end{split}
\label{eq:totsemiactconfgauge}\eeq
where we included the matter action. Combining the $\rho$ and $\phi$ equations leads to (see Appendix \ref{app:eoms} for details)
\beq 0=2\left(e^{-2\phi}-\frac{aN}{24}\right)\partial_{-}\partial_{+}(\rho-\phi)\;.\eeq
Thus, $(\rho-\phi)$ is a free field, as in classical CGHS, unless the dilaton takes the critical value 
\be
\phi_{c}=\frac{1}{2}\ln\left(\frac{24}{aN}\right)\,.
\ee
Away from the critical value $\phi_{c}$, notice that the BPP model ($a=0$) exactly reproduces the classical CGHS relation (\ref{eq:freefieldeq}). 

The metric equations of motion for the full action (\ref{eq:fullactRSTBPP}) are
\beq
\begin{split}
\langle T_{\mu\nu}\rangle&=2e^{-2\phi}\left[\nabla_{\mu}\nabla_{\nu}\phi-g_{\mu\nu}\Box\phi+g_{\mu\nu}(\nabla\phi)^{2}-g_{\mu\nu}\lambda^{2}\right]\\    
&+\frac{N}{12}\left[(a-1)(g_{\mu\nu}\Box\phi-\nabla_{\mu}\nabla_{\nu}\phi)+(1-2a)\left((\nabla_{\mu}\phi)(\nabla_{\nu}\phi)-\frac{1}{2}g_{\mu\nu}(\nabla\phi)^{2}\right)\right]\;.
\end{split}
\label{eq:meteomsfullsemi}\eeq
As in the semi-classical CGHS model, the expectation value of the quantum stress-tensor $\langle T_{\mu\nu}\rangle$ is defined as the metric variation of the Polyakov action alone (\ref{eq:TmunuPoly}). Using the identities (\ref{eq:usegoeidconf}), in conformal gauge
\beq 
\begin{split} 
\label{eq:EoM_Tpp}
&\langle T_{\pm\pm}\rangle=\left(2e^{-2\phi}+\frac{N}{12}(1-a)\right)[\partial^{2}_{\pm}\phi-2(\partial_{\pm}\rho)(\partial_{\pm}\phi)]+\frac{N}{12}(1-2a)(\partial_{\pm}\phi)^{2}\;,\\
&\langle T_{\pm\mp}\rangle=-\left(2e^{-2\phi}+\frac{N}{12}(1-a)\right)\partial_{+}\partial_{-}\phi+4e^{-2\phi}(\partial_{+}\phi)(\partial_{-}\phi)+\lambda^{2}e^{2(\rho-\phi)}\;,
\end{split}
\eeq
with stress-tensor components (\ref{eq:Tpmanom}) --- (\ref{eq:Tmmappv2}) and state dependent functions $t_{\pm}(w^{\pm})$ (\ref{eq:tpmcongen}). Using these stress-tensor components, the constraint (\ref{eq:qstressrelnorm}) becomes 
\beq 
\begin{split}
 0&=\left(2e^{-2\phi}+\frac{N}{12}(1-a)\right)\left[2(\partial_{\pm}\rho)(\partial_{\pm}\phi)-\partial^{2}_{\pm}\phi\right]+\frac{N}{12}(2a-1)(\partial_{\pm}\phi)^{2}-\frac{N}{12}[(\partial_{\pm}\rho)^{2}-(\partial_{\pm}^{2}\rho)]+\langle:T_{\pm\pm}:\rangle\;,   
\end{split}
\label{eq:constraintRSTBPPv1}\eeq
where $\langle:T_{\pm\pm}:\rangle\equiv \langle \Psi|:T_{\pm\pm}:|\Psi\rangle$. 

To extract explicit analytic solutions of the semi-classical model, it proves useful to introduce field variables \cite{Cruz:1995zt} 
\beq \label{eq:liouvilledef}
\begin{split}
&\Omega\equiv \sqrt{\frac{12}{N}}e^{-2\phi}+\sqrt{\frac{N}{12}}a\phi\;,\\
&\chi\equiv \Omega-\sqrt{\frac{N}{12}}(\phi-\rho)=\sqrt{\frac{12}{N}}e^{-2\phi}+\sqrt{\frac{N}{12}}(a-1)\phi+\sqrt{\frac{N}{12}}\rho\;.
\end{split}
\eeq
The action (\ref{eq:totsemiactconfgauge}) then takes the form of a two-dimensional Liouville theory (see Appendix \ref{app:eoms} for details)
\beq I=\frac{1}{\pi}\int_{\mathcal{M}}dw^{+}dw^{-}\left[(\partial_{-}\Omega)(\partial_{+}\Omega)-(\partial_{-}\chi)(\partial_{+}\chi)+\lambda^{2}e^{2\sqrt{\frac{12}{N}}(\chi-\Omega)}+\frac{1}{2}\sum_{i=1}^{N}(\partial_{+}f_{i})(\partial_{-}f_{i})\right]\;.\label{eq:BPPRSTactLiouv}\eeq
Evidently, the interpolating model has restored the Kruskal gauge symmetry; specifically, $\Omega=\chi$ amounts to $\rho=\phi$. Varying with respect to $\chi$ and $\Omega$ gives, respectively, 
\beq 
\begin{split}
\partial_{+}\partial_{-}\chi=-\sqrt{\frac{12}{N}}\lambda^{2}e^{2\sqrt{\frac{12}{N}}(\chi-\Omega)}\;,\quad \partial_{+}\partial_{-}\Omega=-\sqrt{\frac{12}{N}}\lambda^{2}e^{2\sqrt{\frac{12}{N}}(\chi-\Omega)}\;,   
\end{split}
\label{eq:eomschiOmega}\eeq
from which it directly follows $\partial_{+}\partial_{-}(\chi-\Omega)=0$, such that $(\chi-\Omega)$ behaves as a free field. Further, in Liouville variables the constraint (\ref{eq:constraintRSTBPPv1}) is compactly expressed as
\beq 0=-(\partial_{\pm}\chi)^{2}+(\partial_{\pm}\Omega)^{2}+\sqrt{\frac{N}{12}}\partial^{2}_{\pm}\chi+\langle:T_{\pm\pm}:\rangle\;.\label{eq:constraintRSTBPPLiouv}\eeq
Finally, note that the scalar curvature $R=8e^{-2\rho}\partial_+\partial_-\rho$ in terms of the Liouville variables reads 
\be \label{eq:scalarcurvatureL}
R=\frac{8 e^{-2\rho}}{\Omega'}\left(\partial_+\partial_-\chi-\frac{\Omega''}{\Omega'^2}\partial_+\Omega\partial_-\Omega\right)\,,
\ee
where $\Omega'\equiv\frac{d\Omega}{d\phi}=\sqrt{\frac{N}{12}}a-2\sqrt{\frac{12}{N}}e^{-2\phi}$. 

\subsection{Exact quantum black hole solutions}\label{sec:QBHsolutions}

\noindent  Let us now construct exact solutions to the  semi-classical interpolating model \eqref{eq:fullactRSTBPP}. These  configurations depend on the vacuum state of the quantum matter fields, encoded in the functions $t_{\pm}(w^{\pm})$ introduced above. We will focus on exact solutions for two different vacuum states: Hartle-Hawking and Boulware vacua. 
 
In general, the solutions will be expressed in terms of the Liouville fields $(\Omega,\chi)$.  From the free field equation $\partial_+\partial_-(\chi - \Omega) = 0$, the general solution obeys
\beq
\chi - \Omega = \sqrt{\frac{N}{48}} \left( f_+(w^+) + f_-(w^-) \right)\, ,
\eeq
for chiral functions $f_{\pm}$ and 
arbitrary lightcone coordinates $\{w^{\pm}\}$. Substituting this free field into the equations of motion (\ref{eq:eomschiOmega}) gives
\beq \partial_{+}\partial_{-}\chi=-\sqrt{\frac{12}{N}}\lambda^{2}e^{f_{+}+f_{-}}=\partial_{+}\partial_{-}\Omega\;.\eeq
Integrating these equations, together with the constraint (\ref{eq:constraintRSTBPPLiouv}) gives general solutions of the form
\beq 
\begin{split}
 &\Omega=\mathcal{C}-\sqrt{\frac{12}{N}}h_{+}(w^{+})h_{-}(w^{-})-\sqrt{\frac{12}{N}}(F_{+}(w^{+})+F_{-}(w^{-}))\;,\quad \chi=\Omega+\sqrt{\frac{N}{48}}(f_{+}+f_{-})\;,
\end{split}
\eeq
for integration constant $\mathcal{C}$, $h_{\pm}(w^{\pm})\equiv \lambda \int dw^{\pm}e^{f_{\pm}}$, and 
\beq F_{\pm}(w^{\pm})=\int^{w^{\pm}}dw_{1}^{\pm}\int^{w_{1}^{\pm}}dw_{2}^{\pm}\langle :\hspace{-1mm} T_{\pm\pm}(w_{2}^{\pm})\hspace{-1mm}:\rangle\;.\eeq

The residual gauge freedom associated with conformal transformations is encoded in the arbitrary chiral functions $f_\pm(w^\pm)$. A convenient gauge choice is the Kruskal gauge, where $f_\pm = 0$, and hence $\Omega = \chi$. As before, when working in Kruskal gauge, we set $\{w^{\pm}\}=\{x^{\pm}\}$. In Kruskal gauge, the solution for $\Omega$ is 
\beq
\label{eq:Omega}
\Omega(x^+, x^-) = \sqrt{\frac{12}{N}}\left(\frac{M}{\lambda} - \lambda^2 x^+ x^- + F^+(x^+) + F^-(x^-)\right),
\eeq
and functions $F_{\pm}$ become 
\beq \label{eq:constraint-liouville}
F_{\pm}(x^{\pm}) = \frac{N}{12} \int^{x^{\pm}} dx^{\pm}_1 \int^{x^{\pm}_1} dx^{\pm}_2\, \,  t_{\pm}(x^{\pm}_2)\,,
\eeq
where $t_{\pm}(x^{\pm})$ are the state-dependent functions introduced in \eqref{eq:normordgen}. This solution can be directly obtained by integrating the constraint equations \eqref{eq:constraint-liouville}. The integration constant $M/\lambda$ is chosen such that in the classical limit $N\to 0$ we recover the CGHS eternal black hole mass.

From the Liouville fields, we may invert \eqref{eq:liouvilledef} to directly obtain solutions for $(\phi,\rho)$. Explicitly,
\beq
\begin{split}
&\phi =\sqrt{\frac{12}{N}}\frac{\Omega}{a}+\frac{1}{2}W_{-1}\left(-\frac{24}{a N}e^{-\sqrt{\frac{48}{N}}\frac{\Omega}{a}}\right)\, \, ,\\
&\rho=\phi+\sqrt{\frac{12}{N}}(\Omega-\chi)\;,
\end{split}
\eeq
where $W_{-1}(z)$ is the Lambert function with index $-1$. When $a=\frac{1}{2}$, we recover the solutions to the RST model (which will be relevant in Section \ref{sec:numericalqBHs}), while $a=0$ will return the BPP model solutions,
\beq
\phi^{(a=0)}=\frac{1}{2}\ln\left(\sqrt{\frac{12}{N}}\Omega^{-1}\right)\, .
\label{eq:bppphi}\eeq

 In the following sections, we will compute the asymptotic mass for the semi-classical solutions above. In this context, the asymptotic expansion of the dilaton field $\phi$ for large $\Omega$ will be needed,
\beq \label{eq:largeOmega}
\phi\sim -\frac{1}{2}\ln\left(\sqrt{\frac{N}{12}}\Omega\right)-\frac{a}{2\Omega} \sqrt{\frac{N}{48}}\ln\left(\sqrt{\frac{N}{12}}\Omega\right)+\mathcal{O}(\Omega^{-2})\, .
\eeq
Furthermore, it will be convenient to work in Eddington-Finkelstein coordinates $\{\sigma^{\pm}\}$. Since $\phi$ and $\Omega$ are scalars, they transform trivially $\phi(\sigma^+,\sigma^-)=\phi(x^+,x^-)$ and  $\Omega(\sigma^+,\sigma^-)=\Omega(x^+,x^-)$. For the conformal factor we apply the anomalous transformation given in \eqref{eq:rhosigma-rhox} so that 
\beq \label{eq:rhosigma-rhoxbis}
\rho(\sigma^+,\sigma^{-})=\rho(x^+,x^{-})+\frac{\lambda}{2}(\sigma^+-\sigma^-)= \rho(x^+,x^-)+\lambda\sigma^1\, ,
\eeq
where in the last line we used $2\sigma^1=(\sigma^+-\sigma^-)$ and we recall again $x^{\pm}=\pm \lambda^{-1}e^{\pm\lambda \sigma^{\pm}}$.

\subsubsection{Hartle-Hawking vacuum}

\noindent Let us particularize to the Hartle-Hawking state, i.e., where $t_{\pm}(x^{\pm})=0$ in Kruskal coordinates. Consequently, the Liouville field (\ref{eq:Omega}) becomes
\beq
\label{eq:omegaH}
\Omega^{(\text{HH})}=\sqrt{\frac{12}{N}}\left(\frac{M}{\lambda}-\lambda^2 x^+x^-\right)\,.
\eeq
For general $a$, it is worth going back to the original variables $\phi$ and $\rho$. Inverting \eqref{eq:liouvilledef} in Kruskal gauge, we find the dilaton has the form 
\beq \label{eq:phiHH}
\phi^{(\text{HH})}(x^{\pm})= \frac{12}{a N}\left(\frac{M}{\lambda}-\lambda^2 x^+x^-\right)+\frac{1}{2}W_{-1}\left(-\frac{24}{a N}e^{- \frac{24 }{aN}\left(\frac{M}{\lambda}-\lambda^2 x^+x^-\right)}\right)\;.
\eeq
Meanwhile, in Eddington-Finkelstein coordinates,
\be
\label{eq:phibrHH}
\phi^{(\text{HH})}(\sigma^{\pm})=\frac{12}{aN}\left(e^{2 \lambda \sigma^1}+\frac{M}{\lambda}\right)+\frac{1}{2} W_{-1}\left(-\frac{24}{a N}e^{-\frac{24}{a N}\left(e^{2 \lambda \sigma^1}+\frac{M}{\lambda}\right)}\right)\,,
\ee
where $2\sigma^1=\sigma^+-\sigma^-$. Note that for $a\neq 0$ the linear dilaton vacuum is not among the solutions \eqref{eq:phibrHH}.\footnote{To wit, take $M=0$:
\be
\label{eq:ldvHH}
\phi^{(\text{HH},M=0)}(\sigma)=\frac{12}{a N}e^{2 \lambda \sigma}+\frac{1}{2}W_{-1}\left(-\frac{24}{aN}e^{-\frac{24}{aN}e^{2\lambda\sigma}}\right)\,.
\ee
Asymptotically, i.e., $\sigma^1\rightarrow\infty$, this goes like $-\lambda\sigma^1-\frac{a N}{24}e^{-2\lambda\sigma^1}+\mathcal{O}\left(e^{-4\lambda\sigma^1}\right)$.
} Thus, the BPP model is the only one among the 1-parameter family of analytically solvable semi-classical models that has the linear dilaton vacuum as a solution.

For completeness, we also determine the function $\xi(x^+,x^-)=\xi_+(x^+)+\xi_-(x^-)$, which obeys the equation of motion given in \eqref{eq:tpmcongen}, as it will be useful in the following sections. In this gauge we find
\be \label{eq:xiHHvacum}
\xi_{\pm}=c_{\pm}-\ln\left(\pm \lambda( x^{\pm}\pm x_0^{\pm})\right)\;,
\ee
where the constants $c_{\pm}$ and $x_0^{\pm}$ are, for now, arbitrary.
Using the transformation law \eqref{eq:transformxi} with $\{z^{\pm}\}=\{\sigma^{\pm}\}$ and $\{w^{\pm}\}=\{x^{\pm}\}$, the function $\xi_{\pm}$ in Eddington-Finkelstein coordinates $(\sigma^{\pm})$ is expressed as
\beq
\xi_{\sigma^{\pm}}=c_{\pm}+\frac{1}{2}(\pm\lambda\sigma^{\pm})-\ln\left(e^{\pm \lambda\sigma^{\pm}}+\lambda e^{\pm\lambda \sigma^{\pm}_0}\right)\;.
\label{eq:xisigHH}\eeq

Lastly, let us explicitly evaluate the stress-energy tensor components \eqref{eq:Tppappv2} and \eqref{eq:Tmmappv2}. Using the dilaton \eqref{eq:phiHH} we find in Kruskal coordinates 
\be
\label{eq:NEC_HH}
\langle \text{HH}|T_{x^{\pm}x^{\pm}}|\text{HH}\rangle=\frac{N\lambda^2}{48}\left(\frac{24}{aN}\right)^2\frac{-1+W_{-1}\left(-\frac{24}{aN}e^{-\frac{24}{aN}(\frac{M}{\lambda}-\lambda^2x^+x^-)}\right)}{\left(1+W_{-1}\left(-\frac{24}{aN}e^{-\frac{24}{aN}(\frac{M}{\lambda}-\lambda^2x^+x^-)}\right)\right)^3}\frac{(-\lambda^2 x^+x^-)^2}{(\pm\lambda x^{\pm})^2}\,.
\ee
Meanwhile, in Eddington-Finkelstein coordinates, we attain
\be
\langle \text{HH}|T_{\sigma^{\pm}\sigma^{\pm}}|\text{HH}\rangle=\frac{12\lambda^2}{a^2N}e^{4\lambda\sigma^1}\frac{-1+W_{-1}\left(-\frac{24}{aN}e^{-\frac{24}{aN}(\frac{M}{\lambda}+e^{2\lambda\sigma^1})}\right)}{\left(1+W_{-1}\left(-\frac{24}{aN}e^{-\frac{24}{aN}(\frac{M}{\lambda}+e^{2\lambda\sigma^1})}\right)\right)^3}\,.
\ee
At the horizon, where $\sigma^1\rightarrow -\infty$, we see $\langle \text{HH}|T_{\sigma^{\pm}\sigma^{\pm}}|\text{HH}\rangle\rightarrow 0$. Alternatively, asymptotically far away, $\sigma^1\rightarrow\infty$, we have $\langle T_{\sigma^{\pm}\sigma^{\pm}}(\sigma^+,\sigma^-)\rangle\rightarrow\frac{N\lambda^2}{48}=\frac{N\pi^2 T^2}{12}$ which corresponds to a thermal bath at temperature $T=\frac{\lambda}{2\pi}$. 

Thus far we have left the interpolating parameter $a$ generic. The case $a=0$ (BPP) has some features worth commenting on. Introducing  $\upsilon=e^{2 \lambda \sigma^1}+\frac{M}{\lambda}$, for small $a$ the Lambert function and, consequently, the dilaton (\ref{eq:phiHH}), have the following expansions
\be
W_{-1}\left(-\frac{24}{aN}e^{-\frac{24}{aN}\upsilon}\right)=-\frac{24}{aN}\upsilon-\ln(\upsilon)\left(1+\frac{aN}{24}\frac{1}{\upsilon}+\mathcal{O}\left(a^2\right)\right)\Rightarrow\phi^{(\text{HH})}(\sigma^{\pm})=-\frac{1}{2}\ln(\upsilon)\left(1+\frac{aN}{24}\frac{1}{\upsilon}+\mathcal{O}\left(a^2\right)\right)\,.
\ee
 Thus, when $a=0$ and $M=0$ one recovers the linear dilaton vacuum solution , $\phi^{(\text{HH},a=0,M=0)}=-\lambda\sigma^1$. However, we saw that $\langle T_{\sigma^{\pm}\sigma^{\pm}}(\sigma^+,\sigma^-)\rangle=\frac{N\lambda^2}{48}$ at asymptotic infinity. Therefore, in this case the linear dilaton vacuum  describes a system in thermal equilibrium \cite{Bose:1995pz}.

A way to interpret this result is to include the $I_{a}$ action in with the matter stress-tensor \cite{Kim:1995jta}
\be
\label{eq:Tmatter}
T_{\mu \nu}^{\text{matter}}=-\frac{2\pi}{\sqrt{-g}}\frac{\delta (I_{\text{Poly}}+I_{a})}{\delta g^{\mu\nu}}=\langle T_{\mu\nu}\rangle + T^{(a)}_{\mu \nu}\,.
\ee
The second term,
\be
\label{eq:Ta}
T_{x^\pm x^\pm}^{(a)}=-\frac{N}{12}(1-a)\left(\partial_{\pm}^2 \phi-2\left(\partial_{ \pm} \rho\right)\left(\partial_{ \pm} \phi\right)\right)-\frac{N}{12}(1-2 a)\left(\partial_{ \pm} \phi\right)^2\,,
\ee
can be read directly from the right hand side of \eqref{eq:EoM_Tpp}. In Kruskal gauge ($\phi=\rho$) and for $a=0$, one finds $T_{x^\pm x^\pm}^{(a=0)}=-\langle T_{\mu\nu}\rangle$. Consequently, $T_{\mu \nu}^{\text{matter}}=0$, rendering the linear dilaton vacuum a solution with non-zero radiation.

\subsubsection{Boulware vacuum}

\noindent In Kruskal coordinates, the Boulware vacuum has $t_{\pm}(x^{\pm})=\frac{1}{4(x^{\pm})^{2}}$, and the Liouville field \eqref{eq:Omega} is easily found to be
\beq
\label{eq:omegaB}
\Omega^{(\text{B})}=\sqrt{\frac{12}{N}}\left(\frac{M}{\lambda}-\lambda^2 x^+x^--\frac{N}{48} \ln(-\lambda^2 x^+x^-)\right)\,.
\eeq
 Inverting \eqref{eq:liouvilledef}, the dilaton field in Kruskal gauge is
\beq \label{eq:phiB}
\phi^{(\text{B})}(x^{\pm})= \frac{12}{a N}\left(\frac{M}{\lambda}-\lambda^2 x^+x^-\right)-\frac{1}{4a}\ln(-\lambda^2x^+x^-)+\frac{1}{2}W_{-1}\left(-\frac{24}{a N}e^{- \frac{24 }{aN}\left(\frac{M}{\lambda}-\lambda^2 x^+x^-\right)+\frac{1}{2a}\ln(-\lambda^2x^+x^-)}\right)\,.
\eeq
Meanwhile, in Eddington-Finkelstein coordinates, we have
\be
\label{eq:phiBsigma}
\phi^{(\text{B})}(\sigma^{\pm})= \frac{12}{aN}\left(e^{2\lambda \sigma^1}+\frac{M}{\lambda}-\frac{N}{24}\lambda \sigma^1\right)+\frac{1}{2} W_{-1}\left(-\frac{24}{aN}e^{-\frac{24}{aN} \left(e^{2\lambda \sigma^1}+\frac{M}{\lambda}-\frac{N}{24}\lambda \sigma^1\right)}\right)\, ,
\ee
where $2\sigma^1=\sigma^+-\sigma^-$. The linear dilaton vacuum is a solution only for $a=0$ or $a=1/2$ (BPP or RST) and $M=0$.\footnote{Note that for $a=1/2$, we can use the property $x=W_{-1}(x e^x)$ in \eqref{eq:phiBsigma} to easily recover the linear dilaton vacuum solution.} 

Let us now determine $\xi(x^+,x^-)=\xi_+(x^+)+\xi_-(x^-)$, which obeys the equation of motion given in \eqref{eq:tpmcongen}. In Kruskal gauge, we have $t_{\pm}=\frac{1}{4(x^{\pm})^2}$ and we get 
\be \label{eq:xiBvacum}
\xi_{\pm}=c_{\pm}-\frac{1}{2}\ln\left(\pm \lambda x^{\pm}\right)
\ee
where the constants $c_{\pm}$ are arbitrary. Using the transformation law \eqref{eq:transformxi}, we can write $\xi_{\pm}$ in Eddington-Finkelstein coordinates $(\sigma^{\pm})$,
\bea
\xi_{\sigma^\pm}&=&c'_\pm-\ln(\pm\lambda(\sigma^\pm \pm\sigma_0^\pm))\;,
\eea
for some new coefficient $c'_{\pm}$.

Lastly, the stress-energy tensor components \eqref{eq:Tppappv2} and \eqref{eq:Tmmappv2} in Kruskal coordinates are
\bea
\label{eq:NEC_B}
\langle \text{B}|T_{x^{\pm}x^{\pm}}(x^{\pm})|\text{B}\rangle=&-&\frac{N\lambda^2}{48 (\lambda x^{\pm})^2}\left[\left(1-\frac{1}{a(1+W_{-1}\left(f\right))}\right)\right.\nonumber\\
&-&\left.\left(\frac{48}{aN}\right)^2\left(\frac{N}{48}+\lambda^2 x^+x^-\right)^2\frac{1}{\left(1+W_{-1}\left(\Upsilon\right)\right)^2}\left(\frac{1}{4}-\frac{1}{2}\frac{1}{1+W_{-1}\left(\Upsilon\right)}\right)\right]\,,
\eea
where 
\be
\Upsilon=-\frac{24}{aN}e^{-\frac{24}{aN}\left(\frac{M}{\lambda}-\lambda^2x^+x^-\right)+\frac{1}{2a}\ln(-\lambda^2x^+x^-)} \,.
\ee
Near the horizon, $x^+x^-=0$, $\Upsilon\rightarrow 0$, $W_{-1}(\Upsilon)\rightarrow -\infty$ and $\langle \text{B}|T_{x^{\pm}x^{\pm}}(x^{\pm})|\text{B}\rangle\rightarrow-\frac{N}{48 (x^{\pm})^2}$.  It is straightforward to write down the quantum stress-tensor in  Eddington-Finkelstein coordinates $\langle \text{B}|T_{\sigma^{\pm}\sigma^{\pm}}(x^{\pm})|\text{B}\rangle$, however, it is cumbersome and not particularly revealing. Performing an asymptotic expansion of the Lambert function we find $\langle \text{B}|T_{x^{\pm}x^{\pm}}(x^{\pm})|\text{B} \rangle\rightarrow 0+\mathcal{O}(e^{-4\lambda\sigma^1})$ as $\sigma^1\rightarrow\infty$.

\subsection{Energy conditions and singularities}

\noindent Alone, semi-classical gravitational field equations allow for any spacetime to be a solution sourced by some stress-tensor. To preclude the existence of ``exotic'' spacetimes with undesirable properties, one further imposes energy conditions \emph{ad hoc}. Such energy conditions are also core assumptions in the foundational singularity theorems and cosmic censorship conjectures of (semi-)classical gravity. In this regard, the pointwise null energy condition (NEC), $\langle T_{\mu\nu}\rangle \ell^{\mu}\ell^{\nu}\geq0$ for all null vectors $\ell^{\mu}$, and averaged null energy condition (ANEC), 
\be
\int_{\gamma}\langle T_{\mu\nu} \rangle \ell^{\mu}\ell^{\nu} \geq 0 \,,
\ee
for all complete null geodesics $\gamma$ with tangent $\ell^{\mu}$, standout (cf. \cite{Kontou:2020bta} for a review). The reason is that they are the hardest conditions to violate and, in the case of the pointwise conditions, all other conditions imply the NEC. Famously, the NEC is the condition used in Penrose's singularity theorem \cite{Penrose:1964wq} and Hawking's area theorem \cite{Hawking:1971vc} but is necessarily violated in the presence of Hawking radiation, even for dilatonic black holes \cite{Alexandre:2024htk}. Of interest is also the weaker achronal ANEC (AANEC) where $\gamma$ is taken to be an achronal null geodesics,\footnote{An achronal null geodesic is a null geodesic where there are no two points connected by a timelike curve.} where its self-consistent form is free of counterexamples in semi-classical gravity. 

Below we evaluate both the NEC and (half-)ANEC, calculating the expectation value of $\langle T_{\mu\nu}\ell^{\mu}\ell^{\nu}\rangle$ in the Hartle-Hawking and Boulware states for a null observer along the null coordinate $x^-$. In the Hartle-Hawking vacuum there is no NEC violation while in the Boulware vacuum both the NEC and the half-ANEC are violated. 

We will then analyze the singularity structure of the quantum black holes in both vacua. We do not apply singularity theorems that show the existence of incomplete geodesics, but instead we directly compute the Ricci scalar and examine the nature of the curvature singularities. We find that naked singularities cannot appear in the semi-classical regime for the Hartle-Hawking vacuum. In the case of the Boulware vacuum, however, a naked singularity can form in the large mass regime indicating a violation of the weak cosmic censorship conjecture.

\subsubsection{NEC and ANEC}

\begin{figure}[t!]
    \centering
    \includegraphics[width=0.45\linewidth]{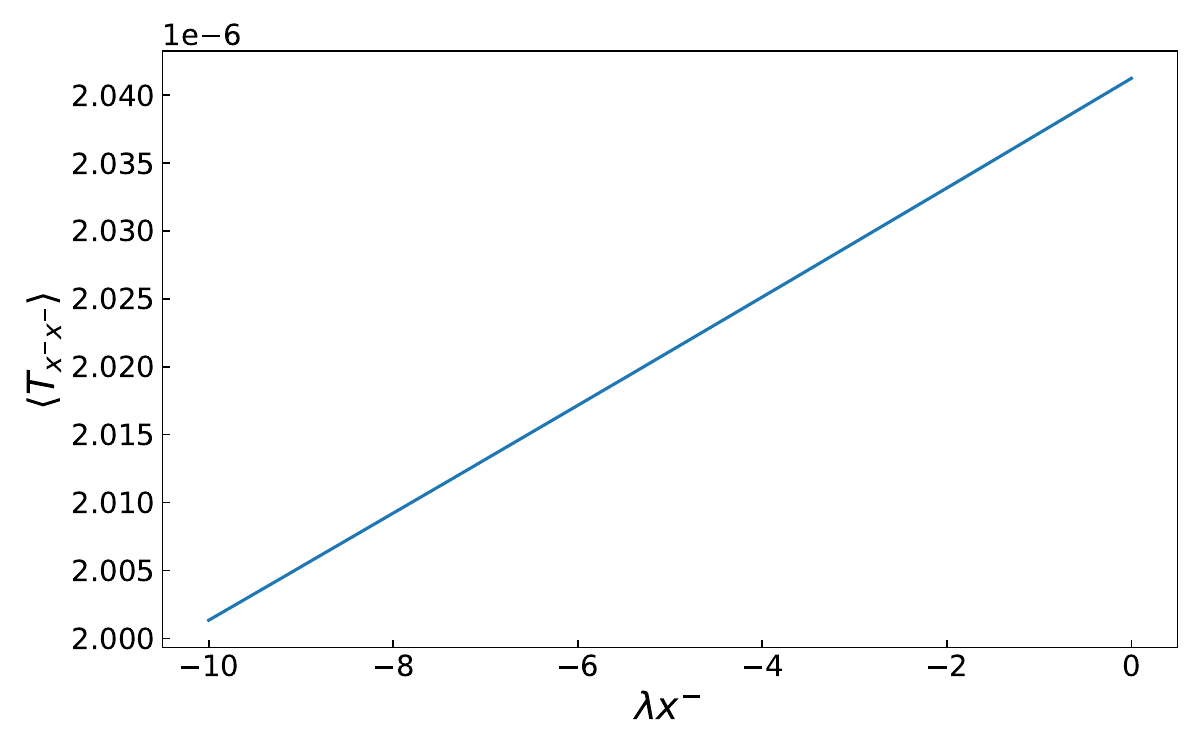}
    \includegraphics[width=0.45\linewidth]{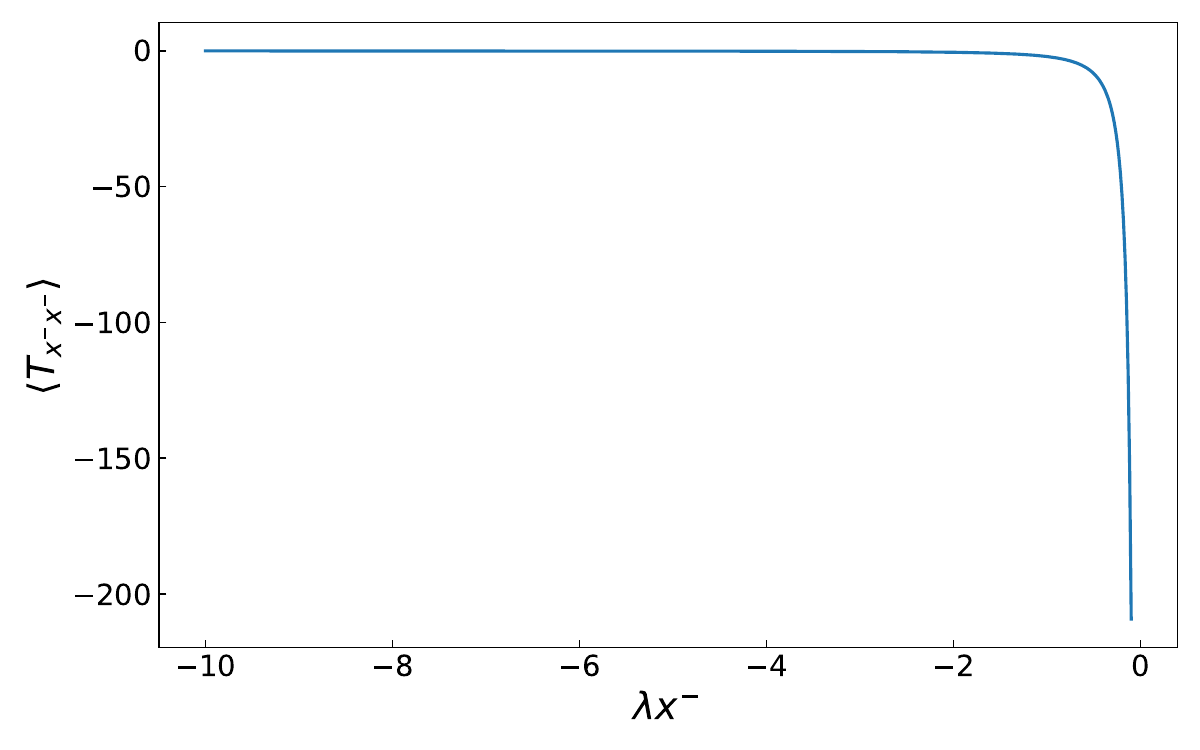}
    \caption{NEC for the Hartle-Hawking (left) and Boulware (right) vacua as a function of $\lambda x^-$ for $N=100$, $M/\lambda=1000$ and $a=1/2$ at  $\lambda x^+=1$.}
    \label{fig:TmmHB}
\end{figure}

\noindent For the Hartle-Hawking vacuum, the NEC$_{\text{HH}}$ is given by \eqref{eq:NEC_HH}. Since the Lambert function $W_{-1}(x)$ is defined between $-1/e$ with value $-1$ and $0$, where it diverges to $-\infty$, it is easy to see that NEC$_{\text{HH}}>0$ for the spacetime region outside the horizon for all $a$. Figure \ref{fig:TmmHB} displays NEC$_{\text{HH}}$ for the RST model, as an example. As expected, at the past horizon $\lambda x^+=0$, we find $\langle T_{x^-x^-}\rangle=0$. 
Since the NEC$_{\text{HH}}$ is never violated, its average will also not be violated such that the ANEC$_{\text{HH}}$ holds. It is possible to distinguish between contributions to \eqref{eq:NEC_HH} coming from the black hole background and the thermal radiation. 
If we exclude the `thermal' contribution as in \cite{Alexandre:2024htk} then the NEC$_{\text{HH}}$ can be violated, depending on $a$.

For the Boulware vacuum, the NEC$_{\text{B}}$ is given by \eqref{eq:NEC_B}. It is easy to see that this expression allows for $\langle T_{\pm\pm}\rangle <0$ for all $a$, violating the NEC. A particular example is presented in Figure \ref{fig:TmmHB} for the RST model. While the NEC may be violated, let us consider the behavior of the  NEC$_{\text{B}}$ averaged along an arbitrary complete null geodesic. A numerical evaluation, displayed in Figure \ref{fig:TmmHB}, indicates a violation of the half-ANEC, i.e., integrating along a null geodesic from asymptotic infinity to the horizon. (It is impossible to evaluate the ANEC in this case as there are no complete null geodesics.) The violation of the half-ANEC is not uncommon for quantum fields \cite{Kontou:2020bta}. In the next section, we show this solution describes a wormhole throat that terminates at a null singularity at $-\lambda^2 x^+x^-=0$. Additionally, we can take the limit $\lambda x^+\rightarrow 0$ in \eqref{eq:NEC_B}, finding $\langle T_{\pm\pm}\rangle\rightarrow -\frac{N}{48}\frac{1}{(x^\pm)^2}$. Integrating between $-\infty$ and a point $x_0^-<0$ near the horizon, we find, for any $a$, $\int_{-\infty}^{x_{0}}\langle T_{--}\rangle dx^{-}\to\frac{N}{48}\frac{1}{x_0^-}$.

\subsubsection{Singularities in semi-classical solutions} 

\noindent The inclusion of backreaction changes the location and the nature of the singularities relative to the classical black hole. Indeed, the Ricci scalar, 
\be
R=\frac{8 e^{-2\rho}}{\Omega'}\left(\partial_+\partial_-\chi-\frac{\Omega''}{\Omega'^2}\partial_+\Omega\partial_-\Omega\right)\,,
\ee
is clearly diverging when $\Omega'\equiv\frac{d\Omega}{d\phi}=0$, or, equivalently, where $\sqrt{\frac{N}{12}}a-2\sqrt{\frac{12}{N}}e^{-2\phi}=0$.
For $a\neq0$, this singularity corresponds to the dilaton value 
\be
\phi_{\text{cr}}^{(a)}=-\frac{1}{2}\ln\left( a \frac{N}{24} \right) \,.
\ee
Note that no such singularity occurs for the BPP model ($a=0$). For $N=0$ we recover the location of the classical black hole singularity.  Let us now analyze the singularity structure of the backreacted geometry when the quantum matter is in the Hartle-Hawking and Boulware states.

\vspace{2mm}

\begin{figure}[t!]
\centering
    \includegraphics[height=5cm]{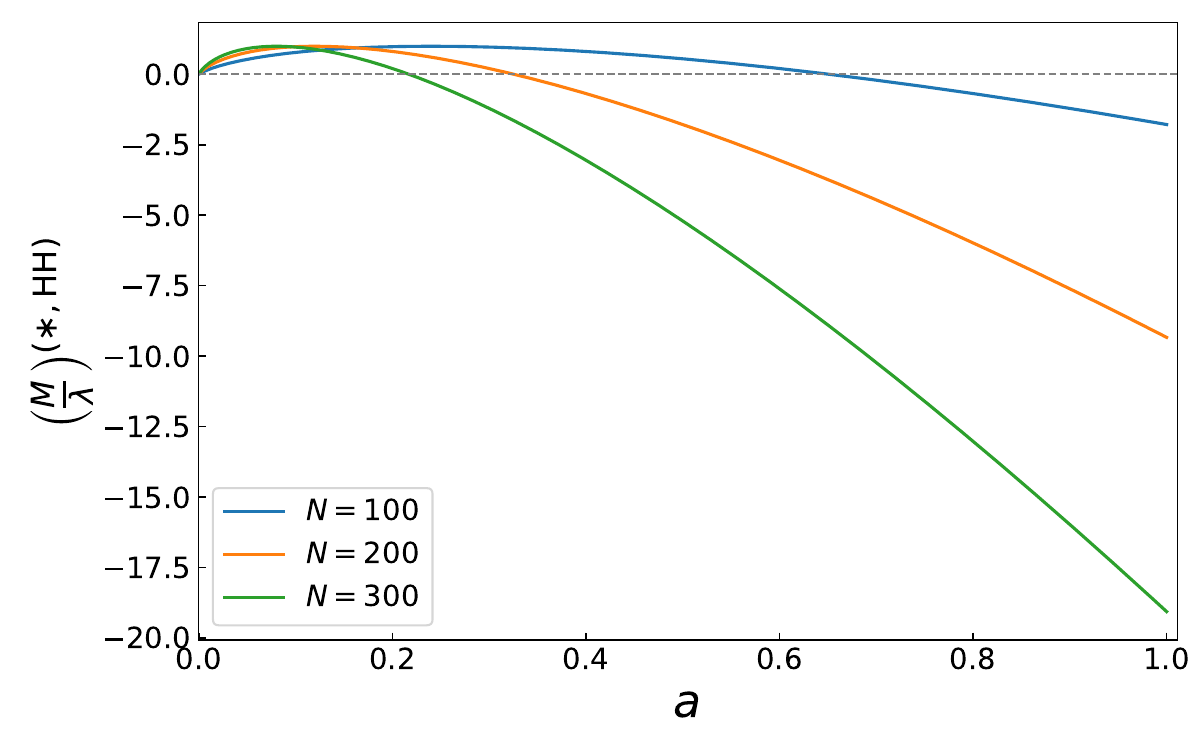}\qquad 
    \includegraphics[height=5cm]{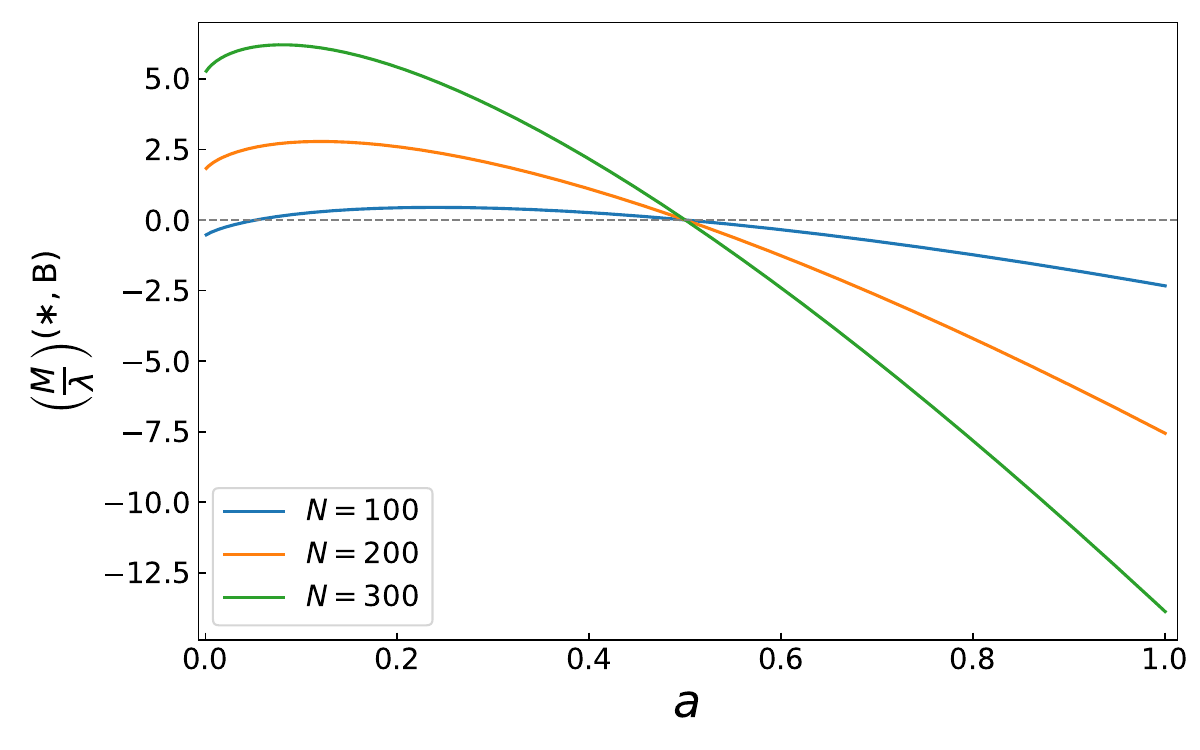}
    \caption{Critical mass $M^{\ast}/\lambda$. \emph{Left.} Hartle-Hawking; $N=100,200,300$ (top, blue; middle orange; bottom black). \emph{Right.} Boulware; $N=100,200,300$ (bottom left, blue; middle left; top left, black).}
\label{fig:critmass}\end{figure}

\noindent $\bullet$ \emph{Hartle-Hawking vacuum.} 
Evaluating $\Omega$ at $\phi_{\rm cr}^{(a)}$ and using the solution for the HH vacuum \eqref{eq:omegaH}, we find the location of the constant $\phi_{\rm cr}^{(a)}$ curve
\be
\label{eq:singHHloc}
\lambda^2x_s^+x_s^-=\frac{M}{\lambda}-\frac{N}{24}a\left(1-\ln\left(\frac{aN}{24}\right)\right)\,.
\ee
This represents a spacelike singularity provided
\be
\label{eq:MinHH}
\frac{M}{\lambda}>\left(\frac{M}{\lambda}\right)^{(\ast,\text{HH})} \,,
\ee
where 
\be
\left(\frac{M}{\lambda}\right)^{(\ast,\text{HH})}\equiv \frac{N}{24}a\left(1-\ln\left(\frac{a N}{24}\right)\right)\,.
\ee
The possible cases are:
\begin{itemize}
\item 
$M/\lambda > (M/\lambda)^{(\ast,\text{HH})}$ : There exists a spacelike singularity behind a horizon.
\item 
$M/\lambda = (M/\lambda)^{(\ast,\text{HH})}$ : Critical case, a null singularity lies at the horizon $x^+_sx^-_s=0$, see \eqref{eq:singHHloc}. 
\item 
$M/\lambda < (M/\lambda)^{(\ast,\text{HH})}$ : Naively, here in principle one could have a timelike naked singularity. It is important to note, however, when the critical mass becomes negative, a naked singularity cannot occur. As shown in Figure \ref{fig:critmass}, the critical mass becomes negative for smaller and smaller values of $a$ as $N$ increases. The maximum value of $\left(M/\lambda \right)^{(\ast,\text{HH})}$ is at $a=24/N$ and gives $\left(M/\lambda \right)^{(\ast,\text{HH})}=1$ at any $N$. However, recall the semi-classical regime of validity \eqref{eq:scvalidity}. For self-consistency, $M/\lambda \gg 1$ and the case $M/\lambda \leq  \left(M/\lambda \right)^{(\ast,\text{HH})}$ is disallowed. Thus, semi-classically such naked singularities do not appear.

\end{itemize}

\noindent $\bullet$ \emph{Boulware vacuum.} In this state the scalar curvature develops two potential singularities. 
The first arises whenever $d\Omega/d\phi=0$ as in the Hartle-Hawking case. The second singularity is associated with the term $\partial_+\Omega^{(B)}\partial_-\Omega^{(B)}\sim 1/(-\lambda^2 x^+x^-)$ at the location of the classical horizon $x^+x^-=0$, which in the Boulware state becomes  a null curvature singularity.

\begin{figure}[t!]
    \centering
    \includegraphics[width=0.5\linewidth]{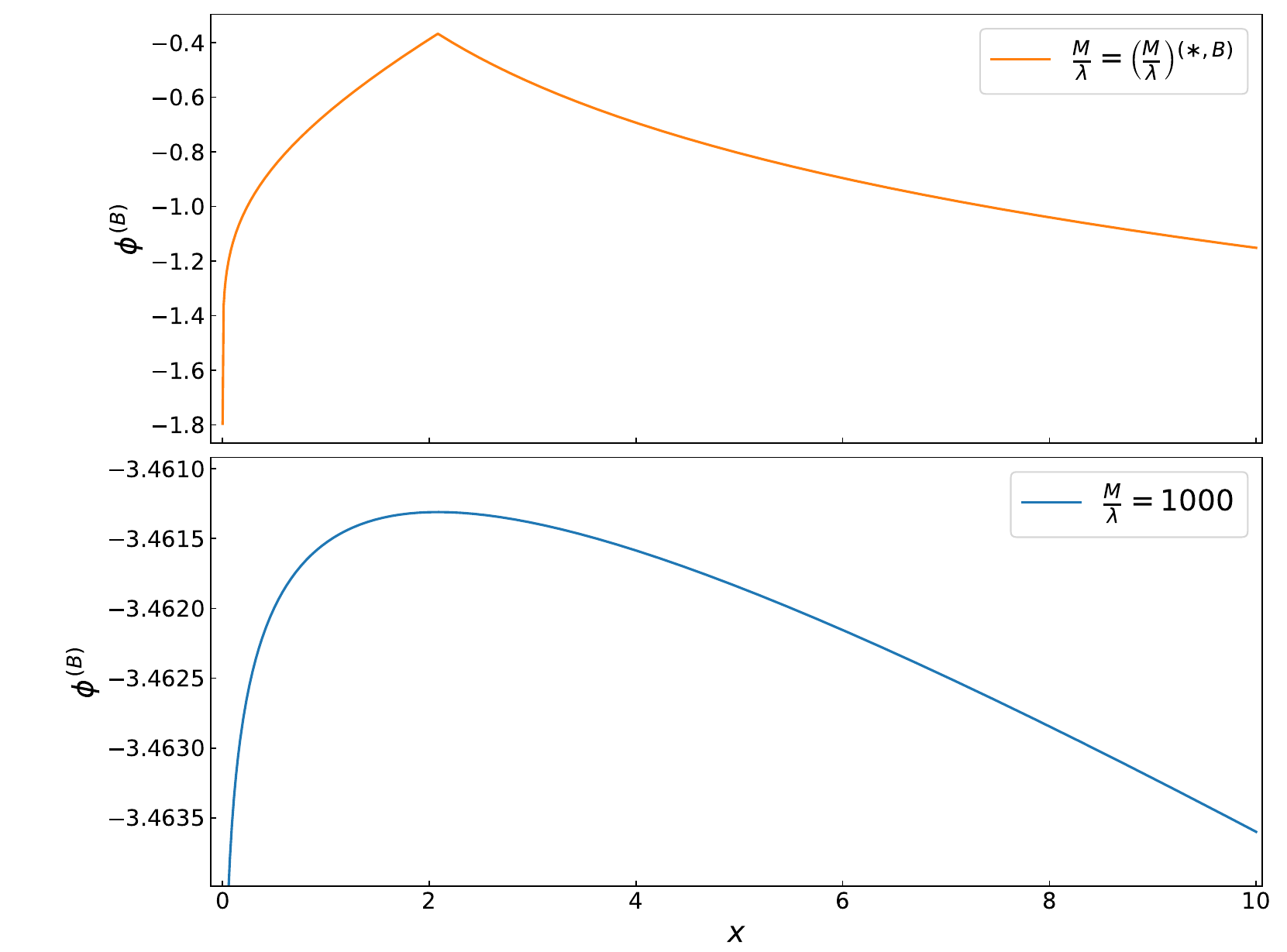}
    \caption{$\phi^{(\text{B})}(x)$ as a function of $x$ for $N=100$, $a=1/2$ and masses $M/\lambda=1000$ and $M^{(*,\text{B})}/\lambda$.}
    \label{fig:phi_B}
\end{figure}

To analyze this structure, it is convenient to study the explicit dilaton solution
\eqref{eq:phiB} as a function of $x\equiv -\lambda^2 x^+x^-$, see Figure \ref{fig:phi_B}. At spatial infinity ($x\to\infty$), the dilaton diverges to $-\infty$ with the asymptotic expansion
\be
\phi^{(\text{B})}(x)=-\frac{1}{2}\ln x-\frac{1}{2x}\!\left(\frac{M}{\lambda}+\Big(\frac{aN}{24}-\frac{N}{48}\Big)\ln x\right)+\mathcal{O}\!\left(\frac{1}{x^2}\right).
\ee
As we move inward, $\phi^{(B)}(x)$ increases and attains a maximum $\phi_m$ at
$x_m=N/48$, after which it decreases again toward $-\infty$ as $x\to 0$. 
Explicitly,
\be
\phi_m=\frac{1}{4a}\left(1-\ln\left(\frac{N}{48}\right)\right)+\frac{12}{aN}\frac{M}{\lambda}+\frac{1}{2}W_{-1}\left(\frac{2}{a}\left(\frac{12}{N}\right)^{1-\frac{1}{2a}}e^{-\frac{24}{aN}\left(\frac{N}{48}+\frac{M}{\lambda}\right)}\right)\,.
\ee
The value $\phi_m$ represents the minimal radius of the throat of a wormhole from a higher-dimensional interpretation \cite{Potaux:2021yan,delRio:2024fxn}. 
Thus, the geometry resembles a wormhole throat. The second branch of the wormhole, however, is replaced by a null singularity at the classical horizon $x=0$, which lies at a finite affine distance.

The singularity structure depends on the relative position of $\phi_{\text{cr}}^{(a)}$ and the maximal value $\phi_m$. The point $\phi_{\text{cr}}^{(a)}$ corresponds to the minimum of $\Omega(\phi)$, with value $\Omega_{\rm cr}=\Omega(\phi_{\rm cr}^{(a)})$. The solution $\Omega^{(\text{B})}(x)$, however, need not reach this value. The extremum of $\Omega^{(\text{B})}(x)$ occurs at $x_m=N/48$, consistently with the location of $\phi_m$. Imposing the condition
$\Omega_{\rm cr}=\Omega^{(\text{B})}(x_m)$
fixes a critical mass 
\be
\label{eq:Mstar}
\left(\frac{M}{\lambda}\right)^{(\ast,\text{B})}=\frac{aN}{24}\left(1-\ln\left(\frac{aN}{24}\right)\right)-\frac{N}{48}\left(1-\ln\left(\frac{N}{48}\right)\right) \,.
\ee

The possible cases are then:
\begin{itemize}
    \item $M/\lambda >(M/\lambda)^{(\ast,\text{B})}$: Here $\phi_{\text{cr}}^{(a)}>\phi_m$, so the critical value is never reached. In this case the horizon never forms so the geometry is not describing a black hole. Instead we can say it describes a wormhole throat that terminates at the null singularity at $x=0$. 
    \item $M/\lambda=(M/\lambda)^{(\ast,\text{B})}$: In this critical case, $\phi_{\text{cr}}^{(a)}=\phi_m$ and the curvature remains finite at $\phi_{\text{cr}}^{(a)}$. We can see this by doing an expansion of $R$ for $\phi\sim\phi_{\text{cr}}^{(a)}+\delta \phi$ and $\Omega(\phi_{\text{cr}}^{(a)}+\delta \phi)=\Omega^{(\text{B})}(x_m+\delta x)$. We find 
    \be
    R\sim\left(\frac{12}{aN}\right)\sqrt{\frac{8}{a}}\lambda^2\,, \quad \text{at} \quad \phi_{\text{cr}}^{(a)}
    \ee
    removing the singularity. However, the null singularity at $x=0$ remains, see Figure \ref{fig:phi_B}.
    \item $M/\lambda<(M/\lambda)^{(\ast,\text{B})}$: Now $\phi_{\text{cr}}^{(a)}<\phi_m$, so the solution runs into the singular point at finite $\phi$, yielding a naked singularity. As we see in Fig.~\ref{fig:critmass} this can occur only for $0 < a<1/2$ as for $a>1/2$ the critical mass is negative. The maximum value of the critical mass is when $a=24/N$, is equal to 
    \be
    \left(\frac{M}{\lambda}\right)^{(\ast,\text{B})}_{\max}=1-\frac{N}{48}\left(1-\ln{\left(\frac{N}{48}\right)}\right) \,.
    \ee
    Numerically, we can compute when this critical value is larger than $N$ to verify it lies in the semi-classically valid regime (\ref{eq:scvalidity}). For $a=24/N$, the value of $N>1$ where the critical mass is equal to $N$, up to an order one factor, occurs for $N\sim 10^{23} - 10^{200}$. Thus, $(M/\lambda)^{(\ast,B)}_{\text{max}}>M/\lambda \gg N$,  is only for exceedingly large masses and $a$ is exceedingly small. However, note that $a$ need not be exceedingly small in general. That is, there exist values $0<a\ll1/2$ such that $(M/\lambda)^{(\ast,B)}>M/\lambda \gg N$, though $N$ must still be very large.  
\end{itemize}

To summarize, we find that naked singularities cannot occur in the Hartle-Hawking vacuum in the semi-classical regime. In the Boulware vacuum, however, we have a wormhole geometry consistent with~\cite{Potaux:2021yan,delRio:2024fxn}. Additionally, we find naked singularities can occur in a regime of very large masses, or black holes with exceedingly large entropy. Technically, this indicates a violation of weak cosmic censorship. 

Naked singularities in the RST and interpolating models have been observed before \cite{Russo:1992ax,Bose:1995pz,Cruz:1995zt,Solodukhin:1995te}, in static and dynamical geometries. In particular, if one follows the semi-classical dynamics of an evaporating CGHS black hole without imposing boundary conditions in the strong-coupling region, the endpoint singularity is naked; future evolution is not uniquely determined without imposing extra data. To circumvent the singularity, one can impose reflecting boundary conditions along a critical line where the effective coupling becomes strong \cite{Russo:1992ax,Russo:1992yh}. Such boundary conditions excise the singular region, and the black hole will evaporate completely. We will revisit our finding in the discussion Section \ref{sec:discussion}.

\section{Mass and entropy of quantum black holes} \label{sec:massentqbhs}

\noindent Here we analyze the mass and entropy of the exact quantum black hole solutions described in Section \ref{sec:QBHsolutions}. We do so by extending the quasi-local methods in the classical context to compute the quasi-local Brown-York energy and its asymptotic limit. We then evaluate the quantum black hole entropy, showing that Wald entropy equals the sum of the gravitational and fine grained quantum matter entropies, namely, the generalized entropy. Lastly, we examine the quasi-local thermodynamics of the quantum black hole. 

\subsection{Quasi-local and asymptotic mass}

\noindent We apply the quasi-local method as we did in the CGHS model to evaluate the ADM mass for the semi-classical extension. Now the total action is given by $I_{\text{CGHS}}+I_{\text{Poly}}+I_{a}$.  The  quasi-local Brown-York stress-energy tensor is (see Appendix \ref{app:energies} for details) 
\beq
\begin{split}
\tau^{\mu\nu}
&=\gamma^{\mu\nu}n^{\alpha}\partial_{\alpha}\phi\left(\frac{N}{12\pi}(a-1)-\frac{2}{\pi}e^{-2\phi}\right)+\frac{N}{24\pi}\gamma^{\mu\nu}n^{\alpha}\partial_{\alpha}Z\,.
\end{split}
\label{eq:BYstresstot}\eeq
resulting in the quasi-local Hamiltonian (without counterterms)
\beq
\begin{split}
H_{\zeta}&=\oint_{\partial\Sigma}\epsilon_{\partial\Sigma}g_{\mu\nu}\zeta^{\mu}u^{\nu}\left[n^{\alpha}\partial_{\alpha}\phi\left(\frac{N}{12\pi}(a-1)-\frac{2}{\pi}e^{-2\phi}\right)+\frac{N}{24\pi}n^{\alpha}\partial_{\alpha}Z\right]\;.
\end{split}
\label{eq:quasiHamfull}\eeq

Unsurprisingly, the quantum corrections do not ameliorate any of the long-distance divergences. To see this, recall that in Eddington-Finkelstein coordinates, we have $\zeta^{\mu}=\delta^{\mu}_0=e^{\rho}u^{\mu}$ and $n^{\mu}=e^{-\rho}\delta^{\mu}_1$, such that $\mathcal{N}=e^{\rho}$. Using this, the total quasi-local Hamiltonian becomes
\be
\label{eq:Hzeta}
H_{\zeta}=\oint_{\partial\Sigma}\epsilon_{\partial\Sigma}\left[\partial_1\phi\left(\frac{2}{\pi}e^{-2\phi}-\frac{N}{12\pi}(a-1)\right)-\frac{N}{24\pi }\partial_1 Z\right]\;.
\ee
Consider, for example, the HH vacuum solution (\ref{eq:phiHH}) with $x_{0}^{\pm}=0$, 
Asymptotically, 
\be
\label{eq:Hxidiverg}
\lim_{\sigma^{1}\to\infty}H_{\zeta}\approx\frac{1}{\pi}\oint_{\partial\Sigma}\epsilon_{\partial\Sigma}\left[-2\lambda e^{2\lambda\sigma^1}+\frac{N}{12\pi}(a-2)\lambda+\mathcal{O}\left(e^{-2\lambda\sigma^1}\right)\right]\,.
\ee
Thus, the leading order divergence in $\sigma^{1}$ is same as in the CGHS model;  the quantum corrections do not introduce additional divergences.

Consequently, we need to introduce local counterterms at the level of the action to regularize these divergences. Given the above divergence, a natural first guess is to the use the local counterterm for the classical CGHS action (\ref{eq:localctCGHS}). We show in Appendix \ref{app:Waldformalism}, however, this is insufficient to remove all divergences--- except for the BPP model ($a=0$). The local CGHS counterterm (due to quantum corrections to the dilaton) produces subleading divergences linear in $\sigma^{1}$.
 
By inspecting $H_{\zeta}$ for the solution of the semi-classical theory in the Hartle-Hawking and the Boulware vacua, we find that the following counterterm removes the divergences
\be
\label{eq:Ict_semi}
I_{\text{ct}}=-\frac{1}{\pi}\int_{B}\sqrt{\gamma}(2\lambda) e^{-2\phi}+\frac{1}{\pi}\int_{B}\sqrt{\gamma} \frac{N\lambda}{24}(1-2a+\mathfrak{c}) \phi\,,
\ee
where $\mathfrak{c}$ is a state-dependent constant
\be
\mathfrak{c}_{\text{HH}}=-1\,,\qquad\mathfrak{c}_{\text{B}}=0\,.
\ee
In Appendix \ref{ap:HJ}, we derive this counterterm using the Hamilton-Jacobi method, generalizing \cite{Grumiller:2007ju} to the semi-classical context (and differing from the counterterm used in \cite{Solodukhin:1995te}). 

With the above counterterm, we find that we add to the stress-tensor (\ref{eq:BYstresstot}) and Hamiltonian (\ref{eq:quasiHamfull}) the following
\beq 
\begin{split} 
&\tau^{\mu\nu}_{\text{ct}}=-\gamma^{\mu\nu}\left(\frac{2\lambda}{\pi}e^{-2\phi}-\phi\frac{N\lambda}{24\pi}(1-2a+\mathfrak{c})\right)\;,\\
&H_{\zeta}^{\text{ct}}=-\oint_{\partial\Sigma}\epsilon_{\partial\Sigma}g_{\mu\nu}\zeta^{\mu}u^{\nu}\left(\frac{2\lambda}{\pi}e^{-2\phi}-\phi\frac{N\lambda}{24\pi}(1-2a+\mathfrak{c})\right)\;.
\end{split}
\label{eq:ctstressandHam}\eeq
Altogether, the total Hamiltonian including counterterms is 
\beq 
\begin{split}
 H_{\zeta}^{\text{total}}&=\oint_{\partial\Sigma}\epsilon_{\partial\Sigma}g_{\mu\nu}\zeta^{\mu}u^{\nu}\biggr\{-\frac{2}{\pi}e^{-2\phi}(n^{\alpha}\partial_{\alpha}\phi+\lambda)+\frac{N}{24\pi}n^{\alpha}\partial_{\alpha}Z+\frac{N}{12\pi}\left[(a-1)n^{\alpha}\partial_{\alpha}\phi+\frac{\lambda}{2}(1-2a+\mathfrak{c})\phi\right]\biggr\}\;.
\end{split}
\label{eq:fullscHam}\eeq
We will demonstrate below that this results in finite asymptotic energy. Additionally, one may always add to the counterterm Lagrangian a constant. We will return to this point momentarily.

To calculate the ADM mass, we evaluate $H_{\zeta}$ on-shell and take the limit $\sigma^1\rightarrow\infty$, as done for the classical CGHS black hole. The on-shell solutions depend on the state of the quantum matter field, leading to state dependent asymptotic mass. Specifically, we find for Hartle-Hawking state  \eqref{eq:phiHH} and Boulware state \eqref{eq:phiB} that the ADM mass goes like
\beq
\label{eq:ADMHH_H}
M_{\text{ADM}}^{(\text{HH})}=\lim_{\sigma^1\rightarrow\infty} H_{\zeta}=\frac{M}{\pi}-\frac{N\lambda}{6\pi}\;,
\eeq
\beq
M_{\text{ADM}}^{(\text B)}=\lim_{\sigma^1\rightarrow\infty} H_{\zeta}=\frac{M}{\pi}-\frac{N\lambda}{24\pi}\,.
\eeq
Although we find a negative constant being subtracted, note that the ADM mass is never negative when we insist being in the semi-classical regime of validity (\ref{eq:scvalidity}). Further, while the ADM mass is model independent, the mass is different depending on the vacuum state. We have additional freedom at the level of the counterterm (\ref{eq:Ict_semi}) to add a (state-dependent) constant so that the ADM mass in both vacua can be made equal to the Boulware value. Specifically, consider adding the counterterm $I^{\mathfrak{c}_{0}}_{\text{ct}}=\frac{1}{\pi}\int_{\partial\mathcal{M}}dy\sqrt{\gamma}(2 \mathfrak{c}_0)$. For the Hartle-Hawking state, this will modify the ADM mass to
\beq
\label{eq:ADMHH_Hv2}
M_{\text{ADM}}^{(\text{HH})}=\lim_{\sigma^1\rightarrow\infty} H_{\zeta}=\frac{M}{\pi}-\frac{N\lambda}{6\pi}-\mathfrak{c}_{0}\;.
\eeq
Thus, for $\mathfrak{c}_0=-\frac{N\lambda}{8}$, the ADM mass in the HH and Boulware vacua agree. Further, we can adjust $\mathfrak{c}_{0}=-N\lambda/6\pi$ and $\mathfrak{c}_{0}=-N\lambda/24\pi$ for the Hartle-Hawking and Boulware states, respectively, and see that the ADM mass is precisely equal to the value of the classical CGHS black hole.

As in the classical case, we evaluate the quasi-local energy seen by an observer at the finite timelike boundary $B$ for the semi-classical black hole. Let us carry out this procedure when the quantum matter is in the Hartle-Hawking state. The Brown-York energy density for the semi-classical black hole follows directly from the Hamiltonian (\ref{eq:fullscHam}),
\be
E_{\text{BY}}=\frac{2}{\pi}e^{-2\phi}\left(n^{\mu}\partial_{\mu}\phi+\lambda\right)-\frac{N}{24\pi}n^{\mu}\partial_{\mu}Z+\frac{N}{12\pi}\left[(1-a)n^{\mu}\partial_{\mu}\phi+a\lambda\phi\right]\;.
\ee
We use the solutions in the HH state in Eddington-Finkelstein coordinates and Kruskal gauge to write the auxiliary field in terms of $\phi$ (\ref{eq:xisigHH}), $Z=2\rho-2\xi=2\phi+4\lambda\sigma^1-2c_{\xi}$. For this solution, the lapse in conformal gauge is\footnote{Here we inverted the exact solution \eqref{eq:phibrHH} to write $\sigma^1$ in terms of the $\phi$
$$
\sigma^1=-\frac{\phi}{\lambda}+\frac{1}{2\lambda}\ln\left(1-e^{2\phi}\left(\frac{M}{\lambda}-\frac{aN}{12}\phi\right)\right)\,,
$$
and used that in Eddington-Finkelstein coordinates $e^{-2\rho}=1+\frac{M}{\lambda}e^{-2\lambda\sigma^{1}}$.}
\be
\label{eq:lapse}
\mathcal{N}=-g_{\mu\nu}\zeta^{\mu}u^{\nu}=e^{\rho}=\sqrt{1-e^{2\phi}\left(\frac{M}{\lambda}-\frac{aN}{12}\phi\right)}\,.
\ee
And the Brown-York energy at position $\phi_B$ is
\beq
\label{eq:EBY_semi}
E_{\text{BY}}=\frac{2\lambda}{\pi}e^{-2\phi_{B}}(1-\mathcal{N})\left(1-\frac{N}{12}\frac{e^{2\phi_B}}{\mathcal{N}(1-\mathcal{N})}\left( 1-\frac{a\phi_{B}\mathcal{N}}{2}\right)\right)\;.\eeq
The semi-classical Brown-York energy is equal to the classical energy \eqref{eq:BYeneternalCGHSv2} plus a quantum correction, reducing to the classical energy in the limit $N\to0$. 

\subsection{Entropy}

 \noindent Here we analyze the entropy of quantum black hole solutions to the semi-classical interpolating model. 

\subsubsection{Wald entropy}

\noindent For arbitrary diffeomorphism invariant gravity theories, the entropy for stationary black holes is quantified by the N\"other charge associated with the Killing vector field $\zeta^\mu$ generating the bifurcate horizon of the black hole, namely, the Wald entropy functional \cite{Wald:1993nt}, denoted $S_{\text{Wald}}$. When the Killing vector field is normalized to have unit surface gravity $\kappa=1$, Iyer and Wald showed that the Wald functional takes the explicit form  \cite{Wald:1993nt,Iyer:1994ys},
\beq S_{\text{Wald}}=-2\pi\int_{H}dA\frac{\partial \mathcal{L}}{\partial R^{\mu\nu\rho\sigma}}\epsilon_{\mu\nu}\epsilon_{\rho\sigma}\;,\label{eq:IyerWaldentgen}\eeq
where $\epsilon_{\mu\nu}$ is the binormal satisfying $\epsilon_{\mu\nu}\epsilon^{\mu\nu}=-2$, $dA$ the infinitesimal area element of the codimension-2 Killing horizon $H$, and $\mathcal{L}$ is the Lagrangian (density) of the theory.  Using $\frac{\partial R}{\partial R_{\mu\nu\rho\sigma}}=\frac{1}{2}(g^{\mu\rho}g^{\nu\sigma}-g^{\mu\sigma}g^{\nu\rho})$, we see for any diffeomorphism-invariant Lagrangian whose curvature dependence is linear, $\mathcal{L}_{\mathcal{X}}=\sqrt{-g}\mathcal{X}R$ for some scalar function $\mathcal{X}$ independent of background curvature, the Iyer-Wald entropy (\ref{eq:IyerWaldentgen}), in two-dimensions  evaluates to
\beq
S^{\mathcal{X}}_{\text{Wald}}=4 \pi \mathcal{X}\Big|_{H}\;.
\eeq
where used that in two-dimensions the codimenision-2 integral is replaced with evaluating the integrand at the bifurcation point of the horizon. 

Applied to the semi-classical interpolating model, the Wald entropy receives contributions each from the CGHS term \eqref{eq:CGHSLor}, the Polyakov term written in its local (auxiliary-field) form \eqref{eq:Polyactv2}, and the  RST/BPP term \eqref{eq:IbRSTBPPact}. Altogether, the Wald entropy of the semi-classical interpolating models is 
\beq
S_{\text{Wald}}=\Big(2 e^{-2 \phi}+\frac{N}{12} Z+\frac{N}{6}(a-1) \phi\Big) \Big|_H\;.
\label{eq:Swaldgenint}\eeq
 A similar treatment for the Wald entropy in the RST model ($a=1/2$) was first examined in \cite{Myers:1994sg}. Unsurprisingly, the horizon entropy is modified due to quantum backreaction effects, here encoded the dilaton $\phi$ and auxiliary field $Z$; turning off quantum effects, we recover the entropy of the classical CGHS model. Momentarily we will provide an interpretation to the additional terms proportional to $N$. 

On-shell, the auxiliary field is $Z=2\rho-2\xi$ via \eqref{eq:solnZconfg}. In Kruskal gauge, where $\rho=\phi$, the Wald functional remarkably evaluates to
\beq 
\begin{split} 
S_{\text{Wald}}&=2e^{-2\phi_{H}}+\frac{aN}{6}\phi_{H}-\frac{N}{6}\xi_{H}\\
&=\frac{2M}{\lambda}-\frac{N}{6}\xi_{H}\;,
\end{split}
\label{eq:Swaldsemiv2}\eeq
where to arrive to the second line we substituted in the general solution \eqref{eq:phiHH} evaluated on the horizon, using the property $e^{-W_{-1}(z)}=W_{-1}(z)/z$, for $z=-\frac{24}{aN}e^{-24M/aN\lambda}$. Thus, the semi-classical Wald functional evaluates to the entropy of the \emph{classical} black hole, plus a correction proportional to the function $\xi_{H}$ \cite{Solodukhin:1995te,Alexandre:2024htk}. 
We will provide an interpretation of this second term momentarily.

It is not \emph{a priori} clear what kind of `entropy' the Wald functional represents. Indeed, originally the Wald functional was the quantity that replaced the horizon area term in the covariant derivation of first law black hole mechanics \cite{Wald:1993nt}.  At this stage, the Wald functional has not yet been imbued with a thermodynamic interpretation. Such an interpretation is state-dependent. In particular, the Boulware vacuum does not yield a thermal density matrix when restricted to the exterior of a black hole, and thus, strictly speaking, the Wald functional in the Boulware vacuum does not have a thermodynamic interpretation. Meanwhile, the Hartle-Hawking vacuum is a thermal state, and hence the Wald functional (\ref{eq:Swaldgenint}) evaluated in the Hartle-Hawking vacuum (via the solutions $\phi$ and $Z$) is a genuine thermodynamic entropy. Consequently, the Wald entropy in the Hartle-Hawking vacuum is expected to have a statistical interpretation.

\subsubsection{Microcanonical action and entropy}

\noindent In Section \ref{sec:CGHSbhs} we explicitly derived the (quasi-local) thermodynamics of the eternal classical CGHS black hole via a canonical partition function $\mathcal{Z}(\beta)$, for appropriate inverse temperature $\beta$, expressed as a Euclidean path integral (\ref{eq:canonpartfunc}). 

In ordinary statistical thermodynamics, the (microcanonical) entropy at a fixed energy $E_{0}$ is directly given by the logarithm of the microcanonical density of states or partition function $W(E_{0})$. Brown and York \cite{Brown:1992bq} (see also \cite{Brown:1989fa}), recognized that, for theories of gravity, the density of states can be cast as a Euclidean path integral over field configurations at a fixed energy. This is possible because in gravity, the total energy of a system is entirely characterized by gravitational field variables at the boundary.

The difference between the canonical and microcanonical partition functions is what the functional integral is weighted by, respectively dubbed the (Euclidean) ``canonical'' and ``microcanonical'' actions. Specifically, the density of states is heuristically expressed as
 \beq W(E_{0})=\int \mathcal{D}\psi \hspace{1mm} e^{-I_{\text E}^{\text{mc}}[\psi]}\;,\label{eq:mcdengen}\eeq
 for microcanonical action $I_{E}^{\text{mc}}$. The form of the microcanonical action can be deduced to leading order in a saddle-point approximation. To wit, recall the canonical and microcanonical partition functions are related by a Laplace integral transform $\mathcal{Z}(\beta)=\int dE_0 W(E_0)e^{-\beta E_{0}}$. In a stationary phase approximation,  $\ln Z(\beta)\approx \ln W (E_0)-\beta E_{0}$. This is nothing more than the Legendre transform between the thermodynamic potentials characterizing the canonical and microcanonical ensembles, $-\beta F=S_{\text{mc}}-\beta E_{0}$, for 
 canonical free energy $-\beta F(\beta)=\ln Z(\beta)$ and microcanonical entropy $S_{\text{mc}}(E_{0})=\ln W(E_{0})$. Expressing $\mathcal{Z}(\beta)$ as a Euclidean path integral, such that in a saddle-point approximation $\ln Z(\beta)\approx-I_{\text E}^{\text{can}}$, then the microcanonical and canonical actions are also related by a Legendre transform 
 \beq I^{\text{mc}}_{\text E}|_{\text{on-shell}}=I_{\text E}^{\text{can}}|_{\text{on-shell}}-\beta E_{0}\;.\eeq
 The on-shell actions are evidently equivalent for vanishing energy $E_{0}=0$ or in the infinite temperature $\beta\to0$ limit.
 
 The upshot here is, on-shell, the microcanonical action is equal to the entropy. Applying this to the classical CGHS black hole (\ref{eq:thermovarCGHS}), the on-shell microcanonical action coincides with the classical Wald entropy. This is an example of a more general result due to Iyer and Wald \cite{Iyer:1995kg}. For an arbitrary diffeomorphism invariant gravity theory on a Euclidean manifold $\mathcal{M}_{E}$ with a timelike Killing symmetry, generated by $\zeta=\partial_t$, e.g., the Euclidean section of a stationary black hole, the on-shell microcanonical action is equal to the Wald entropy,
 \beq I^{\text{mc}}_{\text E}|_{\text{on-shell}}=-S_{\text{Wald}}\;.\label{eq:ImcSwald}\eeq
 This can be understood as a path integral derivation of the Wald entropy functional for stationary black holes in the microcanonical ensemble.\footnote{The \emph{off-shell}  Euclidean microcanonical action is given by a Legendre-like transform of the (canonical) Euclidean action involving the Lagrangian form $L$ in Euclidean signature and the Noether charge $Q_{\zeta}$~\cite{Iyer:1995kg}
 $$I^{\text{mc}}_{\text E}=-i\left(\int_{\mathcal{M}_{\text E}}\hspace{-2mm}L-\int_{\partial \mathcal{M}_{\text E}}\hspace{-3mm}dt\wedge Q_{\zeta}\right)\;.\label{eq:ImcoffshellBHs}$$
 }
 Though the original Iyer-Wald derivation was accomplished for pure higher-derivative theories of gravity, the arguments apply \emph{mutatis mutandis} for arbitrary two-dimensional dilaton theories of gravity. The on-shell relation (\ref{eq:ImcSwald}) was extended to causal diamonds for arbitrary theories, but especially 2D semi-classical dilaton models in  \cite{Pedraza:2021ssc}. 

A direct derivation of the horizon entropy in an arbitrary gravity theory, developed for Gauss-Bonnet gravity by Ba\~{n}ados-Teitelboim-Zanelli (BTZ) \cite{Banados:1993qp}, is to evaluate the Gibbons--Hawking--York term on the boundary of an infinitesimal disk $D_{\epsilon}$ of radius $\epsilon$ orthogonal to punctures in the Euclidean spacetime corresponding to a bifurcate Killing horizon in Lorentzian signature. This prescription was shown to be equivalent to the microcanonical action in \cite{Brown:1995su}, and the Noether charge formalism in \cite{Iyer:1995kg}. This method can be easily extended to the case of arbitrary two-dimensional dilaton theories of gravity (cf. \cite{Pedraza:2021ssc} for bifurcate horizons of causal diamonds). 

To this end, consider the boundary terms (in Euclidean signature) of the semi-classical action (\ref{eq:fullactRSTBPP}), evaluated at the bifurcate horizon 
\beq
I_{\text E}^{\text{GHY}}=-\frac{1}{\pi}\int_{0}^{\beta}dt_E\sqrt{\gamma}K\left(e^{-2\phi}+\frac{N}{24}Z+\frac{N}{12}(a-1)\phi\right)\;.
\eeq
The last two terms in the integrand correspond to the GHY surface terms needed to make the (localized) Polyakov and $I_{a}$ actions have a well-posed variational problem. 
To proceed, let us work in Euclideanized Rindler coordinates, 
\beq ds^{2}_{E}=e^{2\rho}\left[(\lambda X)^{2}dT_{E}^{2}+dX^{2}\right]\;,\label{eq:eucrindsc}\eeq
where, as in the classical CGHS black hole Euclidean time, $T_{E}\sim T_{E}+\beta$ for $\beta=2\pi/\lambda$ to remove the conical singularity at $X=0$. In the Hartle-Hawking state, the Liouville field (\ref{eq:omegaH}) and dilaton are (\ref{eq:phiHH}), 
\beq \Omega^{(\text{HH})}=\sqrt{\frac{12}{N}}\left(\frac{M}{\lambda}-X^{2}\right)\;,\quad \phi^{(\text{HH})}(X)=\frac{12}{aN}\left(\frac{M}{\lambda}-\lambda^{2}X^{2}\right)+\frac{1}{2}W_{-1}\left(-\frac{24}{aN}e^{-\frac{24}{aN}(\frac{M}{\lambda}-\lambda^{2}X^{2})}\right)\;.\label{eq:HHsolnsRind}\eeq
Then, the GHY term evaluates to\footnote{It is useful to know $\sqrt{\gamma}=\lambda Xe^{\rho}$ and $K=e^{-\rho}X^{-1}(1+X\partial_{X}\rho)$. Note further $\sqrt{\gamma}K|_{X=0}\to \kappa$ for surface gravity $\kappa$.}
\beq 
\begin{split} 
I_{\text E}^{\text{GHY}}&=-\frac{1}{\pi}\int_{0}^{\beta}dT_{E}\lambda(1+X\partial_{X}\rho)\left(\sqrt{\frac{N}{12}}\Omega^{(\text HH)}-\frac{N}{12}\xi\right)\biggr|_{X=0}\\
&=-\frac{2M}{\lambda}+\frac{N}{6}\xi|_{H}=-S_{\text{Wald}}\;,
\end{split}
\eeq
where we assumed Kruskal gauge $\rho=\phi$ and that solutions $\xi$ are independent of Rindler time $T_{E}$. We see clear agreement with the Wald entropy (\ref{eq:Swaldsemiv2}).

\subsubsection{Wald entropy and generalized entropy}

\noindent When evaluated for the quantum black hole in the Hartle-Hawking vacuum, the Wald entropy is equal to the thermal entropy of the classical CGHS black hole, plus a correction, cf. (\ref{eq:Swaldsemiv2}). That the classical CGHS entropy arises is a result of a striking cancellation when passing to Kruskal gauge.  To better understand the meaning of the correction, consider only the Wald functional associated with the Polyakov action, evaluated on-shell $Z=2\rho-2\xi$,
\beq S_{\text{Wald}}^{\text{Poly}}=\frac{N}{6}\rho_{H}-\frac{N}{6}\xi_{H}\;.\eeq
Further,  recall the von Neumann entropy of a two-dimensional CFT with central charge $N$ in vacuum over an interval $[(x_{1},y_{1}),(x_{2},y_{2})]$ in a curved background in conformal gauge $ds^{2}=-e^{2\rho(x,y)}dx dy$ is (see \emph{e.g.}, \cite{Fiola:1994ir})
 \beq 
 \begin{split}
 S_{\text{vN}}&=\frac{N}{6}\ln\left[\frac{1}{\delta_{1}\delta_{2}}(x_{2}-x_{1})(y_{2}-y_{1})e^{\rho(x_{1},y_{1})}e^{\rho(x_{2},y_{2})}\right]=\frac{N}{6}(\rho(x_{1},y_{1})+\rho(x_{2},y_{2}))+\frac{N}{6}\ln\left[\frac{1}{\delta_{1}\delta_{2}}(x_{2}-x_{1})(y_{2}-y_{1})\right]\;.
 \end{split}
 \label{eq:vnentgen}\eeq
 Here $\delta_{1,2}$ are independent UV regulators located at the endpoints of the interval. Since the auxiliary $Z$ field is a stand in for a collection of external conformal matter fields. Let us now describe how the Polyakov contribution to the Wald entropy takes the form of a fine-grained von Neumann entropy associated with the quantum matter fields encoded in the auxiliary field $Z$. 

 Focus on solutions in the Hartle-Hawking vacuum (a similar line of reasoning holds for the Boulware state).  We saw above that the general solution to $\xi$ in Kruskal coordinates \eqref{eq:xiHHvacum} has 
\beq
\xi=c_+ + c_- -\ln(+\lambda( x^++x_0^+))- \ln(-\lambda (x^- - x_0^-))\, .
\eeq
We still have freedom to fix the constants of integration $c_{\pm}$. We will do this in two ways. Consider first setting $c_{+}+c_{-}=\ln(-\lambda^{2}\delta_{+}\delta_{-})$. In so doing, we find
\beq \xi =-\ln\left[\frac{1}{\delta_{+}\delta_{-}}(x^{+}+x_{0}^{+})(x^{-}-x_{0}^{-})\right]\;,\label{eq:xisolnv1}\eeq
such that
\beq  \label{eq:Spoly01} S^{\text{Poly}}_{\text{Wald}}=\frac{N}{6}\rho_{H}(x^{+},x^{-})+\frac{N}{6}\ln\left[\frac{1}{\delta_{+}\delta_{-}}(x^{+}+x_{0}^{+})(x^{-}-x_{0}^{-})\right]\biggr|_{H}\;.\eeq
The second term clearly has the form of the logarithmic contribution to the von Neumann entropy (\ref{eq:vnentgen}), for $(x_{2},x_{1})=(x^{+},-x_{0}^{+})$ and $(y_{2},y_{1})=(x^{-},x_{0}^{-})$, with $x_{0}^{\pm}$ are some generic endpoints for the interval. Evaluating at the horizon bifurcation point $(x^{+}=x^{-}=0)$, one may follow the reasoning of \cite{Fiola:1994ir} (see also \cite{Alexandre:2024htk}), such that 
\beq
\label{eq:Swald_1}
S_{\text{Wald}}
=\frac{2M}{\lambda}+\frac{N}{6} \ln \left(\frac{-x_{0}^{+} x_{0}^{-}}{\delta^2}\right)\, ,
\eeq
where we used Kruskal gauge $\rho=\phi$, and identified UV regulators at the endpoints, $\delta_{+}=\delta_{-}\equiv\delta$.

Cast like this, the second term has the interpretation of the entropy of a thermal bath \cite{Fiola:1994ir}. Indeed, working in Eddington–Finkelstein coordinates, then
\beq
\label{eq:Sthermal}
\frac{N}{6} \ln \left(\frac{-x_{0}^{+} x_{0}^{-}}{\delta^2}\right)=\frac{N}{6} \lambda \left((\sigma^+-\sigma^-)_0- (\sigma^+-\sigma^-)_\delta\right)=\frac{N}{6} 2\lambda (\sigma_{0}-\sigma_\delta)=2\pi \frac{N}{6} T_{\text{H}} L=S_{\text{thermal}}\, ,
\eeq
where $2(\sigma_{0}-\sigma_\delta)=L$  can be thought as the volume (in one space dimension) of a box of size $L$,
\beq
\label{eq:Stot}
S_{\text{Wald}}=\frac{2M}{\lambda}+ S_{\text{thermal}}\,.
\eeq
Thus, from this perspective, the Wald entropy is equal to the sum of the classical CGHS horizon entropy and the entropy of a gas of thermal radiation at a temperature $T_{\text{H}}$. 

Let us now take another perspective. Note we actually have enough freedom with the integration constants such that we may append the solution (\ref{eq:xisolnv1}) via the constant $-\rho(x_{0}^{+},x_{0}^{-})$.\footnote{While our presentation here has us choose our constants of integration in an \emph{ad hoc} manner, a justification can be given by first solving the equation of motion for $Z$ in flat space (where $\rho=0$), choosing $\xi$ to obey Dirichlet boundary conditions, and then performing a Weyl transformation to a curved background with $\rho\neq0$. In so doing, one sees that $Z_{\text{flat}}$ can be can be written as
a two-point correlation function of primary operators $\partial_{Z}$. For more, see the discussion around Eq. (3.20) of \cite{Pedraza:2021cvx}.} This would mean the entire Polyakov contribution to the Wald entropy is identified with the von Neumann entropy of a two-dimensional CFT, 
\beq S^{\text{Poly}}_{\text{Wald}}=S_{\text{vN}}\;.\eeq
Consequently, the total Wald entropy is
\beq S_{\text{Wald}}=2e^{-2\phi}+\frac{N}{6}(a-1)\phi+S_{\text{vN}}=S_{\text{grav}}+S_{\text{vN}}\;,\eeq
to be evaluated on the horizon.  Here, we point out, we are not explicitly invoking the Kruskal gauge symmetry such that $\rho=\phi$. Rather, we have isolated a ``gravitational'' contribution to the entropy (i.e., the first two terms coming from the CGHS and RST/BPP-like term), and a fine-grained matter contribution. Thus, backreaction effects have modified the entropy in two ways: one acquires a correction to the gravitational contribution and a quantum matter contribution. The gravitational correction is merely a consequence of demanding to work with an analytically solvable model. The main takeaway here is that the semi-classical Wald entropy is the sum of a gravitational entropy and a fine-grained matter entropy, namely, the \emph{generalized entropy} \cite{Bekenstein:1974ax}, 
\beq S_{\text{Wald}}=S_{\text{gen}}\;.\eeq
This same equivalence was observed for semi-classical models of JT gravity \cite{Pedraza:2021cvx,Svesko:2022txo}, and appears to only hold in the case of two-dimensional theories, a consequence of being able to capture all of the backreaction effects with the Polyakov contribution. Further, in lieu of the equivalence of the on-shell Euclidean microcanonical action and the Wald entropy (\ref{eq:ImcSwald}), we see that, presently,  $I^{\text{mc}}_{E}|_{\text{on-shell}}=-S_{\text{gen}}$. 

Note that the gravitational contribution to the Wald entropy is dependent on the interpolating parameter $a$. In particular, for any $a$, $S_{\text{grav}}$ evaluated on the horizon is not equal to the classical CGHS horizon entropy. Rather, for solutions (\ref{eq:HHsolnsRind}) (for $a\neq0$),
\beq 
\begin{split} 
&2e^{-2\phi_{H}}=-\frac{aN}{12}W_{-1}\left(z\right)\;,\\
&\frac{N}{6}(a-1)\phi_{H}=\frac{2M}{\lambda}-\frac{2M}{a\lambda}+\frac{N(a-1)}{12}W_{-1}(z)
\end{split}
\eeq
with $z=-\frac{24}{aN}e^{\frac{-24 M}{aN\lambda}}$. Combined, the gravitational contribution to the Wald entropy evaluated at the horizon is
\beq S_{\text{grav}}=\frac{2M}{\lambda}-\frac{2M}{a\lambda}-\frac{N}{12}W_{-1}(z)\;.\eeq
This shows that for $a=1$, the classical contribution to the entropy vanishes (despite $S_{\text{grav}}$ formally going like the Wald functional for the CGHS term) leaving only contributions due to quantum backreaction effects. Meanwhile, for $a=0$, where $\phi_{H}^{(a=0)}=\frac{1}{2}\ln(\lambda/M)$ (\ref{eq:bppphi}), we see
\beq S_{\text{grav}}^{(a=0)}=\frac{2M}{\lambda}-\frac{N}{12}\ln(\lambda/M)\;,\eeq
where one always has a quantum correction to the classical CGHS result. Here the gravitational entropy will always be positive when we work in the semi-classical regime of validity (\ref{eq:scvalidity}).

\subsection{Quasi-local thermodynamics}

 \noindent Here we analyze the quasi-local thermodynamics of the eternal quantum black hole solutions to the semi-classical interpolating model for quantum fields in the Hartle-Hawking state. We again work in Euclideanized Rindler coordinates (\ref{eq:eucrindsc}).
 As in the analysis for the classical black hole, we consider an observer located at a finite timelike Dirichlet boundary $B$. The associated inverse Tolman temperature \eqref{eq:tolmanT} is $\beta_{\text{T}}=\beta_\text{H}\mathcal{N}_B$, for inverse Hawking temperature $\beta_{\text{H}}$ (the periodicity of the Euclidean time circle) and lapse $\mathcal{N}_B=\mathcal{N}(\phi_B)$ given in \eqref{eq:lapse}. Though the periodicity remains $\beta_{\text{H}}=2\pi/\lambda$, the Tolman temperature is quantum corrected since the lapse receives quantum corrections.

We can write the semi-classical Brown-York energy \eqref{eq:EBY_semi} in terms of $\beta_{\text{T}}$ and the `area' $A_{B}\equiv e^{-2\phi_B}$,
\be
E=-4A_{B}\frac{(\beta_{\text{T}}-\beta_{\text{H}})}{\beta_{\text{H}}^{2}}-\frac{aN}{12\beta_{\text{H}}}\ln(A_{B})-\frac{N}{3}\frac{1}{\beta_T}\,.
\ee
Clearly, when we turn off quantum corrections $N\to0$, we recover the classical energy (\ref{eq:thermovarCGHS}). 

Inverting the lapse, we find the relation
\beq \frac{M}{\lambda}=A_{B}\frac{(\beta_{\text{H}}^{2}-\beta_{\text{T}}^{2})}{\beta_{\text{H}}^{2}}-\frac{aN}{24}\ln(A_{B})\;.\eeq
The semi-classical regime of validity (\ref{eq:scvalidity}) then implies
\beq
\label{eq:scvalidtemp}
\text{Semi-classical regime:}\quad \frac{A_{B}}{N\beta_{\text{H}}^{2}}\gg \frac{1}{(\beta_{\text{H}}^{2}-\beta_{T}^{2})}\left(1-\frac{a}{24}\ln(A_{B})\right)\;\Longrightarrow \;T^{2}\gg \frac{T_{H}^{2}}{\left[1-\left(\frac{N}{A_{B}}+\frac{aN}{24A_{B}}\ln(A_{B})\right)\right]}\;.\eeq
Evidently, for the temperature $T$ to be positive and real, we require 
\be
\label{eq:T2positive}
\frac{A_{B}}{N}>1+\frac{a}{24}\ln(A_{B})
\ee
We see that for the BPP model ($a=0$), the semi-classical regime coincides with the region where the

Consequently, the entropy (\ref{eq:Stot}) as a function of $A_{B}$ and $T$ is
\beq S(T,A_{B})=2A_{B}\left(1-\frac{\lambda^{2}}{4\pi^{2}T^{2}}\right)-\frac{aN}{12}\ln(A_{B})+S_{\text{thermal}}\;.\eeq
It is easy to show $\Delta S\equiv S(T,A_{B})-S_{\text{thermal}}$ is non-negative for all semi-classically allowed temperatures (\ref{eq:scvalidtemp}). We can further recast the thermal entropy (\ref{eq:Sthermal}) in terms of $A_{B}$ and $\beta_{T}$,
\beq S_{\text{thermal}}=\frac{N}{3}\lambda\sigma_{B}^1=\frac{N}{6}\ln\left[A_{B}-\left(\frac{M}{\lambda}+\frac{aN}{24}\ln(A_{B})\right)\right]=\frac{N}{6}\ln\left(A_{B}\frac{\beta_{\text{T}}^{2}}{\beta_{\text{H}}^{2}}\right)\;.\eeq
Previously we expressed the thermal gas entropy (\ref{eq:Sthermal}) in terms of a fixed length $L$, yet here we have expressed the entropy as a function of the thermodynamic data $(A_{B},\beta_{\text{T}})$. This is consistent by requiring $A_{B}$ change accordingly as the temperature is tuned, as vice-versa.  Together, 
\beq S(T,A_{B})=2A_{B}\left(1-\frac{\lambda^{2}}{4\pi^{2}T^{2}}\right)+\frac{(2-a)N}{12}\log(A_{B})-\frac{N}{3}\ln\left(\frac{2\pi T}{\lambda}\right)\;.\eeq
Inverting the energy and temperature, we can express the entropy as a function of $A_{B}$ and $E$, though the analytic expression is not particularly illustrative.

The canonical free energy of the quantum black hole is 
\beq F=E-\frac{1}{\beta_{\text{T}}}S=-2A_{B}\frac{(\beta_{\text{T}}-\beta_{\text{H}})^{2}}{\beta_{\text{H}}^{2}\beta_{\text{T}}}+\frac{N}{12\beta_{\text{H}}\beta_{\text{T}}}\left[a(\beta_{\text{H}}-\beta_{\text{T}})\ln(A_{B})-2\beta_{\text{H}}\left(2+\ln\left(\frac{A_{B}\beta_{\text{T}}^{2}}{\beta_{\text{H}}^{2}}\right)\right)\right]\;.\eeq
Further, the quantum corrected surface pressure (\ref{eq:surfpress}) is easily computed to be
\beq \sigma\equiv-\left(\frac{\partial E}{\partial A_{B}}\right)_{\hspace{-1mm}S}=2\frac{(\beta_{\text{T}}-\beta_{\text{H}})^{2}}{\beta_{\text{T}}\beta^{2}_{\text{H}}}+\frac{N}{12A_{B}}\frac{(2-a)\beta_{\text{H}}+a\beta_{\text{T}}}{\beta_{\text{H}}\beta_{\text{T}}}
\;.\label{eq:Qsurfpress}\eeq
With these thermodynamic variables we find they obey the following first law
\beq \delta E=T\delta S_{\text{gen}}-\sigma \delta A_{B}\;,\label{eq:firstlawthermoQ}\eeq 
where we emphasize that here the classical entropy has been replaced by the generalized entropy, consistent with \cite{Pedraza:2021cvx,Svesko:2022txo} and the literature on three-dimensional quantum black holes. 

\begin{figure}[t!]
\centering
\includegraphics[width=0.5\linewidth]{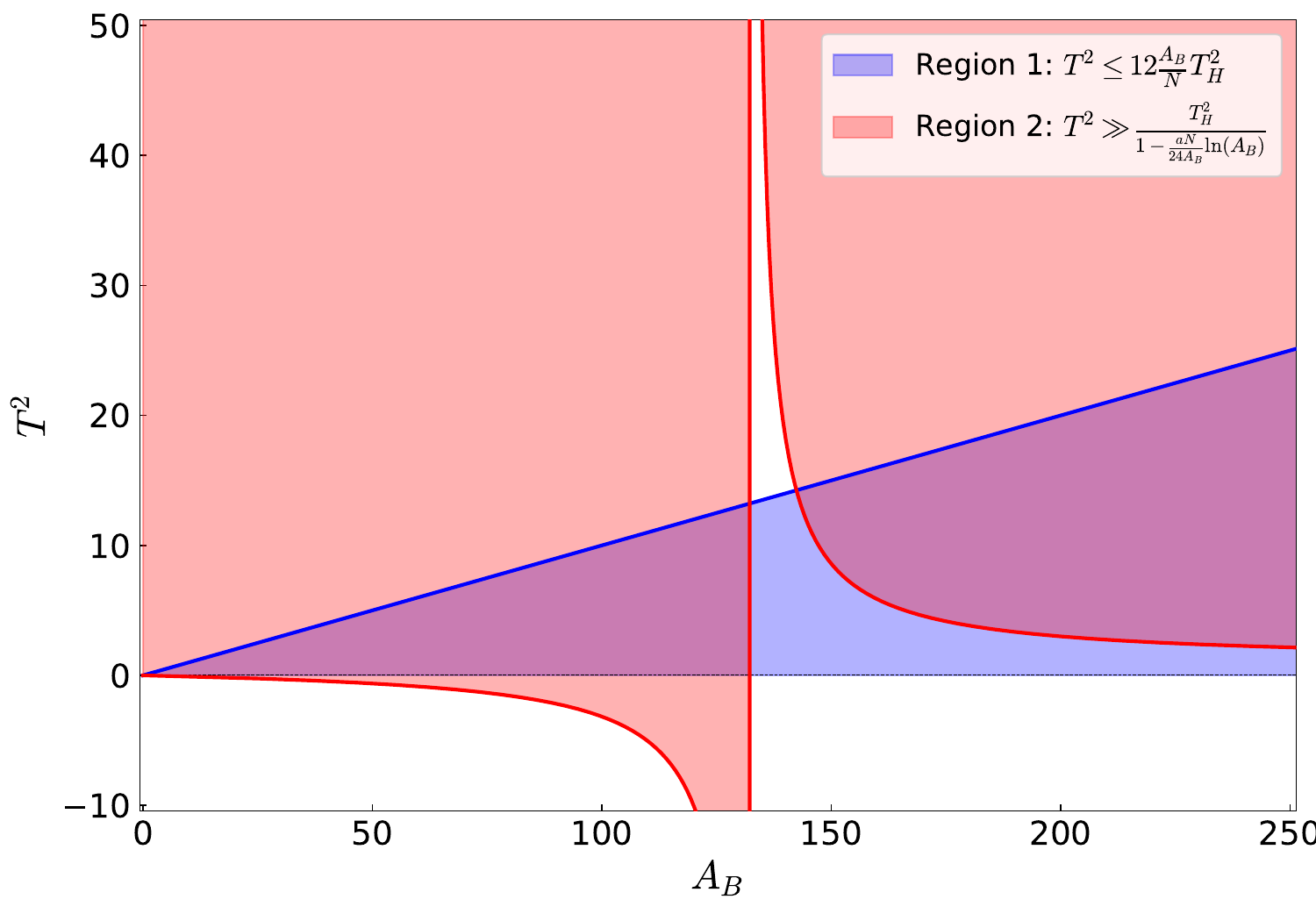}
    \caption{Temperature $T$ as a function of $A_B$ based on the lower bound \eqref{eq:scvalidtemp} (regions in red) and the upper bound \eqref{eq:heatcapvalid} (regions in blue). Overlapping regions indicate the parameter space where the quantum black holes are thermally stable and are semi-classical valid. Here $a=1/2$ (RST), $N=120$ and $T_H/\lambda=1$.}
    \label{fig:Tregions}
\end{figure}

The heat capacity at fixed $A_{B}$ (equivalently, fixed $\phi_{B}$) is
\beq C_{A_{B}}=-\beta_{\text{T}}^{2}\left(\frac{\partial E}{\partial\beta_{\text{T}}}\right)_{\hspace{-1mm} A_{B}}=T\left(\frac{\partial S}{\partial T}\right)_{\hspace{-1mm}A_{B}}=4A_{B}\frac{\beta_{\text{T}}^{2}}{\beta_{\text{H}}^{2}}-\frac{N}{3}\;.\eeq
Thus, the heat capacity is shifted from the classical black hole heat capacity \eqref{eq:heatcapCGHS} by an overall (large) negative number. Non-negativity of $C_{A_{B}}$, and hence a sensible canonical partition function, requires an upper bound for the Tolman temperature, 
\beq C_{A_{B}}\geq0\; \;\;\Longrightarrow \;\; T^2\leq 12\frac{A_{\text{B}}}{N}T^2_{\text{H}}\;.\label{eq:heatcapvalid}\eeq
There always exists a range of temperatures for which both \eqref{eq:scvalidtemp} and (\ref{eq:heatcapvalid}) are satisfied, i.e., there exists a window for thermally stable quantum black holes, see Figure \ref{fig:Tregions}.

When the boundary $B$ tends to asymptotic infinity, $\beta_{\text{T}}\rightarrow\beta_{\text{H}}$ and $A_{\text{B}}\rightarrow\infty$, and $ C_{A_{B}}$ tends to positive infinity, as for the classical system. Alternatively, it is standard to work with Casimir subtracted quantities (see, e.g., \cite{Frassino:2024bjg}); specifically, we would subtract the quasi-local energy of empty space in the presence of backreacting quantum fields. With respect to the Casimir subtracted quantities, we likewise find thermally stable quantum black holes. 

\section{Numerical quantum black holes}\label{sec:numericalqBHs}

\noindent Pre-dating the analytically solvable models, such as RST and BPP, quantum black hole solutions to semi-classical CGHS were analyzed numerically, e.g., \cite{Hawking:1992cc, Birnir:1992by,Susskind:1992gd,Piran:1993tq,Hayward:1995sr}. While their geometric pictures were shown to share the same qualitative features as the exact models, the horizon thermodynamics were less explored. In this section we numerically construct eternal quantum black holes in the semi-classical CGHS model and analyze their thermal behavior, where, in particular, we find the entropy of the analytic and numerical black holes differ in a notable way. 

\subsection{The model}

\noindent We aim to obtain black hole solutions --including their mass and entropy-- from the full semi-classical CGHS model, which incorporates quantum corrections via the Polyakov action. To this end, we consider the action
\be
I = I_{\text{CGHS}} + I_{\text{Poly}} + (1-\epsilon)\, I_{\text{RST}}\;,
\ee
where the $(1-\epsilon)$ term allows us to interpolate between the solvable RST model (for $\epsilon=0$) and the pure semi-classical CGHS model (for $\epsilon=1$), in which the classical symmetry is broken. In our analysis, we focus primarily on the $\epsilon=1$ case, using the RST model as a consistency check on our numerical strategy.

We work in conformal gauge $ds^2=-e^{2\rho} dy^+ dy^-$
with generalized Kruskal coordinates $(y^+,y^-)$, defined in Appendix \ref{app:coordsys}. These coordinates prove convenient for our numerical analysis as we will demonstrate. The generalized Kruskal coordinates $y^{\pm}$ are related to the usual Eddington–Finkelstein coordinates as
\beq \lambda \sigma^{\pm}=\pm\alpha \ln(\pm\lambda  y^{\pm})\mp k\;.\label{eq:genKruscoordmain}\eeq
The explicit values of the parameters $\alpha$ and $k$ that define $(y^+,y^-)$ will be found \emph{a posteriori} after solving the equations of motion from the asymptotic expansion of the solutions.

For these coordinates, the equations of motion for $\rho$ and $\phi$, respectively, read  
\bea\label{eq:eom01-numerics}
\frac{4\pi}{\sqrt{-g}}\frac{\delta I}{\delta g^{\pm \mp}} =0 \quad \to  \qquad 0&=&e^{-2\phi}\left(2\partial_+\partial_-\phi-4(\partial_+\phi)( \partial_-\phi)-\lambda^2 e^{2\rho}\right)-\frac{N}{12}\partial_+\partial_-\big(\rho-(1-\epsilon)\,\frac{\phi}{2}\big)\, ,\\
\frac{\delta I}{\delta \phi}=0\quad \to\qquad 0&=&e^{-2\phi}\left(-4\partial_+\partial_-\phi+4\partial_+\phi\partial_-\phi+2\partial_+\partial_-\rho + \lambda^2 e^{2\rho}\right)+\frac{N}{24}(1-\epsilon)\partial_+\partial_-\rho\, . \label{eq:eom02-numerics}
\eea
Furthermore, varying with respect to the diagonal components of the metric leads to the following constraint equations
\be \label{eq:constraint-numerics}
\frac{4\pi}{\sqrt{-g}}\frac{\delta I}{\delta g^{\pm \pm}} =0\quad \to \qquad 0 =\big(e^{-2\phi}+\frac{N}{48}(1-\epsilon)\big)\left(4\partial_{\pm}\rho \partial_{\mp}\phi-2\partial_{\pm}\partial_{\pm}\phi\right)-\frac{N}{12}(\partial_\pm \rho\partial_{\pm} \rho -\partial_{\pm}\partial_{\pm}\rho)+t_{\pm}(y^{\pm})\, .
\ee
As discussed in subsection \ref{subsubsec:normalorderedSET}, different choices of $t_{\pm}$ correspond to different vacua. From the definition of normal ordering, the vacuum state $|0_y\rangle$ for the $y$-observer is the state such that the expectation value of the normal ordered (non-covariant) energy-momentum tensor in $(y^+,y^-)$ coordinates \eqref{eq:normordgen} vanishes,
\be \left\langle 0_y\right|: T_{ \pm \pm}\left(y^{ \pm}\right):\left|0_y\right\rangle=0, \quad \Leftrightarrow \quad t_{ \pm}\left(y^{ \pm}\right)=0\, .
\ee
In these coordinates, the functions $t_{\pm}$  for the Boulware and  Hartle-Hawking vacua can be obtained applying the transformation law in \eqref{eq:tpmtrans}, 
namely
\bea \label{eq:boulwarevacuum01}
t_{\pm}^{\text{B}}(y^{\pm})&=& \frac{1}{2}\{\sigma^{\pm},y^\pm\}=\frac{1}{4(y^{\pm})^2}\, ,\\
t_{\pm}^{\text{HH}}(y^{\pm})&=& \frac{1}{2}\{x^{\pm},y^\pm\}=-\frac{\alpha^2-1}{4(y^{\pm})^2}\, ,
\eea
We can therefore conclude the vacuum state for the static observer in generalized Kruskal coordinates  is not the Boulware vacuum since the Schwarzian derivative  does not vanish. It is neither the vacuum state for the static observer in Kruskal coordinates unless $\alpha=1$. Instead, it describes a thermal state at temperature\footnote{We can see this by using the asymptotic solutions for the metric we present below to evaluate the stress-energy tensor. As $\lambda \sigma\rightarrow\infty$, then $\langle {T}_{\sigma^\pm\sigma^\pm}\rangle=\frac{N\lambda^2}{48\alpha^2}=N\frac{\pi^2}{12}T^2\,$ which corresponds to thermal bath at temperature $T=\frac{\lambda}{2\pi \alpha}$.}
\be\label{eq:bath-temperature}
T=\frac{\lambda}{2\pi\alpha}\, .
\ee
For $\alpha=1$ we see $T=T_{\text{H}}$. From now on, we will refer to these states as Boulware and Hartle-Hawking, respectively.

Since we are interested in static solutions, it is convenient to rewrite the problem in terms of the static, dimensionless variable,
\be 
y = -\lambda^2\, y^+y^-\,.
\ee
In terms of this variable, we massage the equations of motion \eqref{eq:eom01-numerics}, \eqref{eq:eom02-numerics} and \eqref{eq:constraint-numerics}, such that the constraint equations depend only on first-order derivatives, and obtain the following second-order differential equations for $\phi$ and $\rho$ 
\bea
\label{eq1}
\left( \frac{N}{48}(\epsilon-1) e^{2 \phi}+1\right)\Bigl(\phi^{\prime}+y\, \phi^{\prime \prime}\Bigr)+\left(\frac{N}{48}(\epsilon+1)e^{2 \phi}-1 \right)\Bigl(\rho^{\prime}+y\, \rho^{\prime \prime}\Bigr)&=&0\, ,\\
\label{eq2}
\left(\frac{1}{8}\left(\frac{N}{12}\right)^2(\epsilon-1)^2 e^{4 \phi}-\frac{N}{12} (\epsilon+1) e^{2 \phi}+2\right)\Bigl(\phi^{\prime}+ y\, \phi^{\prime \prime}\Bigr)-\left(\frac{N}{48}(\epsilon+1) e^{2 \phi}-1\right)\Bigl(e^{2 \rho}-4y\, (\phi^{\prime})^{2}\Bigr)&=&0\, ,
\eea
together with the constraint
\be
\label{constraint}
-\frac{N}{12} \rho^{\prime}\left(y \rho^{\prime}+1\right)-\frac{N}{24} (\epsilon-1)\left(2 y \rho^{\prime}+1\right) \phi^{\prime}+e^{-2\phi}(e^{2 \rho}+2 \phi^{\prime}\left(2 y \rho^{\prime}-2 y \phi^{\prime}+1\right))+ Y(y)=0\, .
\ee
Here we have defined
\beq Y\equiv\frac{t_{+}(y^+)}{\lambda^2(\lambda y^{-})^2}= \frac{t_{-}(y^-)}{\lambda^2(\lambda y^{+})^2}\;.\eeq
For simplicity,  moving forward we take $Y=0$, corresponding to the vacuum state for the static observer in generalized Kruskal coordinates (\ref{eq:genKruscoordmain}). We note, however, that $\alpha$ and $k$ are not yet fixed. Once we characterize the asymptotic behavior of the numerical solutions $\phi$ and $\rho$, the constants $\alpha$ and $k$ will be determined, as we show below.

In our numerical treatment we focus on regular solutions at $y=0$, i.e., where both $\phi$ and $\rho$ admit a series expansion around $y=0$.\footnote{We note that since $\rho$ is not a scalar function, it can diverge in other coordinates. However, we assume $y^\pm$ are such that {\it both} $\phi$ and $\rho$ are finite. In other words, fixing $\rho=\rho_{H}$ implicitly fixes $y^\pm$.} We then fix the values of $\phi$ and $\rho$ at the horizon $y=0$,
\be \label{eq:BCaty0}
\phi(0)=\phi_H \,,  \qquad  \rho(0)=\rho_{H}\, .
\ee
The first derivatives $\phi'(0)$ and $\rho'(0)$ can be determined using the equations of motion and imposing regularity of solutions at the horizon \cite{Birnir:1992by}. We find
\bea \label{eq:derivatives}
\phi'(0)&=&\frac{2 e^{2 \rho_{H}} \left(-4+\frac{N}{12} (\epsilon +1) e^{2 \phi_{H}}\right)}{16+(\frac{N}{12}) ^2
   (\epsilon -1)^2\, e^{4 \phi_{H}}-\frac{2N}{3}  (\epsilon +1) e^{2 \phi_{H}}}\, ,\\
   \rho'(0)&=&-\frac{2 e^{2 \rho_{H}} \left(4+\frac{N}{12}  (\epsilon -1) e^{2 \phi_{H}}\right)}{16+(\frac{N}{12}) ^2
   (\epsilon -1)^2\, e^{4 \phi_{H}}-\frac{2N}{3}  (\epsilon +1) e^{2 \phi_{H}}}\, . \label{eq:derivatives2}
\eea

A few more preliminary remarks are in order.  
First, upon inserting \eqref{eq1} and \eqref{eq2} into the Ricci scalar $R=8e^{-2\rho}\partial_+\partial_-\rho$, it is straightforward to show there is a curvature singularity when the dilaton takes the value $\phi_c=-\frac{1}{2}\ln\left(\frac{N}{12}\right)$. Then, to find static black hole solutions that asymptotically tend to the linear dilaton vacuum, we require $\phi_{H}<\phi_c$. If $\phi_H>\phi_c$, then $\phi(y)$ will approach that value again if we want to have an analytically flat solution \cite{Hawking:1992cc}. So there would be another singularity. 

Finally, in our numerical analysis we fix $N=120$ and focus on the range
$\phi_{H} \in (-10,-1.25)$, since
$\phi_c \simeq -1.15$.
The semi-classical condition \eqref{eq:scvalidity}, namely $M/\lambda > N$,
imposes an upper bound on $\phi_{H}$.
In particular, using the large-$|\phi_{H}|$ approximation described below (see \eqref{eq:2mlargephi0}), we find $\phi_{H} < \phi_{\mathrm{sc}}$,
where $\phi_{\mathrm{sc}}$ is defined by the condition $M/\lambda = N$.
Explicitly,
\[
\phi_{\mathrm{sc}}
=
\frac{1}{2}\left(49 + W_{-1}\!\left(-\frac{48}{N e^{49}}\right)\right),
\]
which evaluates to $\phi_{\mathrm{sc}} \simeq -2.45$ for $N=120$, as displayed in
Figure \ref{fig:entropy}. Nevertheless, we also explore the region
$\phi_c > \phi_{H} > \phi_{\mathrm{sc}}$ as a way to test the robustness of
the large-$|\phi_{H}|$ approximation.
As for $\rho_{H}$, we either fix $\rho_{H} = 0$ or set $\rho_{H} = \phi_{H}$.
Note that shifts in $\rho_{H}$ correspond to a rescaling of the $y^\pm$
coordinates, as follows from \eqref{eq2} and from our asymptotic expansions of $\phi$ and $\rho$, as we now describe.

\subsection{Asymptotic expansions, mass and entropy}

\noindent Before presenting our numerical results, we carry out a preliminary analysis to derive their expected asymptotic behavior. In particular, for large $y$,  assuming $e^{2\phi}\to 0$, one finds  (\ref{eq1}) reduces to the free field equation \eqref{eq:freefieldeq}. In other words, for large $y$, one recovers the residual gauge symmetry described in \eqref{eq:freefieldeq}, which in terms of $y$ reads $y\,\partial_y\, (y \, \partial_y(\phi-\rho))=\mathcal{O}(y^{-2\alpha})$. The general solution to this equation is
\be
\label{eq:asygauge}
\phi-\rho= b_0\ln(y)+b_1+\mathcal{O}(y^{-2\alpha})\,.
\ee
Notice the coefficients $b_0$ and $b_1$ fix the conformal gauge asymptotically. For example, for  $b_0=b_1=0$, $\phi=\rho$ and we recover the Kruskal gauge  asymptotically. From the relationship between Kruskal and extended Kruskal coordinates $x^{\pm}=\pm e^{-k}(\pm y^{\pm})^{\alpha}$ and the corresponding transformation law for $\rho$ in \eqref{eq:rhotranslaw}, the integration constants in \eqref{eq:asygauge} are $b_0=\frac{1-\alpha}{2}$ and $b_1=k-\ln(\alpha)$. Since we are fixing the boundary conditions at $y=0$, the values of $k$ and $\alpha$ are determined numerically.

To better understand our numerical solution, it is worth to study the expected asymptotic behavior of our solutions. Given \eqref{eq:asygauge}, and the asymptotic expansion of the known analytic solutions for the RST model,\footnote{The asymptotic expansion of the RST solutions in the Kruskal gauge \eqref{eq:phiHH}, ($\rho=\phi$), is \[\phi(x)=-\frac{\ln x}{2}-\frac{1}{2x}\left(\frac{M}{\lambda}+\frac{N}{24}\ln x\right)+\mathcal{O}(x^{-2})\] with $x=-\lambda^2x^+x^-$.} we generalize the asymptotic expansion for $\phi$ and $\rho$ to be, e.g. \cite{Hawking:1992cc}
\bea
\label{eq:asyphi}
\phi &=&   -\frac{\alpha}{2}\ln y +k\quad \,\,\,  -\frac{1}{y^\alpha}\left(c_1+c_2 \ln y\right)+\sum_{n=2}^{\infty} y^{-n\alpha} \left(c^{\,(n)}_1+ c^{\,(n)}_2 \ln y + \cdots +c^{\,(n)}_{n+1} (\ln y)^n \right)\, ,\\
\rho &=&   -\frac{1}{2}\ln y +\ln(\alpha)- \frac{1}{y^\alpha}\left( c_1+ c_2 \ln y\right) +\sum_{n=2}^{\infty} y^{-n\alpha} \left(\bar c^{\,(n)}_1+ \bar c^{\,(n)}_2 \ln y + \cdots +\bar c^{\,(n)}_{n+1} (\ln y)^n \right) \,. \label{eq:asyrho} 
\eea
We have verified \eqref{eq:asyphi} and \eqref{eq:asyrho} solve the equations of motion, \eqref{eq1} and \eqref{eq2}, order by order, for large $y$. Note that the leading order corresponds to the linear dilaton vacuum in $y^{\pm}$ coordinates. The constant $c_1$ is a free parameter related to the mass of the black hole. The constant $c_2$ is fixed from the constraint equation. For our vacuum choice $Y=0$, it reads
\be
\label{eq:c2}
c_2=e^{2k}\frac{N}{96\alpha}\,.
\ee
For completeness, we note that for the Boulware vacuum \eqref{eq:boulwarevacuum01}, we would have $c_2=0$. 
 To study the entropy and the mass of the black hole solutions, it suffices to consider these expansions up to order $\mathcal{O}(y^{-\alpha})$, so we can safely ignore the terms with $n\geq2$ in the expansions as long as $y$ is large enough. With these expansions in mind, the numerical analysis becomes clear. We fix the {\it input} parameters $\phi_{H}$, $\rho_{H}$ and $N$. We then solve the equations with boundary conditions at $y=0$, see \eqref{eq:BCaty0} and \eqref{eq:derivatives}. From the numerical solutions, we extract the {\it output} parameters $\alpha$, $k$ and $c_1$ using eqs. \eqref{eq:asyphi} and \eqref{eq:asyrho}. Schematically, 
\begin{align}
(N, \phi_{H},\rho_{H}) \leftrightarrow (\alpha, k,c_1). \nonumber
\end{align}
Once we get them, we can build composite quantities, such as the mass and the entropy. 

\subsubsection{Mass, entropy and temperature }

\noindent The mass and the entropy of the black hole can be constructed from the {\it inputs} and the {\it outputs} of the model. For the mass, we can compute both the black hole mass $M$ and the ADM mass $M_{\text{ADM}}$. Let us start with the ADM mass  using the quasi-local Hamiltonian $H_{\zeta}$ \eqref{eq:Hzeta},
\be
H_{\zeta}=\oint_{\partial\Sigma}\epsilon_{\partial\Sigma}\left[\partial_1\phi\left(\frac{2}{\pi}e^{-2\phi}+\frac{N}{24\pi}(1-\epsilon)\right)-\frac{N}{24}\partial_1 Z\right]\;.
\ee
In the thermal state we are considering,  $t_{y^{\pm}}=0$ and we can use the results from the previous sections and conclude that functions $\xi_{\pm}$ are given by 
\be \label{eq:xiHHvacum_num}
\xi_{\pm}=c_{\pm}-\ln\left(\pm \lambda( y^{\pm}\pm y_0^{\pm})\right)\;.
\ee
Changing to Eddington-Finkelstein coordinates,  $\lambda\sigma^{\pm}=\pm\alpha\ln(\pm\lambda y^{\pm})\mp k$, we have
\be
\xi_{\sigma^{\pm}}=c_{\pm}+\frac{k}{2\alpha}-\frac{\ln(\alpha)}{2}
\pm\frac{\lambda}{2\alpha}\sigma^{\pm}-\ln\left(e^{\frac{k\pm\lambda\sigma^{\pm}}{\alpha}}+\lambda y_0^{\pm}\right)\;.
\ee
For the simpler case $y_0^{\pm}=0$, we have
\be
\xi_{\sigma^{\pm}}=c_{\pm}-\frac{k}{2\alpha}-\frac{\ln(\alpha)}{2}
\mp\frac{\lambda}{2\alpha}\sigma^{\pm}\,,
\ee
and $\xi_{\sigma^+}+\xi_{\sigma^-}=c_++c_--\ln(\alpha)-\frac{k+\lambda\sigma^1}{\alpha}$. To remove the divergences of $H_{\zeta}$, we use the same counterterms that we found in the analytical case. Particularizing for the asymptotic solutions \eqref{eq:asyphi} and \eqref{eq:asyrho}, and tuning the coefficient of the linear counterterm to cancel the divergences, we find   
\be
\label{eq:Ict_num}
I_{\text{ct}}=-\frac{1}{\pi}\int_{B}\sqrt{\gamma}(2\lambda) e^{-2\phi}-\frac{1}{\pi}\int_{B}\sqrt{\gamma} \frac{4c_2\lambda}{\alpha}e^{-2k}\phi+\frac{1}{\pi}\int_{B}dt\sqrt{-\gamma}C_0\,.
\ee
Then, the ADM mass is
\beq
\label{eq:ADMHH_Hnum}
M_{\text{ADM}}=\lim_{\sigma^1\rightarrow\infty} H_{\zeta}=\frac{2\lambda c_1}{\pi}e^{-2k}+\frac{4\lambda c_2 e^{-2k}}{\pi\alpha}(k-1)-\frac{N\lambda}{24\pi}(1-\epsilon)-\frac{N\lambda}{12\pi\alpha}-\frac{C_0}{\pi}\;.
\eeq
If we particularize this expression for the RST model
, i.e., $\epsilon=0$,  in Kruskal gauge we have $c_1=\frac{M}{2\lambda}$, $c_2=\frac{N}{96}$, $k=0$ and $\alpha=1$, and we recover \eqref{eq:Ict_semi} and \eqref{eq:ADMHH_H} for the Hartle-Hawking state
\bea
I_{\text{ct}}&=&-\frac{1}{\pi}\int_{B}\sqrt{\gamma}(2\lambda) e^{-2\phi}-\frac{1}{\pi}\int_{B}\sqrt{\gamma} \frac{N\lambda}{24}\phi+\frac{1}{\pi}\int_{B}dt\sqrt{-\gamma}C_0\;,\\
\label{eq:ADMrstagain}M_{\text{ADM}}&=&\frac{M}{\pi}-\frac{N\lambda}{6\pi}-\frac{C_0}{\pi}\,.
\eea
For the semi-classical CGHS model, $\epsilon=1$, if we replace $c_2$ by the expression found from the consistency of the asymptotic expansion \eqref{eq:c2} gives
\bea
I_{\text{ct}}&=&-\frac{1}{\pi}\int_{B}\sqrt{\gamma}(2\lambda) e^{-2\phi}-\frac{1}{\pi}\int_{B}\sqrt{\gamma} \frac{N\lambda}{24\alpha^2}\phi+\frac{1}{\pi}\int_{B}dt\sqrt{-\gamma}C_0\;,\\
\label{eq:adm_num}M_{\text{ADM}}&=&\frac{2\lambda c_1}{\pi}e^{-2k}+\frac{N\lambda}{24\pi\alpha^2}k-\frac{N\lambda}{12\pi\alpha}\left(1+\frac{1}{2\alpha}\right)-\frac{C_0}{\pi}\,.
\eea

Comparing \eqref{eq:adm_num} with $M_{\text{ADM}}$ for the RST model \eqref{eq:ADMrstagain}, we can identify the first two terms of this equation with the parameter $M$ (note that in general, for the RST, $k=\phi_H-\rho_H\neq 0$, as we show below). In fact, this can be seen from the previous particularization of $c_1$ and $c_2$ for the RST. Thus, we identify the classical black hole mass as
\be
\label{eq:adm}
\frac{M}{\lambda}= 2c_1e^{-2k}+k\frac{N}{24\alpha^2}\,.
\ee
We can isolate this contribution of the classical CGHS black hole by appealing to the pseudotensor method (cf. \eqref{eq:subtractedBHmass}). To do so,  
we first write the asymptotic expansions \eqref{eq:asyphi} and \eqref{eq:asyrho} in Eddington-Finkelstein coordinates ($y=-\lambda^2y^+ y^-=e^{2(\lambda\sigma+k)/\alpha}$) 
\bea
\phi_{\sigma^1} &=&\phi_y=-\lambda\sigma^1-\left(c_1+\frac{2kc_2}{\alpha}\right)e^{-2(k+\lambda\sigma^1)}-\frac{2c_2}{\alpha}\sigma^1 e^{-2(k+\lambda\sigma^1)}+\mathcal{O}\left(e^{-4\lambda\sigma^1}\right) \\ 
\label{eq:rhonumsigma}\rho_{\sigma^1} &=&\rho_y + \frac{\lambda\sigma^1+k}{\alpha}-\ln(\alpha)\,,
\eea
where the sub-indices $_{\sigma^1,y}$ refer to the conformal factor in Eddington–Finkelstein and extended Kruskal coordinates, respectively. From these expressions, we can identify $r_{\text{semi-CGHS}}^{(\text{HH})}=\frac{2c_2}{\alpha} e^{-2k}$, and evaluate the black hole mass using \eqref{eq:subtractedBHmass}, obtaining precisely \eqref{eq:adm}.


Before computing the entropy, is worth mentioning that since we are studying static solutions, the black hole should be in thermal equilibrium with the thermal bath \eqref{eq:bath-temperature}. Therefore, its temperature is
\be \label{eq:temperature-simple-RST}
T_\text{H}=\frac{\lambda}{2\pi\alpha}\, .
\ee
For the entropy, we use Wald's prescription \eqref{eq:Swaldgenint} and obtain 
\beq
S_{\text{Wald}}=\bigg(2e^{-2\phi}+\frac{N}{12}Z-(1-\epsilon)\frac{N}{12}\phi\bigg)\bigg|_H\,,
\eeq
where $Z=2\rho-2\xi$ on-shell. The auxiliary field $\xi_{\pm}$ is given by \eqref{eq:xiHHvacum_num}. Following our discussion around \eqref{eq:xiHHvacum} we choose the integration constants $c_++c_-=\ln(-\lambda^2\delta_+\delta_-)$ such that, when we evaluate $\xi$ at the horizon bifurcation point ($y^{+}=y^{-}=0$), we find  
\be 
\xi|_H=-\ln\left(-\frac{y_0^+y_0^-}{\delta^2}\right)=-\frac{\lambda}{\alpha}((\sigma^+_0-\sigma^+_{\delta})-(\sigma^-_0-\sigma^-_{\delta}))=-\frac{2\lambda\sigma_{\text{max}}}{\alpha}=-\frac{\lambda L}{\alpha}.
\ee
In the last step we have written the result in Eddington-Finkelstein coordinates and have identified $(\sigma^+_0-\sigma^+_{\delta})-(\sigma^-_0-\sigma^-_{\delta})=2\sigma_{\max}=L$ with the volume of a one-dimensional box of size $L$. Inserting this result into the Wald entropy we find 
\be
\label{eq:Stot_num}
S_{\text{Wald}}=2e^{-2\phi_{H}}-(1-\epsilon)\frac{N}{12}\phi_{H}+\frac{N}{6}\rho_{H} + \frac{N}{6} \frac{\lambda L}{\alpha}\, .
\ee
As in the previous section, we can see that the last term above can be interpreted as the entropy of a bath of thermal radiation at temperature $T=\lambda/(2\pi\alpha)$. In our numerical analysis, we will be interested in the difference 
\be
\label{eq:entropy}
\Delta S=S_{\text{Wald}}-S_{\text{thermal}}=2e^{-2\phi_{H}}-\frac{N(1-\epsilon)}{12}\phi_{H}+\frac{N}{6}\rho_{H}\,.
\ee
This expression interpolates between the RST (for $\epsilon=0$) and the semi-classical CGHS $(\epsilon=1)$ models. Note also that this expression has both explicit and implicit dependence on $\epsilon$ since $\alpha$ and $k$ implicitly depend on the specific model and thus depend on $\epsilon$. 

\noindent $\bullet$ {\it RST model in generalized Kruskal coordinates.} In the RST model, it is not difficult to find the exact values for the {\it outputs} $(\alpha,k,c_1)$ in terms of the {\it inputs} $(N,\phi_{H},\rho_{H})$. We note that, in this case, the asymptotic relation between $\phi$ and $\rho$ in \eqref{eq:asygauge} becomes exact for all $y$. In this case, we can solve \eqref{eq1} and \eqref{eq2} analytically, with initial conditions $\phi_{H}$ and $\rho_{H}$ at $y=0$, together with the regularity conditions \eqref{eq:derivatives} and \eqref{eq:derivatives2}, that in terms of the {\it inputs} reads
\be\label{eq:RSTphi0rho0}
\phi=\phi_{H}+\frac{24}{N}e^{-2\phi_{H}}\left(1+e^{2\rho_H}y\right)+\frac{1}{2}W_{-1}\left(-\frac{48}{N}e^{-2\phi_{H}-\frac{48}{N}e^{-2\phi_{H}}(1+e^{2\rho_{H}}y)}\right)\, .
\ee
Finally, expanding this result for large $y$, we immediately get
\be
\alpha=1\, \qquad  \text{and} \qquad c_1=\frac{e^{-2\rho_{H}}}{2}+ \frac{N}{48} \, \rho_{H}\,e^{2(\phi_{H}-\rho_{H})}\, .
\ee
Using the asymptotic solution \eqref{eq:asygauge} for generalized Kruskal coordinates with $\alpha=1$,
\be
k=\phi_{H}-\rho_{H}=\phi-\rho\,.
\ee
then the ADM mass \eqref{eq:adm} gives
\be
\frac{M}{\lambda}=e^{-2\phi_{H}}+\frac{N}{24}\phi_{H}\,.
\ee
The temperature is given in \eqref{eq:temperature-simple-RST} with $\alpha=1$, and the relationship between the entropy \eqref{eq:entropy}, and the mass \eqref{eq:adm} reduces to
\be 
\label{eq:DeltaSapp}
\Delta S=2 e^{-2\phi_{H}}+\frac{N}{12}\phi_{H}-\frac{Nk}{6}=\frac{2M}{\lambda}-\frac{Nk}{6}\,.
\ee
In the Kruskal gauge ($k=0$) we recover $\Delta S=\frac{2M}{\lambda}$ in \eqref{eq:Swaldsemiv2}. Note that we can always shift the value of $\rho_{H}$ to match $\phi_{H}$ by rescaling the $x^{\pm}$ coordinates. Our goal in what follows is to see whether this result still holds in the case $\epsilon=1$.  Before exploring this relationship numerically, we will set our expectations using the large $|\phi_{H}|=-\phi_{H}$ expansion.

\subsubsection{Large $|\phi_{H}|$ expansion}

\noindent The limit of large $|\phi_{H}|=-\phi_{H}$ corresponds to the situation $\phi_{H}\ll \phi_c=-\frac{1}{2}\ln\left(\frac{N}{12}\right)$. 
A consistent choice for the small parameter is
\be
\eta\equiv\frac{N}{12}e^{2\phi_{H}}~\ll1~,
\ee
such that, for any given $N$, one can always choose $|\phi_{H}|=-\phi_{H}$ large enough for $\eta$ to be small.
From the equations of motion \eqref{eq1} and \eqref{eq2} it is easy to see that the leading order of this expansion is the classical limit (that is, $N\to 0$). In this limit, the classical solution \eqref{eq:eternalBH} is recovered, but now in terms of $\phi_{H}$ and $\rho_{H}$. To obtain information about the {\it output} parameters $(\alpha, k, c_1)$ we need to go to higher orders. 

To gain further intuition of this expansion, it is convenient to start with the RST model ($\epsilon=0$). In this case, the the large $|\phi_{H}|$ expansion can be readily computed from \eqref{eq:RSTphi0rho0}. We obtain
\be 
\phi=\phi_{H} - \frac{1}{2}\ln \left(1+e^{2\rho_{{H}}}y\right)- \frac{\eta}{8}~\frac{\ln \left(1+e^{2\rho_{H}}y\right)}{1+e^{2\rho_{H}} y} + \mathcal{O}(\eta^2)\, ,
\ee
and $\rho=\phi-\phi_{H} + \rho_{H}$. We then propose the following ansatz for the large $-\phi_{H}$ expansion for arbitrary $\epsilon$
\bea \label{eq:largephi0phi}
\phi= \phi_{H} - \frac{1}{2}\ln \left(1+e^{2\rho_{H}}y\right) + \eta ~ F(y) + \mathcal{O}(\eta^2)\, ,\\
\rho= \rho_{H} - \frac{1}{2}\ln \left(1+e^{2\rho_{H}}y\right) + \eta ~ G(y) + \mathcal{O}(\eta^2)\, . \label{eq:largephi0rho}
\eea
Inserting the ansatz into the equations of motion and imposing $F(0)=G(0)=0$ we obtain a closed expression for $F$ and $G$,
\bea
F(y)&=& \frac{1}{8(1+e^{2\rho_{H}}y)}\Big(-\epsilon \, e^{2\rho_{H}}y+(\epsilon \, e^{2\rho_{H}} y-1)\ln(1+e^{2\rho_{H}}y)\Big)\, , \\
G(y)&=& \frac{1}{8(1+e^{2\rho_{H}}y)}\Big(-2\epsilon \, e^{2\rho_{H}}y-(1+\epsilon)\ln(1+e^{2\rho_{H}}y)\Big)\, .
\eea
Finally, expanding the approximate solutions \eqref{eq:largephi0phi} and \eqref{eq:largephi0rho} for large  $y$, and comparing them with the asymptotic expansions \eqref{eq:asyphi} and \eqref{eq:asyrho} we directly get
\bea
\alpha&=&1-\frac{\eta}{4}~\epsilon+\mathcal{O}(\eta^2)\, ,\\
k&=&\phi_{H}-\rho_{H}+\frac{\eta}{8}~\epsilon~(2\rho_{H}-1) +\mathcal{O}(\eta^2)\;,\\
c_1&=&\frac{e^{-2\rho_{H}}}{2}+ \frac{\eta}{4}  \Big(\rho_{H}+\epsilon\,(\rho_{H}-1)\Big)~e^{-2\rho_{H}} +\mathcal{O}(\eta^2)\, .
\eea

From this result, we see that for the RST model the $\mathcal{O}(e^{2\phi_{H}})$ already gives us the exact result for $(\alpha,k,c_1)$. For the Polyakov action ($\epsilon=1$), we expect further corrections from higher orders. However, at this order we can state some conclusions regarding the relationship between mass and entropy. From the expressions for the mass \eqref{eq:adm}  and entropy \eqref{eq:entropy} together with the approximated {\it outputs}, we get 
\bea
\label{eq:2mlargephi0}
\frac{2M}{\lambda}&=&2e^{-2\phi_{H}}\left(1+\frac{\eta}{2}\left(\phi_{H}-\frac{\epsilon}{2}\right) + \mathcal{O}(\eta^2)\right)\;,\\
\Delta S&=&2 e^{-2\phi_{H}}\left(1+\frac{\eta}{2}(2\rho_{H} -\left(1-\epsilon)\phi_{H}\right) + \mathcal{O}(\eta^2)\right)\;.
\eea
The difference between $\Delta S$ and $2M/\lambda$ scales as
\be
\label{eq:Sm2M}
\Delta S - \frac{2M}{\lambda}
=
2 e^{-2\phi_{H}}\,\frac{\eta}{4}
\bigl[\epsilon(1+2\phi_{H})-4(\phi_{H}-\rho_{H})\bigr]\,.
\ee
For $\epsilon=0$, and $k=\phi_{H}-\rho_{H}$, we recover the
result obtained in \eqref{eq:DeltaSapp}.
For the same initial conditions $(N,\phi_{H},\rho_{H})$, the entropy of the
semi-classical CGHS model (i.e.\ $\epsilon=1$) differs from that of the
exactly solvable RST model by the term
$e^{-2\phi_{H}}\,\frac{\eta}{2}(1+2\phi_{H})$.
For the latter family of models, the black hole temperature is fixed
by $\lambda$, while in the former case we find that the initial conditions modify
the temperature through the parameter $\alpha$ (see
\eqref{eq:temperature-simple-RST}).
One might therefore attribute the entropy difference to this mismatch in
temperature.
However, expanding $2M/\lambda'$ with $\lambda'=\lambda+\delta\lambda$ and
$\delta\lambda=-\lambda(1-1/\alpha)\simeq\lambda(1-\alpha)$ yields an
effective entropy $2M\alpha/\lambda$.
Comparing both sets of models at the same temperature still leaves a
residual difference, $e^{-2\phi_{H}}\,\eta(1+\phi_{H}).$
The origin of this contribution can thus be traced to the breaking of the
symmetry of the classical action in the semi-classical CGHS model, a symmetry
that is preserved by the parametric family of models.

\subsection{Numerical results}

\noindent In this subsection, we present the numerical results. We focus on static solutions for the semi-classical CGHS model in extended Kruskal coordinates. We begin by describing the numerical integration and the fitting procedure used to match $\phi$ and $\rho$ onto their corresponding asymptotic expansions \eqref{eq:asyphi} and \eqref{eq:asyrho}. This allows us to extract the relevant {\it output} parameters $(\alpha,k, c_1)$ and use them to determine the mass \eqref{eq:adm} and temperature \eqref{eq:temperature-simple-RST} of the black hole. We also investigate the numerical solutions for $\phi$ and $\rho$  and compare them with their analytical counterparts in the RST model and in the large-$|\phi_{H}|$ expansion. Finally, we analyse the relationship between mass and entropy and show that the simple relation that holds in the RST model is not maintained for the semi-classical CGHS model.

\subsubsection{Description of the numerical calculations}

\noindent 
We numerically solve the coupled non-linear differential equations for $\phi$ and $\rho$, given by \eqref{eq1} and \eqref{eq2}, under the constraint \eqref{constraint}, using an adaptive 8th-order Dormand-Prince method with stringent error tolerances ($\sim 10^{-15}$) implemented in \textit{Python}. We set the initial conditions for the variables  $\phi_{H}$ and $\rho_{H}$ at the horizon,\footnote{Because of the numerical problems of the equations at $y=0$, we applied the initial conditions at $y=10^{-8}$.} while the initial values of their derivatives are determined by Eqs. \eqref{eq:derivatives} and \eqref{eq:derivatives2}. The equations were solved in generalized Kruskal coordinates $y$ on a logarithmically-spaced grid spanning $[10^{-8}, 2\times 10^8]$.

To extract the asymptotic coefficients in \eqref{eq:asyphi}, we employ a two-step fitting procedure in the far-field region (more explicitly $y \in [2\times 10^6, 2\times 10^8]$).
First, we fit the difference for the solution of $(\phi-\rho)$ to the logarithmic form
\be
\phi-\rho= m\, \ln(y) + n  + \mathcal{O}(y^{-2\alpha})\,,
\ee
Note that this fitting ensures that the error goes as $\mathcal{O}(y^{-2\alpha})$. Comparing this result with the asymptotic expansion \eqref{eq:asygauge}, we easily get
\be
\alpha=1-2m\, ,\qquad k=n + \ln(1-2 m)\, .
\ee

Once we have $\alpha$ and $k$, we can extract $c_1$. To this end, we use the following two functions
\be \label{eq:rhofit}
(\rho-\bar \rho) \, y^{\alpha}\, , \qquad (\phi-\bar \phi)\, y^{\alpha}\;,
\ee
where 
\be 
\bar \phi= -\frac{\alpha}{2}\ln y +k -c_2\frac{\ln y}{y^\alpha} \, \qquad \bar \rho=  -\frac{1}{2}\ln y +\ln(\alpha)- c_2\frac{\ln y}{y^\alpha}\, ,
\ee
with $c_2$ given in \eqref{eq:c2}, and $\alpha$ and $k$ given by the previous linear fit. According to the asymptotic analysis, these functions go as 
\bea
(\rho-\bar \rho) \, y^{\alpha}&=& -c_1 + \mathcal{O}(y^{-2\alpha})\;,\\
\label{eq:phifit}
(\phi-\bar \phi)\, y^{\alpha} &=& -c_1 + \mathcal{O}(y^{-2\alpha})\;.
\eea
We therefore fit these numerical functions to a constant to obtain two numerical values for $c_1$.

Our final numerical value for $c_1$ is obtained by averaging these two quantities. For the fittings, we employed both Nelder-Mead simplex and L-BFGS-B methods in \textit{Python} to ensure robustness. The error of the fitting, defined as the sum of the square of the difference between the numerical solution and the asymptotic expansion for $y\in[2\times 10^6\,,\,2\times 10^8]$, is the order of $10^{-15}$. Once $\alpha\,,\,k\,,$ and $c_1$ have been calculated, we evaluate the ADM mass and the entropy using \eqref{eq:adm} and \eqref{eq:entropy}, respectively.

\subsubsection{Results for the dilaton and the conformal factor}

\begin{figure}[t!]
    \centering
    \begin{subfigure}[b]{0.45\textwidth}
        \includegraphics[width=\textwidth]{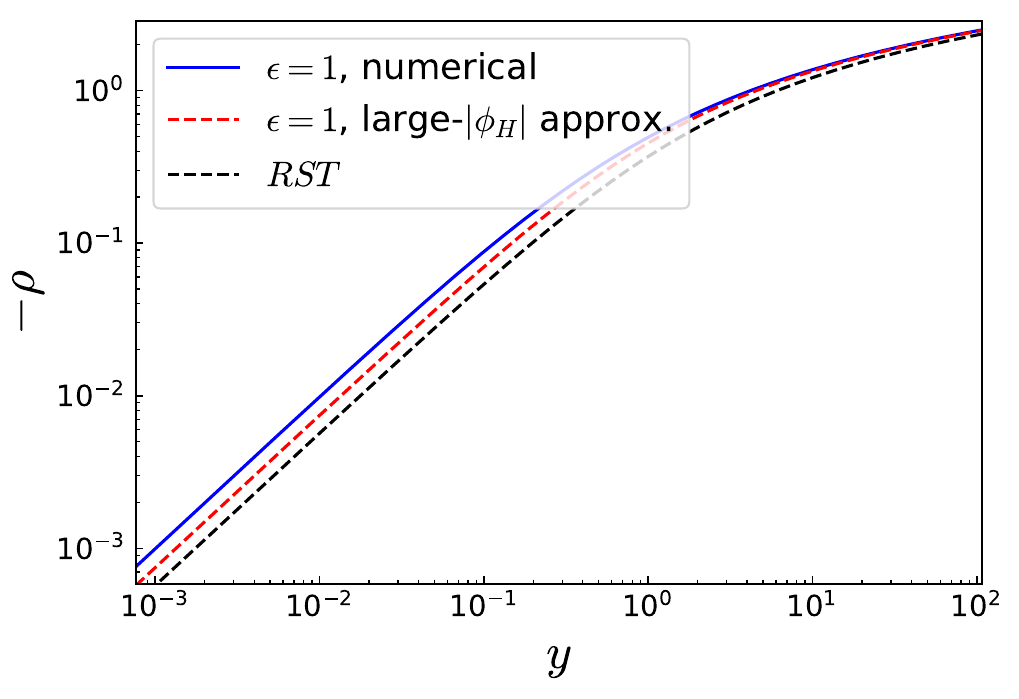}
    \end{subfigure}
    \begin{subfigure}[b]{0.45\textwidth}
        \includegraphics[width=\textwidth]{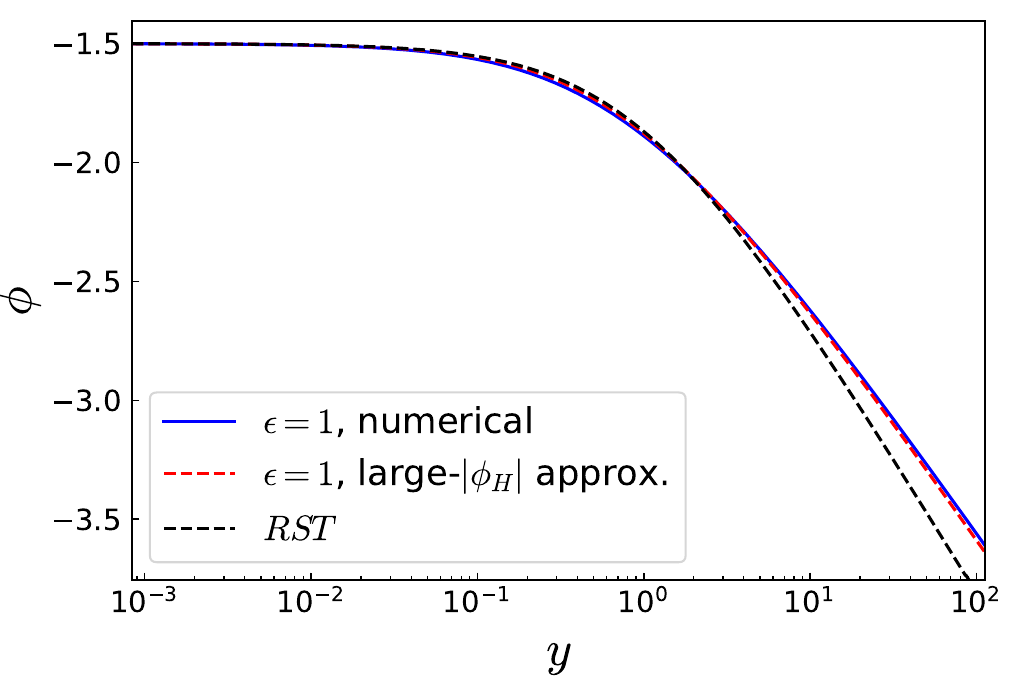}
    \end{subfigure}
    \caption{$\rho(y)$ and $\phi(y)$ for the RST and the $\epsilon=1$, numerical, models. The initial conditions for both models are $\phi_{H}=-1.5$, $\rho_{H}=0$ and $N=120$.}
    \label{fig:phirho_a}
\end{figure}
\begin{figure}[t!]
    \centering
    \begin{subfigure}[b]{0.45\textwidth}
        \includegraphics[width=\textwidth]{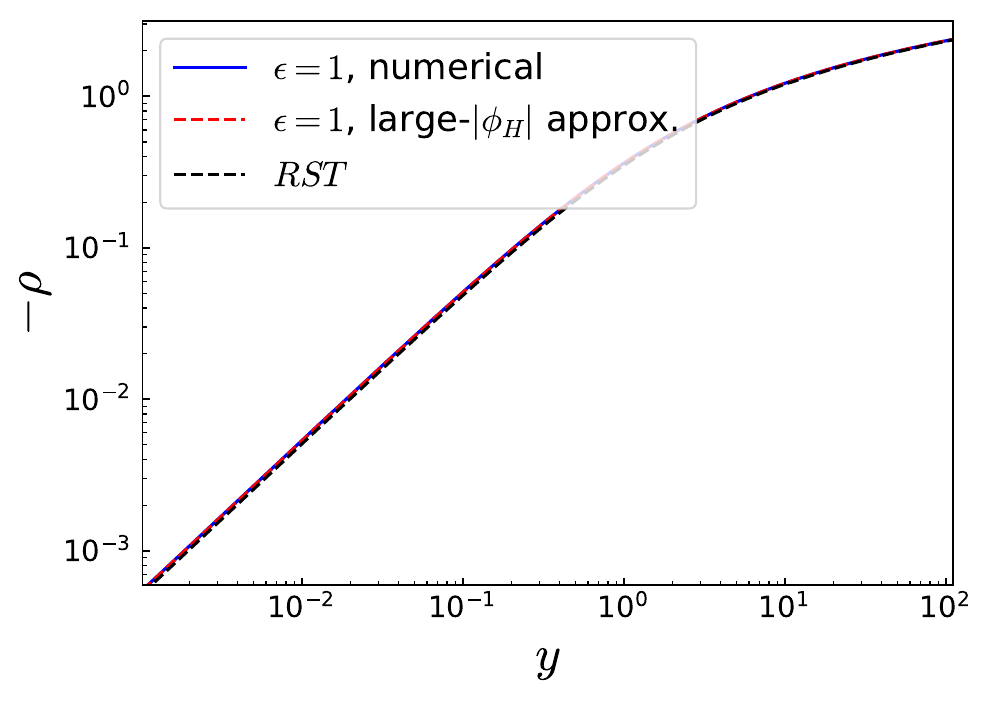}
    \end{subfigure}
    \begin{subfigure}[b]{0.45\textwidth}
        \includegraphics[width=\textwidth]{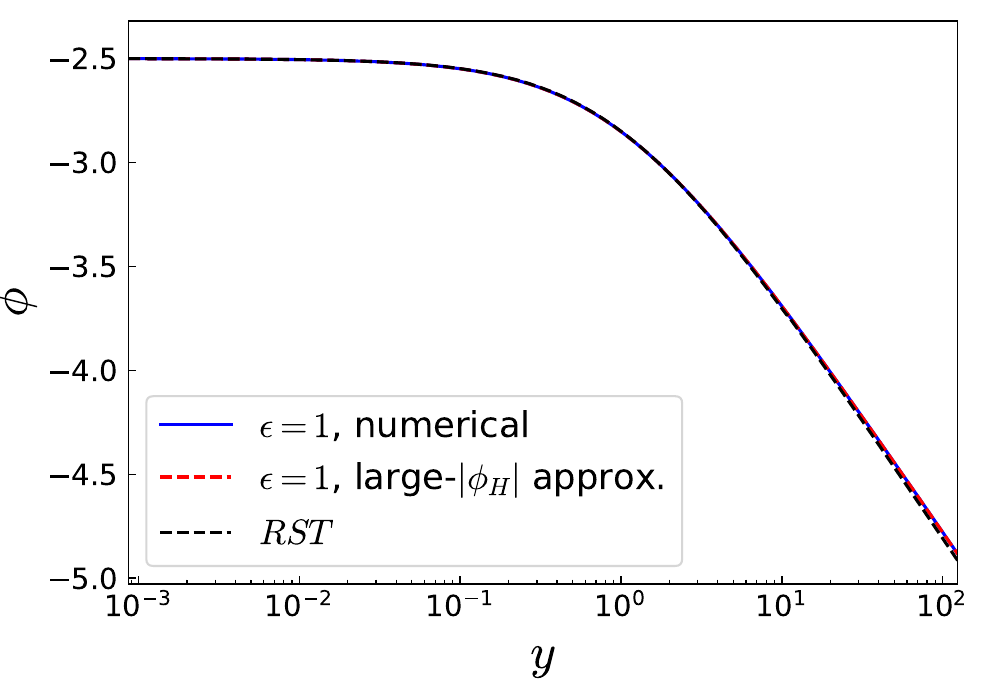}
    \end{subfigure}
\caption{$\rho(y)$ and $\phi(y)$ for the RST and the $\epsilon=1$, numerical, models. The initial conditions for both models are $\phi_{H}=-2.5$, $\rho_{H}=0$, $N=120$. Note that this choice of $\phi_{H}$ and $N$ ensures that we are at the upper bound imposed by the semi-classical condition \eqref{eq:scvalidity}, see the discussion below \eqref{eq:derivatives2}.}
    \label{fig:phirho_b}
\end{figure}
\noindent In Figures \ref{fig:phirho_a} and \ref{fig:phirho_b} we compare the numerical solutions for $\phi(y)$ and $\rho(y)$ for the RST ($\epsilon=0$) and semi-classical CGHS ($\epsilon=1$) models. 
For light black holes, i.e., $\phi_{H}\sim\phi_c$, Figure \ref{fig:phirho_a} shows that $\phi$ and $\rho$ exhibit similar behavior in both the RST and $\epsilon = 1$ models, albeit with differing numerical values. We have included the large $\phi_{H}$ approximation up to order $e^{2\phi_{H}}$, showing a close agreement with the numerical calculations.
For heavy black holes, $\phi_{H}\ll\phi_c$, the back reaction terms in the equations of motion are small and both $\epsilon=0$ and $\epsilon=1$ models tend to the classical solution. Figure \ref{fig:phirho_b} already shows this behavior for $\phi_{H}=-2.5$.

\begin{figure}[t!]
    \centering
      \includegraphics[width=0.45\textwidth]{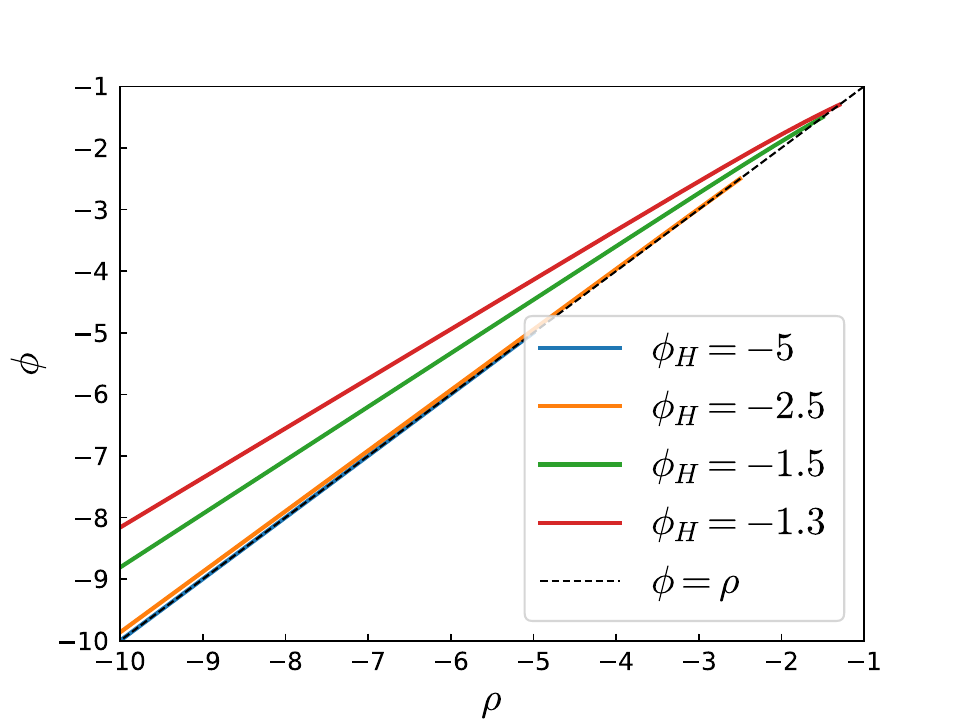}
\caption{Dilaton $\phi$ as a function of $\rho$ for the $\epsilon=1$ numerical model for different initial conditions $\phi_{H}$, with $\rho_{H}=\phi_{H}$ and $N=120$. As a reference, the (black dashed) line $\phi=\rho$ has been included. Note that the highest values of $\phi$ and $\rho$ are reached at the horizon, i.e. the initial conditions, since we solved for the outside of the horizon region.}
    \label{fig:PvsR}
\end{figure}

We can also plot $\phi$ versus $\rho$, as shown in Figure \ref{fig:PvsR} for different values of $\phi_H$ with $\rho_H=\phi_H$. For the RST and the classical CGHS models $\phi=\rho$ with these initial conditions. In semi-classical CGHS, we see that for small values of $|\phi_H|$ there are corrections in the $\rho$ versus $\phi$ dependence. However, as the value of $|\phi_H|$ increases, quantum effects become negligible, and we recover the classical solution where the Kruskal symmetry is reinstated.

\subsubsection{Mass and Entropy}

\begin{figure}[t!]
    \centering
    \includegraphics[width=0.45\textwidth]{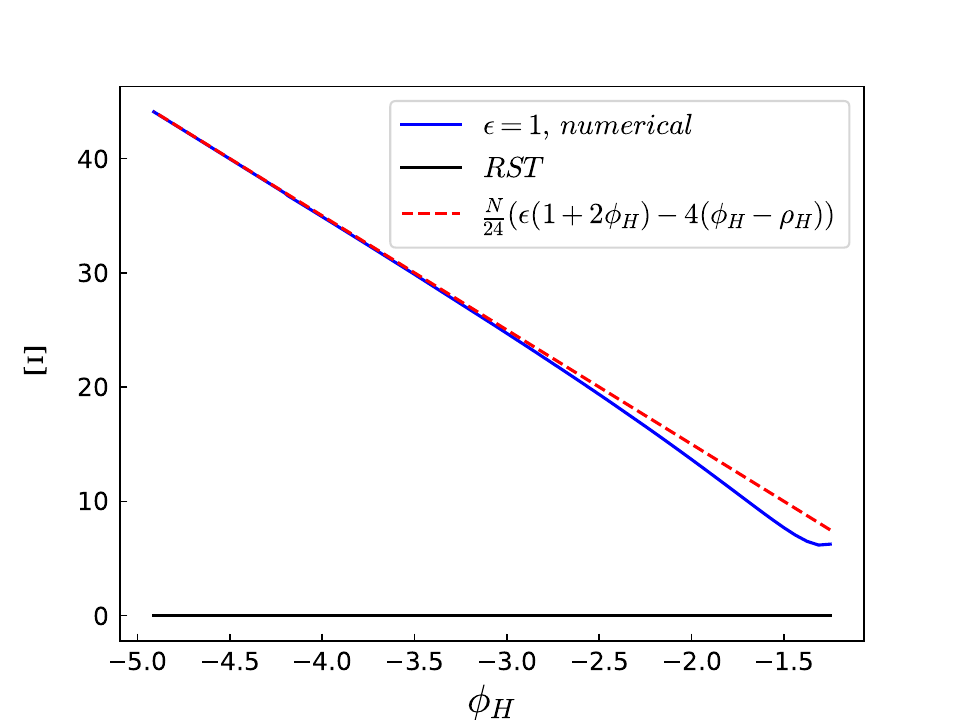}$\qquad$ \includegraphics[width=0.45\textwidth]{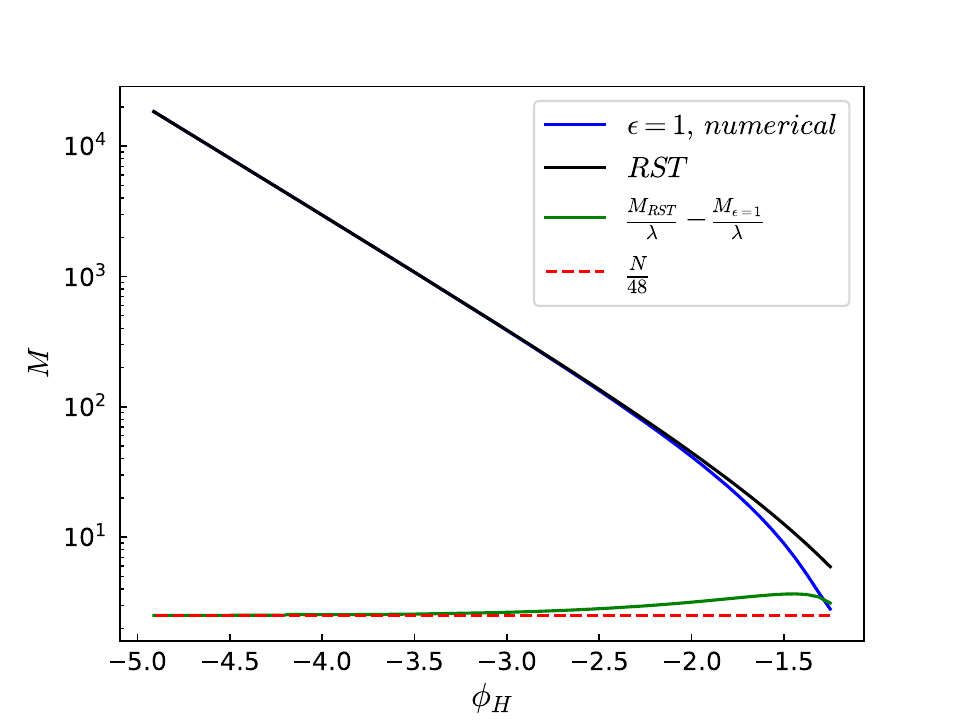}
\caption{On the left-hand plot, comparison of the difference between the entropy and the $\frac{2M}{\lambda}$ quantity, i.e $\Xi$, for the numerical $\epsilon=1$ model, blue line, and the RST, black line. As initial conditions for these calculations, we took $\rho_H=\phi_H$ and $N=120$. Note that $\Xi_{\text{RST}}=0$ since $\Delta S=\frac{2M}{\lambda}$ for $\rho_H=\phi_H$ while $\Xi_{\epsilon=1}$ has a constant plus a linear in $\phi_H$ terms which are well explained by the large-$|\phi_H|$ expansion, the dashed-red line. The comparison of the mass is presented on the right-hand plot. In this case, we added the difference $M_{\text{RST}}/\lambda-M_{\epsilon=1}/\lambda$, green line, and compared against the $\frac{N}{48}$ factor expected from the large-$|\phi_H|$ expansion, red line.}
    \label{fig:entropy}
\end{figure}

\begin{figure}[t!]
    \centering
    \includegraphics[width=0.45\textwidth]{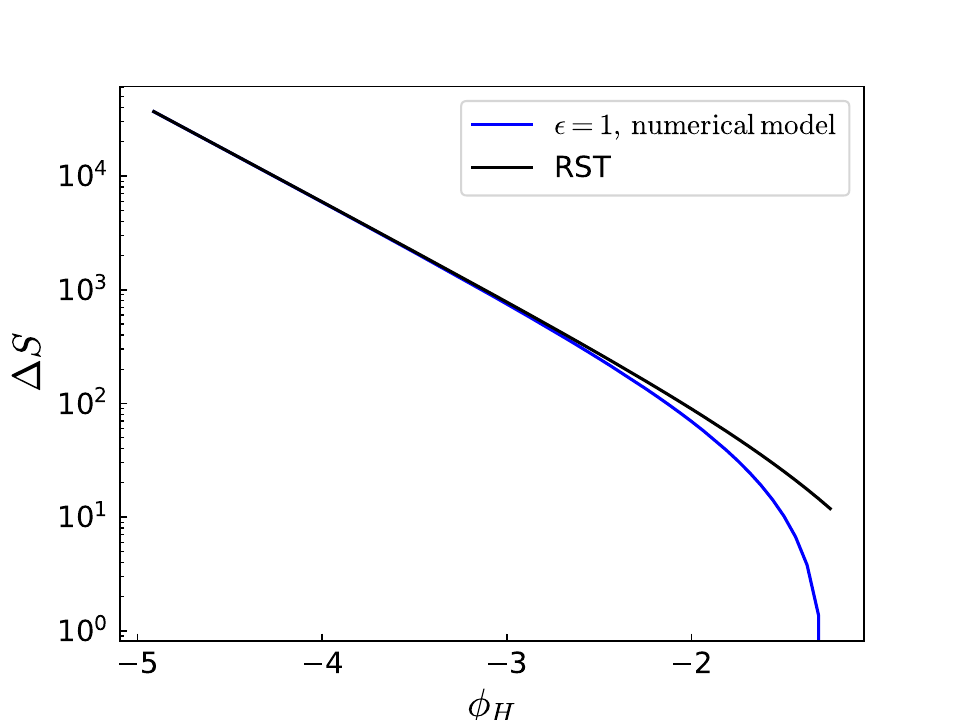}$\qquad$ \includegraphics[width=0.45\textwidth]{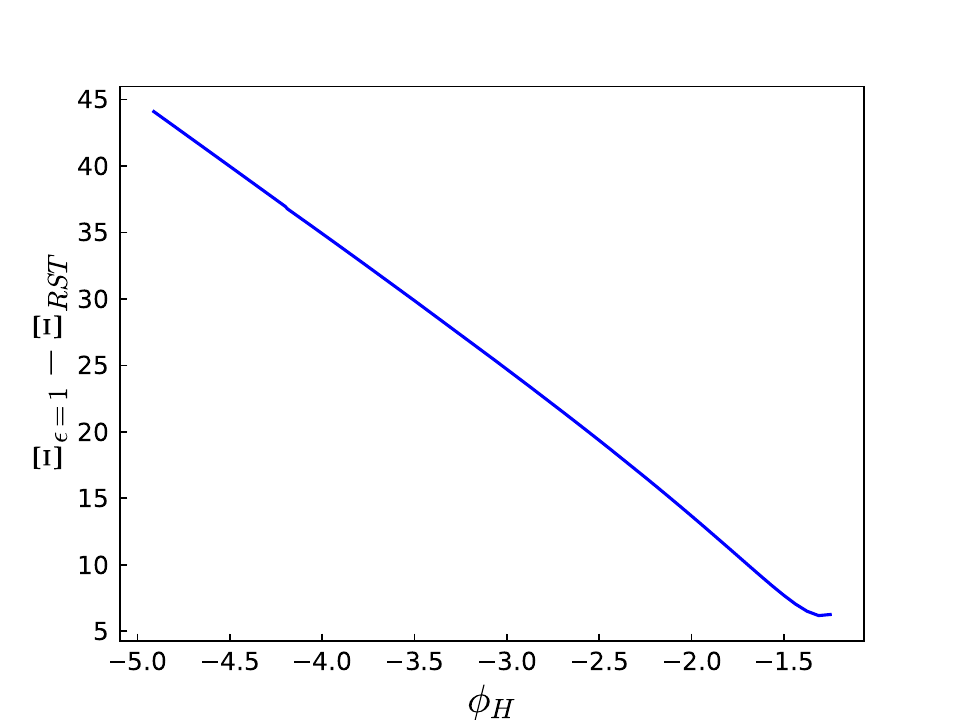}

\caption{Comparison of the entropy for the numerical $\epsilon=1$ and the RST models. In these plots, $N=120$ and $\rho_H=\phi_H$. For large values of $|\phi_H|$, the solution tends to the classical CGHS value, and both the numerical and RST give the same entropy.}
    \label{fig:entropy2}
\end{figure}

\noindent
We can now analyze the effect of adding the RST term in the action on both the ADM mass and the entropy. For this family of models, we found that $\Delta S$, i.e. the Wald entropy minus the thermal contribution, is equal to $\frac{2M}{\lambda}$ in Kruskal gauge, see \eqref{eq:Swald_1}. However, the numerical results shown in Figures \ref{fig:entropy} and \ref{fig:entropy2} for the $\epsilon=1$ model indicate that
\[
\Xi \equiv \left| \Delta S - \frac{2M}{\lambda} \right|
\]
exhibits a linear dependence of the dilaton at the horizon $\phi_{H}$. This behavior is in good agreement with the large-$|\phi_{H}|$ expansion,
\[
\Xi = \frac{N}{24}\!\left[\epsilon\bigl(1+2\phi_{H}\bigr)
      -4\bigl(\phi_{H}-\rho_{H}\bigr)\right]
      + \mathcal{O}(\eta^2)
   = \frac{N}{24}\bigl(1+2\phi_{H}\bigr)
      + \mathcal{O}(\eta^2),
\]
where in the last step we used that $\rho_{H}=\phi_{H}$. The linear term in $\phi_{H}$ appearing in the difference of $\Xi$ between the two models (with $\Xi_{\text{RST}}=0$) can be identified as the entropy contribution arising from the $I_{\text{RST}}$ term, namely $-\frac{N}{12}\phi_{H}$ as shown in \eqref{eq:Stot_num}. The constant term, on the other hand, originates from the fact that the black hole mass $M/\lambda$ differs between the RST model and the semi-classical CGHS model by a factor $\frac{N}{48}$ at first order in the $\eta$-expansion, as shown in \eqref{eq:2mlargephi0}. This is shown on the right-hand side of Figure \ref{fig:entropy}. Therefore, under the same initial conditions $(N, \phi_H,\rho_H)$, the semi-classical CGHS and the RST models lead to different black hole masses $M/\lambda$ by a constant factor that depends on $N$. 

\section{Discussion}
\label{sec:discussion}

\noindent The purpose of this article was two-fold. First, we revisited exact (eternal) quantum black hole solutions to a one-parameter analytically solvable semi-classical extension of classical CGHS gravity, interpolating between the RST and BPP models. We focused on the geometry for quantum matter in Hartle-Hawking and Boulware vacuum states, and the black hole horizon thermodynamics. 

For the geometry we examined in detail the singularity structure of such eternal black holes and the possibility of naked singularities in this semi-classical regime. We found that naked singularities are ruled out in the Hartle-Hawking vacuum for the semi-classical approximation to be self-consistent. In the Boulware vacuum, however, the possibility exists, albeit in a curious regime of the semi-classically viable parameter space, where the black hole must have an exceedingly high entropy. An argument against the existence of such naked singularities is possible when the classical black hole has a higher-dimensional pedigree. In that event, the regime of validity (\ref{eq:scvalidity}) is appended such that the classical entropy is always much less than the ground state entropy, $S_{0}\equiv e^{-2\phi_{0}}$, of the higher-dimensional near-extremal black hole, $e^{-2\phi_{0}}\gg e^{-2\phi_{H}}$ (so that the dynamical dilaton captures only small deviations away from extremality \cite{Pedraza:2021cvx}). Thus, consistency with the higher-dimensional perspective could preclude the large-masses necessary for the existence of such naked singularities. Further, independent of the higher-dimensional perspective, here our analysis is only for eternal black holes. It is plausible that black holes of such large mass would not form in dynamical collapse scenarios, the original context for the weak cosmic censorship conjecture.  Finally, we also found that in the Hartle-Hawking vacuum the null energy condition is not violated, while in the Boulware vacuum both the NEC and the half-ANEC can be violated.

In exploring the horizon thermodynamics, we took a quasi-local perspective, introducing a finite Dirichlet boundary, such that the system of interest was the region bounded by the finite boundary and horizon. We computed the quasi-local Brown-York energy (which asymptotes to the ADM mass) of the quantum black hole using local counterterms we derived via a Hamilton-Jacobi method. To compute the entropy we applied the Iyer-Wald prescription, showing that the semi-classical Wald entropy is equal to the generalized entropy, and the on-shell microcanonical action. This echoes results found in the context of semi-classical JT gravity \cite{Pedraza:2021cvx,Pedraza:2021ssc,Svesko:2022txo}, confirming it holds for a wider class of models. When the Wald functional is evaluated on the horizon for matter in the Hartle-Hawking state, the entropy is equal to the entropy of the classical CGHS black hole, plus a thermal radiation component. we derived a first law for quantum black holes, where the classical entropy variation is replaced by the variation of the generalized entropy. Finally, we found that there exist regions of thermal stability. 

The second aim of this project was to numerically construct eternal quantum black hole solutions to semi-classical CGHS gravity, i.e., the CGHS model coupled to the Polyakov action, where the Kruskal gauge symmetry affording the aforementioned analytic solutions is lost.  Here we find that the semi-classical field equations admit regular eternal black hole solutions which can be obtained numerically by imposing initial conditions at the horizon and integrating outward. The resulting solutions are then matched onto the asymptotic expansion of the solutions, from which the mass can be extracted in a controlled manner. As a non-trivial consistency check, we introduced an interpolating parameter $\epsilon$ connecting the RST model to the semi-classical CGHS model and verify that our numerical construction reproduces the RST results in the appropriate limit. In addition, we developed a large- $|\phi_{H}|$ expansion and find that it is a good approximation to the numerical solutions precisely in the regime where the semi-classical description is reliable.  A key result of our analysis is that the mass–entropy relation characteristic of the RST model is not preserved in semi-classical CGHS gravity.

\vspace{2mm}

\noindent Let us now conclude with some final remarks, and point to future directions.

\vspace{2mm}

\noindent \emph{Dynamical black holes.}  In this article we focused on eternal black holes and studied the Boulware and Hartle-Hawking vacua. Notably, classical CGHS gravity and its semi-classical extensions admit exact constructions of dynamical black holes, i.e., those formed under collapse and evaporate. Such solutions serve as useful proxies to explicitly study aspects of black hole evaporation. It would be interesting to extend our methods to dynamical situations, and examine the Unruh vacuum. We could also, for example, quasi-local versions of Bondi mass, and compute the dynamical entropy \cite{Hollands:2024vbe,Visser:2024pwz} of quantum black holes. It would be particularly interesting to see whether we can extend the connection between microcanonical action and Wald entropy to dynamical settings. 

\vspace{2mm}

\noindent \emph{Entropy of Hawking radiation and island formula.} Two-dimensional dilatonic black holes offer a context to explicitly study the black hole information puzzle. In particular, the island rule, an extremization prescription of the generalized entropy, allows for a direct computation of the fine grained entropy of Hawking radiation. More precisely, the island rule is a special case of the quantum extremal surface (QES) formula
\cite{Hubeny:2007xt,Faulkner:2013ana,Engelhardt:2014gca} 
\beq S_{\text{vN}}(\Sigma_X)=\underset{X}{\text{min}}\,\underset{X}{\text{ext}}\left[\frac{\text{Area}(X)}{4G}+S^{\text{sc}}_{\text{vN}}(\Sigma_{X})\right] .\label{eq:QESformula}\eeq
Here $S_{\text{vN}}(\Sigma_X)$ is the fine-grained von Neumann entropy of $\Sigma_X$ in the full quantum theory, $S^{\text{sc}}_{\text{vN}}$ is the von Neumann entropy of `bulk' quantum fields in the semi-classical approximation, and $\Sigma_{X}$ is a codimension-1 slice bounded by a codimension-2 QES $X$ and a cutoff surface. The island formula regards the scenario when case $\Sigma_{X}$ is disconnected, $\Sigma_{X}=\Sigma_{\text{rad}}\cup I$,
where $\Sigma_{\text{rad}}$ is an asymptotically far region outside the black hole (where radiation is collected) and $I$ is an ``island" with $X=\partial I$, that extends into the black hole interior.

 The island formula  has been derived using the ``replica trick" to compute gravitational R\'enyi entropies in the context of AdS$_{2}$ Jackiw-Teitelboim gravity \cite{Almheiri:2019qdq,Penington:2019kki,Goto:2020wnk}, however, such a derivation is lacking for asymptotically flat black holes (see Appendix A of \cite{Hartman:2020swn} for a discussion). At least for eternal black holes, an alternative derivation of the extremization of generalized entropy follows from minimizing the microcanonical action of an entanglement wedge \cite{Pedraza:2021ssc}. When the black hole is coupled to a bath, islands emerge from maximizing the entropy at fixed energy, consistent with the island formula, and without the employ of any replica trick. While this method was explicitly carried our for (A)dS JT gravity \cite{Pedraza:2021ssc,Svesko:2022txo}, in principle the derivation can be adapted to the dilaton gravity models explored in this article. Indeed, the on-shell microcanonical action for a flat causal diamond, will evaluate to (twice) the generalized entropy. It would be worth performing this computation in detail, as it would provide a first principles derivation of the island formula for asymptotically flat black holes. 

 \vspace{2mm}

\noindent \emph{Horizon thermodynamics in the conformal ensemble.} In this article we placed our 2D black holes inside a finite cavity obeying Dirichlet boundary conditions. There is growing motivation to consider other boundary conditions for finite walls in gravity, notably due to issues of well-posedness of finite Dirichlet boundaries in Lorentzian and Euclidean (3+1)-dimensional general relativity \cite{Anderson:2006lqb,Witten:2018lgb,An:2021fcq}. In particular, conformal boundary conditions, where the conformal class of the induced metric and the trace of the extrinsic curvature of a codimension-1 hypersurface are held fixed, are generally better well-posed. Such boundary conditions have led to a novel kind of quasi-local thermodynamics, namely, conformal thermodynamics \cite{Anninos:2023epi,Anninos:2024wpy,Liu:2024ymn,Anninos:2024xhc,Banihashemi:2024yye}, where the (conformal) Tolman temperature and second fundamental form are held fixed. When performing a dimensional reduction for nearly-extremal black holes in the conformal ensemble, the resulting 2D dilaton theory has a reduced pair of boundary conditions \cite{Galante:2025tnt}. It would be worth carrying out a similar quasi-local analysis for the CGHS model (and its semi-classical extensions) for such boundary conditions, as this would effectively characterize the quasi-local conformal thermodynamics of the four-dimensional parent theory.

\vspace{2mm}

\noindent \emph{A quantum Penrose inequality for 2D quantum black holes.} Though in a small corner of the allowed parameter space, we uncovered the possibility of naked singularities. This suggests a violation of the weak cosmic censorship conjecture (WCCC) in two-dimensional semi-classical gravity \cite{Russo:1992yh}. As such, it would be valuable to examine the Penrose inequality in the current context. In four-dimensional general relativity, the Penrose inequality was devised as a means to test the WCCC \cite{Penrose:1969pc}, and connects the ADM mass of the black hole spacetime with the area of any cross section of the event horizon $A$,
\be
G_{4}M_{\text{ADM}} \geq \sqrt{\frac{A}{16 \pi}} \,,
\label{eq:PenroseinqOG}\ee
with saturation for Schwarzschild black holes. An input in the `derivation' of this inequality is that WCCC holds; any violation of the Penrose inequality is therefore thought to indicate a counterexample to WCCC. Thus far, the inequality (\ref{eq:PenroseinqOG}) is a conjecture, having only been proven in special cases \cite{Huisken01,Bray01,Bray:2007opu}. Further, given that entropy in general relativity is proportional to horizon area, the inequality (\ref{eq:PenroseinqOG}) serves as an entropy bound. 

Given that classical dilaton gravity has black holes which form singularities under collapse, analogs of the WCCC and Penrose inequality are expected. As there is no area of the event horizon in $1+1$ dimensions, a natural candidate for the Penrose inequality for two-dimensional dilatonic black holes is to  replace the right-hand side with the Bekenstein-Hawking entropy \cite{Park:1992sd}, 
\beq M_{\text{ADM}}\geq \lambda e^{-2\phi_{H}} \;.\label{eq:PI2D}\eeq
Saturation occurs for the eternal CGHS black hole, where recall the classical entropy is $S_{\text{CGHS}}=2e^{-2\phi_{H}}$.

It is known that the \emph{classical} Penrose inequality (\ref{eq:PenroseinqOG}) is violated in semi-classical gravity \cite{Bousso:2019var,Bousso:2019bkg}. This led to proposals for semi-classical generalizations \cite{Bousso:2019var,Bousso:2019bkg,Frassino:2024bjg,Hafemann:2025mjf,Frassino:2025buh}. In particular, in \cite{Bousso:2019var,Bousso:2019bkg} the authors proposed a \emph{quantum} Penrose inequality, where the classical entropy was replaced by the semi-classical generalized entropy. A three-dimensional variant of this proposal was found to hold for all known three-dimensional anti-de Sitter quantum black holes, suggesting a robust notion of WCCC extends to semi-classical gravity, even at the level of non-perturbative backreaction \cite{Frassino:2024bjg,Frassino:2025buh}.
It would be worth studying the quantum Penrose inequality for the models examined here. These findings could shed light on the status of the weak cosmic censorship conjecture in semi-classical gravity.

\vspace{5mm}

\noindent\textbf{Acknowledgments.}

\vspace{2mm}

\noindent We are grateful to Jose Navarro-Salas and Watse Sybesma for useful discussions. AS acknowledges Juan Pedraza, Watse Sybesma and Manus Visser for numerous past discussions, in part motivating elements of this article.
 The work of JA is supported by
The Science and Technology Facilities Council (grant No. STFC-ST/X000753/1), DPS is supported by the EPSRC studentship grant EP/W524475/1. The work of SP was funded by the Deutsche Forschungsgemeinschaft (DFG, German Research Foundation) under Germany’s Excellence Strategy – EXC 2094 – 390783311. AS is supported by STFC consolidated grant ST/X000753/1 and is partially funded by the Royal Society under the grant ``Concrete Calculables in Quantum de Sitter''.

\appendix 

\section{Coordinate systems} \label{app:coordsys}

\noindent In this appendix we summarize useful geometric identities and coordinate systems used throughout the main text. 

\vspace{2mm}

In conformal gauge, $ds^{2}=-e^{2\rho}dw^{+}dw^{-}$ for arbitrary lightcone coordinates $w^{\pm}$, it is useful to know the following geometric identities:
\beq 
\begin{split}
&g_{+-}=-\frac{1}{2}e^{2\rho}\;,\quad g^{+-}=-2e^{-2\rho}\;, \quad \sqrt{-g}=\frac{1}{2}e^{2\rho}\;,\quad \Gamma^{+}_{\;++}=2\partial_{+}\rho\;,\quad \Gamma^{-}_{\;--}=2\partial_{-}\rho\;,\\
&R=-2\Box\rho=8e^{-2\rho}\partial_{+}\partial_{-}\rho\;,\quad \Box\Phi=-4e^{-2\rho}\partial_{+}\partial_{-}\Phi\;,\quad (\nabla\Phi)^{2}=-4e^{-2\rho}(\partial_{+}\Phi)(\partial_{-}\Phi)\;,\\
&\nabla_{\pm}\nabla_{\pm}\Phi=\partial^{2}_{\pm}\Phi-2(\partial_{\pm}\rho)(\partial_{\pm}\Phi)\;,\quad \nabla_{\pm}\nabla_{\mp}\Phi=\partial_{\pm}\partial_{\mp}\Phi
\end{split}
\label{eq:usegoeidconf}\eeq
for arbitrary scalar field $\Phi$. Also note $\sqrt{-g}=e^{2\rho}/2$.

\vspace{2mm}

Below we list coordinate systems of the classical vacuum solutions.

\vspace{2mm}

\noindent $\bullet$ \emph{Kruskal coordinates.} The static black hole in Kruskal coordinates ($x^{+},x^{-})$ is 
\beq ds^{2}=-e^{2\rho}dx^{+}dx^{-}\;,\quad e^{-2\phi}=e^{-2\rho}=\frac{M}{\lambda}-\lambda^{2}x^{+}x^{-}\;,\eeq
for $M>0$. The bifurcate `point' of the horizon is located at $x^{+}x^{-}=0$, and has a spacelike curvature singularity at $x^{+}x^{-}=M/\lambda^{3}$. 

\vspace{2mm}

\noindent $\bullet$ \emph{Eddington-Finkelstein coordinates.} The null coordinates $(\sigma^{+},\sigma^{-})$ are related to Kruskal coordinates via $\lambda x^{\pm}=\pm e^{\pm\lambda\sigma^{\pm}}$. Under this coordinate transformation, the geometry has the form
\beq 
ds^{2}=-e^{2\rho}d\sigma^{+}d\sigma^{-}\;, \quad e^{-2\rho}=\left(1+\frac{M}{\lambda}e^{-2\lambda\sigma^{1}}\right)\;,\eeq
and
\beq
\label{eqn:kruskdil}
\begin{split}
&e^{-2\phi}=\frac{M}{\lambda}+e^{\lambda(\sigma^{+}-\sigma^{-})}\Rightarrow\phi=-\lambda\sigma^{1}-\frac{1}{2}\ln\left(1+\frac{M}{\lambda}e^{-2\lambda\sigma^{1}}\right),\;
\end{split}
\eeq
where we note that $\rho$ transforms as a metric component\footnote{Under the coordinate transform $x^{\pm}\to\sigma^{\pm}$, it follows $ds^{2}=-e^{2\lambda \sigma^{1}}e^{2\rho(x^{+}(\sigma^{+}),x^{-}(\sigma^{-}))}d\sigma^{+}d\sigma^{-}\equiv -e^{2\rho(\sigma^{+},\sigma^{-})}d\sigma^{+}d\sigma^{-}$, with $e^{2\rho(\sigma^{+},\sigma^{-})}=\left(1+\frac{M}{\lambda}e^{-2\lambda \sigma^{1}}\right)^{-1}$.} and introduced $\sigma^{\pm}=(\sigma^{0}\pm\sigma^{1})$ for time and spatial coordinates $\sigma^{0}$ and $\sigma^{1}$, respectively. Recall that $(\rho-\phi)$ is a free field with general solution $(\rho-\phi)=w_{+}(x^{+})+w_{-}(x^{-})$. This coordinate frame corresponds to the choice where the gauge functions are $w_{\pm}(x^{\pm})=\pm\sigma^{\pm}$.  When $M=0$ the spacetime describes the (flat, i.e., vanishing Ricci scalar) linear dilaton vacuum, and for $M>0$ the black hole asymptotically approaches the linear dilaton vacuum. In these coordinates the bifurcation point is located at $(\sigma^{+}-\sigma^{-})\propto\sigma^{1}\to-\infty$ (the past horizon lies along $\sigma^{+}\to-\infty$ while $\sigma^{-}\to+\infty$ is the future event horizon), such that these coordinates only cover the exterior of the black hole. 

\vspace{2mm}

\noindent $\bullet$ \emph{Generalized Kruskal coordinates.} Generalized Kruskal coordinates $(y^{+},y^{-})$ are related to Kruskal and Eddington-Finkelstein coordinates via 
\beq
\label{ap:genkrus}
\lambda x^{\pm}= \pm e^{-k} (\pm \lambda y^{\pm})^{\alpha}\, ,\qquad \text{and} \qquad 
\lambda\sigma^{\pm}=\pm \alpha \ln(\pm \lambda y^{\pm})\mp k\, .
\eeq 
where $\alpha$ and $k$ are constants, and where we have assumed $\alpha \geq 1$ and $y^+\geq0$, $y^-\leq0$. The classical geometry reads
\beq 
ds^{2}=-e^{2\rho}dy^{+}dy^{-}\;,  \eeq
with
\beq
\rho=\phi -k +\ln(\alpha)-\frac{(1-\alpha)}{2}\ln(-\lambda^2 y^+y^-)\, ,
\eeq
and where $\phi$ in generalized Kruskal coordinates results in 
\beq
\phi= -\frac{1}{2}\ln\left(\frac{M}{\lambda}+e^{-2k}(-\lambda^2 y^+y^-)^\alpha\right)\, .
\eeq
\vspace{2mm}

\noindent $\bullet$ \emph{Schwarzschild coordinates.} Introducing radial coordinate 
\beq r=\frac{1}{2\lambda}\ln\left(e^{2\lambda\sigma^{1}}+\frac{M}{\lambda}\right)\;,\eeq
the black hole geometry takes a Schwarzschild form
\beq ds^{2}=-f(r)(d\sigma^{0})^{2}+f^{-1}(r)dr^{2}\;,\quad f(r)=\left(1-\frac{M}{\lambda}e^{-2\lambda r}\right)\;,\label{eq:schwargaugeapp}\eeq
while the dilaton is 
\beq e^{-2\phi}=e^{2\lambda r}\;.\eeq
In these coordinates the horizon is located at $r_{H}=\frac{1}{2\lambda}\ln(M/\lambda)$, the positive root of $f(r_{H})=0$. Clearly, for $M<0$, the horizon disappears leaving behind a naked singularity (located at $r_{\text{S}}\to-\infty$, where $\phi\to+\infty$, the region of strong coupling). Asymptotically, $r\to\infty$, the black hole approaches the linear dilaton vacuum, where $\phi\to-\infty$ (weak coupling). 

\vspace{2mm}

\noindent $\bullet$ \emph{Rindler-like coordinates.} Using $x^{\pm}=(x^{0}\pm x^{1})$ for time and space coordinates $\{x^{0},x^{1}\}$, the eternal black hole can be brought to Rindler form (up to the conformal factor)
\beq ds^{2}=e^{2\rho}\left[-(\lambda X)^{2}dT^{2}+dX^{2}\right]\;,\quad e^{-2\rho}=e^{-2\phi}=\frac{M}{\lambda}+\lambda^{2}X^{2}\;,\label{eq:rindlercoordapp}\eeq
via the coordinate transformation $x^{0}=X\sinh(\lambda T)$ and $x^{1}=X\cosh(\lambda T)$. For $M>0$, the horizon is now located at $X=0$, and the spacetime possesses a timelike Killing vector $\xi^{\mu}_{(T)}=\partial^{\mu}_{T}$, i.e., translations in Rindler-time $T$. 

\section{Equations of motion and gauge conditions} \label{app:eoms}

\noindent Here we set additional conventions and briefly derive the Lorentzian equations of motion of the CGHS model and its semi-classical extensions. We then describe the relevant actions in conformal gauge.

\subsection{Equations of motion}

\noindent In this work we focus on two-dimensional gravitational actions of the type
\beq I=L_{0}\int_{\mathcal{M}} d^{2}x\sqrt{-g}\left[W(\Phi)R+U(\Phi)(\nabla\Phi)^{2}-V(\Phi)\right]\;,\label{eq:genactapp}\eeq
for coupling constant $L_{0}$, metric $g_{\mu\nu}$, and scalar field $\Phi$. Specifically, we are interested in the following models:

\vspace{3mm}

\noindent \emph{CGHS} (\ref{eq:CGHSLor}): $\quad$ $L_{0}=\frac{1}{2\pi}$, $W(\phi)=e^{-2\phi}$, $U(\phi)=4e^{-2\phi}$, and $V(\phi)=-4\lambda^{2}e^{-2\phi}$\;,

\vspace{3mm}

\noindent \emph{Polyakov} (\ref{eq:Polyactv2}):  $L_{0}=-\frac{N}{48\pi}$, $W(Z)=-Z$, $U(Z)=\frac{1}{2}$, and $V(Z)=0$,

\vspace{3mm}

\noindent \emph{RST/BPP} (\ref{eq:IbRSTBPPact}): $L_{0}=\frac{N}{24\pi}$, $W(\phi)=(a-1)\phi$, $U(\phi)=(1-2a)$, and $V(\phi)=0$.

\vspace{3mm}

Varying the action (\ref{eq:genactapp}) with respect to the metric yields\footnote{Recall the identity $\delta_{g}(W(\Phi)R)=[WR_{\mu\nu}+(g_{\mu\nu}\Box-\nabla_{\mu}\nabla_{\nu})W]\delta g^{\mu\nu}$, and that in two-dimensions the Einstein tensor $G_{\mu\nu}=0$.}
\beq 
\begin{split}
\frac{1}{\sqrt{-g}}\frac{\delta I}{\delta g^{\mu\nu}}=L_{0}\left[(g_{\mu\nu}\Box-\nabla_{\mu}\nabla_{\nu})W(\Phi)+U(\Phi)\left((\nabla_{\mu}\Phi)(\nabla_{\nu}\Phi)-\frac{1}{2}(\nabla\Phi)^{2}g_{\mu\nu}\right)+\frac{1}{2}V(\Phi)g_{\mu\nu}\right]\;,
\end{split}
\label{eq:gravEOMgenapp}\eeq
while a variation with respect to the scalar field $\Phi$ gives
\beq \frac{1}{\sqrt{-g}}\frac{\delta I}{\delta \Phi}=L_{0}\left[RW'(\Phi)-V'(\Phi)-2U(\Phi)\Box\Phi-U'(\Phi)(\nabla\Phi)^{2}\right]\;.\label{eq:dilaEOMgenapp}\eeq
Here we ignore total derivative terms that arise from performing integration by parts.

In conformal gauge, $ds^{2}=e^{2\rho}dw^{+}dw^{-}$, the general action (\ref{eq:genactapp}) is 
\beq
\begin{split}
 I&=\frac{1}{2}L_{0}\int_{\mathcal{M}}\hspace{-2mm}dw^{+}dw^{-}\left[8W\partial_{+}\partial_{-}\rho-4U(\partial_{+}\Phi)(\partial_{-}\Phi)-e^{2\rho}V\right]\\
&=\frac{1}{2}L_{0}\int_{\mathcal{M}}\hspace{-2mm}dw^{+}dw^{-}\left[-8W'(\partial_{+}\Phi)(\partial_{-}\rho)-4U(\partial_{+}\Phi)(\partial_{-}\Phi)-e^{2\rho}V\right]\;,
\end{split}
\eeq
where we used the geometric identities (\ref{eq:usegoeidconf}), and to arrive to the second line we performed an integration by parts on the first term (dropping a total derivative). The $\rho$ variation is easily found to be
\beq 0=\frac{\delta I}{\delta \rho}=4L_{0}\left(W''(\partial_{-}\Phi)(\partial_{+}\Phi)+W'\partial_{-}\partial_{+}\Phi-\frac{1}{4}e^{2\rho}V\right)\;,\eeq
while the scalar field equation of motion (\ref{eq:dilaEOMgenapp}) in conformal gauge is 
\beq 0=4L_{0}e^{-2\rho}\left[W'\partial_{+}\partial_{-}\rho-e^{2\rho}V'+U\partial_{+}\partial_{-}\Phi+\frac{1}{2}U' (\partial_{+}\Phi)(\partial_{-}\Phi)\right]\;.\eeq

\vspace{2mm}

Let us now particularize to the models of interest.

\vspace{2mm}

\subsubsection{Classical CGHS}

For the classical CGHS action we have the metric variation is\footnote{Here we used the following identity
$$(g_{\mu\nu}\Box-\nabla_{\mu}\nabla_{\nu})e^{-2\phi}=2e^{-2\phi}(\nabla_{\mu}\nabla_{\nu}\phi-g_{\mu\nu}\Box\phi)+4e^{-2\phi}(g_{\mu\nu}(\nabla\phi)^{2}-(\nabla_{\mu}\phi)(\nabla_{\nu}\phi))\;.$$}
\beq \frac{1}{\sqrt{-g}}\frac{\delta I_{\text{CGHS}}}{\delta g^{\mu\nu}}=\frac{4e^{-2\phi}}{4\pi}\left[\nabla_{\mu}\nabla_{\nu}\phi-g_{\mu\nu}\Box\phi+g_{\mu\nu}(\nabla\phi)^{2}-g_{\mu\nu}\lambda^{2}\right]\;.\label{eq:metvarCGHSclassapp}\eeq
Additionally, there is the CGHS classical matter action
\beq I_{\text{mat}}=-\frac{1}{4\pi}\int_{\mathcal{M}} d^{2}x\sqrt{-g}\sum_{i=1}^{N}(\nabla f_{i})^{2}\;.\label{eq:matactCGHSapp}\eeq
Varying with respect to the metric gives 
\beq \frac{1}{\sqrt{-g}}\frac{\delta I_{\text{mat}}}{\delta g^{\mu\nu}}=-\frac{1}{4\pi}\sum_{i=1}^{N}\left[(\nabla_{\mu}f_{i})(\nabla_{\nu}f_{i})-\frac{1}{2}g_{\mu\nu}(\nabla f_{i})^{2}\right]\;.\eeq
The metric equations of motion for the CGHS action coupled to matter, $\frac{1}{\sqrt{-g}}\delta_{g}(I_{\text{CGHS}}+I_{\text{mat}})=0$, recovers Eq. (\ref{eq:CGHSeoms}) with classical matter stress-tensor defined as 
\beq T_{\mu\nu}^{\text{mat}}\equiv-\frac{4\pi}{\sqrt{-g}}\frac{\delta I_{\text{mat}}}{\delta g^{\mu\nu}}\;.\eeq
Meanwhile, the dilaton equations of motion are
\beq 0=\frac{1}{\sqrt{-g}}\frac{\delta I_{\text{CGHS}}}{\delta \phi}=-\frac{2e^{-2\phi}}{2\pi}\left[R+4\lambda^{2}+4\Box\phi-4(\nabla\phi)^{2}\right]\;,\eeq
as in (\ref{eq:CGHSeoms}).

The CGHS action coupled to conformal matter in conformal gauge becomes
\beq I=\frac{1}{4\pi}\int_{\mathcal{M}}\hspace{-2mm} dw^{+}dw^{-}\biggr[8e^{-2\phi}\partial_{+}\partial_{-}\rho-16e^{-2\phi}(\partial_{+}\phi)(\partial_{-}\phi)+4\lambda^{2}e^{2(\rho-\phi)}+2\sum_{i=1}^{N}(\partial_{+}f_{i})(\partial_{-}f_{i})\biggr]\;.\eeq
From which the $\rho$ variation of the CGHS action is
\beq \frac{\delta I_{\text{CGHS}}}{\delta \rho}=\frac{2}{\pi}\left(4e^{-2\phi}(\partial_{-}\phi)(\partial_{+}\phi)-2e^{-2\phi}\partial_{-}\partial_{+}\phi+\lambda^{2}e^{2(\rho-\phi)}\right)\;,\label{eq:rhovarCGHSapp}\eeq
and the dilaton variation gives (\ref{eq:dilamatEOMconf}).

\subsubsection{Semi-classical CGHS}

The semi-classical CGHS model is characterized by the total action 
\beq I=I_{\text{CGHS}}+I_{\text{Poly}}\;.\eeq
The metric variation of the localized Polyakov action gives
\beq \frac{1}{\sqrt{-g}}\frac{\delta I_{\text{Poly}}}{\delta g^{\mu\nu}}=-\frac{N}{48\pi}\left[\nabla_{\mu}\nabla_{\nu}Z-g_{\mu\nu}\Box Z+\frac{1}{2}\left((\nabla_{\mu}Z)(\nabla_{\nu}Z)-\frac{1}{2}(\nabla Z)^{2}g_{\mu\nu}\right)\right]\;,\eeq
while the auxiliary scalar field equation is
\beq 0=\frac{1}{\sqrt{-g}}\frac{\delta I_{\text{Poly}}}{\delta Z}=\frac{N}{48\pi}(R+\Box Z)\;.\eeq
It is standard to define the quantum stress-tensor as
\beq \langle T_{\mu\nu}\rangle\equiv-\frac{2\pi}{\sqrt{-g}}\frac{\delta I_{\text{Poly}}}{\delta g^{\mu\nu}}=\frac{N}{24}\left[\nabla_{\mu}\nabla_{\nu}Z-g_{\mu\nu}\Box Z+\frac{1}{2}\left((\nabla_{\mu}Z)(\nabla_{\nu}Z)-\frac{1}{2}(\nabla Z)^{2}g_{\mu\nu}\right)\right]\;.\label{eq:qstresstenappcov}\eeq
Taking the trace yields the 1-loop conformal anomaly,
\beq g^{\mu\nu}\langle T_{\mu\nu}\rangle=-\frac{N}{24}\Box Z=\frac{N}{24}R\;,\eeq
where the second equality follows from the $Z$ equation of motion. The semi-classical CGHS field equations are easily found from 
\beq \frac{2\pi}{\sqrt{-g}}\frac{\delta I_{\text{CGHS}}}{\delta g^{\mu\nu}}=\langle T_{\mu\nu}\rangle\;,\label{eq:qstresstencovapp}\eeq
recovering (\ref{eq:semiCGHSmetEOM}), upon substituting in the metric variation of the CGHS action (\ref{eq:matactCGHSapp}).

In conformal gauge, the Polyakov action reads
\beq I_{\text{Poly}}=\frac{N}{48\pi}\int_{\mathcal{M}}\hspace{-2mm}dw^{+}dw^{-}\left[4Z\partial_{+}\partial_{-}\rho+(\partial_{+}Z)(\partial_{-}Z)\right]\;,\label{eq:PolyactZapp}\eeq
such that the $\rho$ variation of the Polyakov action is 
\beq \frac{\delta I_{\text{Poly}}}{\delta \rho}=\frac{N}{12\pi}(\partial_{-}\partial_{+}Z)\;.\eeq
Thus, together with (\ref{eq:rhovarCGHSapp}), the $\rho$ variation for semi-classical CGHS gives
\beq 0=\frac{2}{\pi}\left(4e^{-2\phi}(\partial_{-}\phi)(\partial_{+}\phi)-2e^{-2\phi}\partial_{-}\partial_{+}\phi+\lambda^{2}e^{2(\rho-\phi)}\right)+\frac{N}{12\pi}(\partial_{-}\partial_{+}Z)\;.\eeq
Using $Z=2\rho-2\xi$ for $\partial_{+}\partial_{-}\xi=0$, the semi-classical equations of motion (\ref{eq:rhoeomwoutRST}) are recovered. Meanwhile the variation with respect to the auxiliary scalar field $Z$ is
\beq \frac{\delta I_{\text{Poly}}}{\delta Z}=-\frac{N}{12\pi} e^{-2\rho}\left[-\partial_{+}\partial_{-}\rho+\frac{1}{2}\partial_{+}\partial_{-} Z\right]\;,\eeq
from which it is easy to confirm $Z=2\rho-2\xi$.\footnote{Integrating by parts the Polyakov action (\ref{eq:PolyactZapp}) and substituting in $Z=2\rho-2\xi$ returns (\ref{eq:PolylocactCG}).}

Further, the quantum stress-tensor (\ref{eq:qstresstencovapp}) in conformal gauge is
\beq \langle T_{\mu\nu}\rangle=\frac{N}{24}\left[\nabla_{\mu}\nabla_{\nu}Z+\frac{1}{2}(\nabla_{\mu}Z)(\nabla_{\nu}Z)+4g_{\mu\nu}e^{-2\rho}(\partial_{+}\partial_{-}Z)+e^{-2\rho}g_{\mu\nu}(\partial_{+}Z)(\partial_{-}Z)\right]\;,\eeq
with components (recall the final line of identities (\ref{eq:usegoeidconf}))
\beq 
\begin{split}
&\langle T_{\pm\mp}\rangle=-\frac{N}{24}\partial_{\pm}\partial_{\mp}Z\;,\\
&\langle T_{\pm\pm}\rangle=\frac{N}{24}\left[\partial^{2}_{\pm}Z-2(\partial_{\pm}\rho)(\partial_{\pm}Z)+\frac{1}{2}(\partial_{\pm}Z)^{2}\right]\;.
\end{split}
\eeq
Setting $Z=2\rho-2\xi$, the components (\ref{eq:Tpmanom}) --- (\ref{eq:Tmmappv2}) follow, for functions $t_{\pm}(w^{\pm})$ in (\ref{eq:tpmcongen}).

\subsubsection{RST and BPP}

The solvable semi-classical model which interpolates between RST and BPP gravity has the total action
\beq I=I_{\text{CGHS}}+I_{\text{Poly}}+I_{a}\;.\label{eq:fullactRSTBPPapp}\eeq
The metric variation of the generalized RST ($a=1/2$), BPP ($a=0$) term, $I_{a}$, is 
\beq \frac{1}{\sqrt{-g}}\frac{\delta I_{a}}{\delta g^{\mu\nu}}=\frac{N}{24\pi}\left[(a-1)(g_{\mu\nu}\Box\phi-\nabla_{\mu}\nabla_{\nu}\phi)+(1-2a)\left((\nabla_{\mu}\phi)(\nabla_{\nu}\phi)-\frac{1}{2}(\nabla\phi)^{2}g_{\mu\nu}\right)\right]\;.\eeq
The semi-classical metric equations (\ref{eq:meteomsfullsemi}) follow from 
\beq \frac{2\pi}{\sqrt{-g}}\frac{\delta}{\delta g^{\mu\nu}}(I_{\text{CGHS}}+I_{a})=\langle T_{\mu\nu}\rangle\;,\eeq
for quantum-stress tensor (\ref{eq:qstresstenappcov}). 
Varying $I_{a}$ with respect to the dilaton, meanwhile, is 
\beq \frac{1}{\sqrt{-g}}\frac{\delta I_{a}}{\delta \phi}=\frac{N}{24\pi}\left[(a-1)R-2(1-2a)\Box\phi\right]\;.\eeq
Thus, the dilaton equation of motion for the full action (\ref{eq:fullactRSTBPPapp}) is 
\beq 0=\frac{1}{\sqrt{-g}}\frac{\delta I}{\delta \phi}=-\frac{2e^{-2\phi}}{2\pi}\left[R+4\lambda^{2}+4\Box\phi-4(\nabla\phi)^{2}\right]+\frac{N}{24\pi}\left[(a-1)R-2(1-2a)\Box\phi\right]\;.\label{eq:dilaeomfullapp}\eeq

In conformal gauge, the RST/BPP term takes the form
\beq I_{a}=\frac{N}{48\pi}\int_{\mathcal{M}}\hspace{-2mm}dw^{+}dw^{-}\left[8(a-1)\phi \partial_{+}\partial_{-}\rho-4(1-2a)(\partial_{+}\phi)(\partial_{-}\phi)\right]\;.\eeq
The $\rho$ variation gives
\beq \frac{\delta I_{a}}{\delta \rho}=\frac{N}{6\pi}[(a-1)\partial_{-}\partial_{+}\phi]\;.\eeq
Consequently, the $\rho$ equations of the full action (\ref{eq:dilaeomfullapp}) are\footnote{We use $\partial_{+}\partial_{-}(e^{-2\phi})=4e^{-2\phi}(\partial_{+}\phi)(\partial_{-}\phi)-2e^{-2\phi}\partial_{+}\partial_{-}\phi$.} 
\beq 0=\partial_{+}\partial_{-}(e^{-2\phi})+\lambda^{2}e^{2(\rho-\phi)}+\frac{N}{12}\partial_{+}\partial_{-}\left[\rho+(a-1)\phi\right]\;.\label{eq:rhoeomfullconfapp}\eeq
Meanwhile, the dilaton equations for the full action  in conformal gauge gives
\beq 
\begin{split}
e^{-2\phi}\left[2\partial_{+}\partial_{-}\rho+\lambda^{2}e^{2\rho}-4\partial_{+}\partial_{-}\phi+4(\partial_{+}\phi)(\partial_{-}\phi)\right]=\frac{N}{12}\left[(a-1)\partial_{+}\partial_{-}\rho+(1-2a)\partial_{+}\partial_{-}\phi\right]\;,  
\end{split}
\label{eq:dilaeomfullconfapp}\eeq
or, 
\beq 0=2e^{-2\phi}\partial_{+}\partial_{-}(\rho-\phi)+\partial_{+}\partial_{-}(e^{-2\phi})+\lambda^{2}e^{2(\rho-\phi)}-\frac{N}{12}\partial_{+}\partial_{-}[(a-1)\rho+(1-2a)\phi]\;.\eeq
Subtracting the $\rho$ equations (\ref{eq:rhoeomfullconfapp}) from the dilaton equations gives
\beq 0=2e^{-2\phi}\partial_{+}\partial_{-}(\rho-\phi)-\frac{Na}{12}\partial_{+}\partial_{-}(\rho-\phi)\;.\eeq

\subsection{Kruskal gauge: lost and found}

\noindent Let us now briefly comment on the gauge symmetry of the various classical and semi-classical actions considered.
First recall that performing an integration by parts on the $\partial_{+}\partial_{-}\rho$ term (dropping a total derivative) in the classical CGHS action coupled to matter  in conformal gauge gives 
\beq 
\begin{split}
I&
=\frac{1}{\pi}\int_{\mathcal{M}} dw^{+}dw^{-}\biggr[2\partial_{-}(\phi-\rho)(\partial_{+}e^{-2\phi})+\lambda^{2}e^{2(\rho-\phi)}+\frac{1}{2}\sum_{i=1}^{N}(\partial_{+}f_{i})(\partial_{-}f_{i})\biggr]\;.
\end{split}
\label{eq:CGHSrewriteSTapp}\eeq
Written in this way, the CGHS action has the residual gauge symmetry
\beq \delta\phi=\delta\rho=\epsilon e^{2\phi}\;,\label{eq:CGHStrans}\eeq
for $\epsilon$ infinitesimal, and associated conserved current $j^{\mu}=\partial^{\mu}(\phi-\rho)$, obeying $\partial_{\mu}\partial^{\mu}(\phi-\rho)=0$. One may further gauge fix such that $\rho=\phi$, namely, the Kruskal gauge.

Next, recall the total semi-classical action (\ref{eq:fullactRSTBPPapp}) in conformal gauge is 
\beq 
\begin{split}
I&=\frac{1}{\pi}\int_{\mathcal{M}}dw^{+}dw^{-}\biggr[\left(2e^{-2\phi}+\frac{(a-1)N}{6}\phi\right)\partial_{+}\partial_{-}\rho-\left(4e^{-2\phi}+\frac{N}{12}(1-2a)\right)(\partial_{+}\phi)(\partial_{-}\phi)\\
&-\frac{N}{12}(\partial_{+}\rho)(\partial_{-}\rho)+\lambda^{2}e^{2(\rho-\phi)}+\frac{1}{2}\sum_{i=1}^{N}(\partial_{+}f_{i})(\partial_{-}f_{i})\biggr]\;.
\end{split}
\eeq
Integrating the $\partial_{+}\partial_{-}\rho$ term by parts, dropping a total derivative, and rearranging terms gives
\beq 
\begin{split}
I&=\frac{1}{\pi}\int_{\mathcal{M}}dw^{+}dw^{-}\biggr[(\partial_{-}\phi)\partial_{+}\left(2e^{-2\phi}+\frac{N}{12}(2a-1)\phi\right)-(\partial_{-}\rho)\partial_{+}\left(2e^{-2\phi}+\frac{N}{6}(a-1)\phi+\frac{N}{12}\rho\right)\\
&+\lambda^{2}e^{2(\rho-\phi)}+\frac{1}{2}\sum_{i=1}^{N}(\partial_{+}f_{i})(\partial_{-}f_{i})\biggr]\;.
\end{split}
\eeq
We can further massage the first line of the integrand such that the total action becomes\footnote{This follows from adding $\frac{N}{12}\rho-\frac{N}{12}\rho$ in the first parenthetic term and $\frac{N}{12}\phi-\frac{N}{12}\phi$ in the second parenthetic term, combining like terms, and performing an integration by parts twice (neglecting total derivatives).}
\beq 
\begin{split}
I&=\frac{1}{\pi}\int_{\mathcal{M}}dw^{+}dw^{-}\biggr[2\partial_{-}(\phi-\rho)\partial_{+}\left(e^{-2\phi}-\frac{N}{24}(\phi-\rho)+\frac{Na}{12}\phi\right)+\lambda^{2}e^{2(\rho-\phi)}+\frac{1}{2}\sum_{i=1}^{N}(\partial_{+}f_{i})(\partial_{-}f_{i})\biggr]\;.
\end{split}
\label{eq:BPPRSTactionrewriteapp}\eeq
Notice that for the BPP model ($a=0$), the action is again invariant under the same transformation (\ref{eq:CGHStrans}) with the same conserved current as in the classical CGHS model --- a simplification relative to the RST model. Consequently, the BPP equations of motion in the Kruskal gauge, $\rho=\phi$, are precisely the same as the classical CGHS equations, while the constraints are modified by terms due to the Polyakov action \cite{Bose:1995pz}. 

We can further rewrite the total action such that it takes the form of two-dimensional Liouville theory coupled to matter. To see this, expand the first term of the integrand,
\beq 2\partial_{-}(\phi-\rho)\partial_{+}\left(e^{-2\phi}-\frac{N}{24}(\phi-\rho)+\frac{Na}{12}\phi\right)=2\partial_{-}(\phi-\rho)\partial_{+}\left(e^{-2\phi}+\frac{Na}{12}\phi\right)-\frac{N}{12}\partial_{-}(\phi-\rho)\partial_{+}(\phi-\rho)\;.\eeq
Next, introduce two scalar fields $\Omega$ and $\chi$ that obey $(\Omega-\chi)=\sqrt{\frac{N}{12}}(\phi-\rho)$. Consequently, it follows that if 
\beq \Omega\equiv \sqrt{\frac{12}{N}}e^{-2\phi}+\sqrt{\frac{N}{12}}a\phi\;,\eeq
then the action (\ref{eq:BPPRSTactionrewriteapp}) has the form\footnote{By rescaling the Liouville fields $\chi\to\sqrt{\frac{N}{12}}\chi$ and $\Omega\to\sqrt{\frac{N}{12}}\Omega$, and further shifting the rescaled quantities as
$$\chi\to \chi-\frac{1}{4}\ln\left(\frac{N}{3}\right)\;,\quad \Omega\to\Omega+\frac{1}{4}\ln\left(\frac{N}{48}\right)\;,$$
then the action (\ref{eq:BPPRSTactLiouvapp}) takes the form as in Eq. (3.62) of \cite{Strominger:1994tn}. }
\beq I=\frac{1}{\pi}\int_{\mathcal{M}}dw^{+}dw^{-}\left[(\partial_{-}\Omega)(\partial_{+}\Omega)-(\partial_{-}\chi)(\partial_{+}\chi)+\lambda^{2}e^{2\sqrt{\frac{12}{N}}(\chi-\Omega)}+\frac{1}{2}\sum_{i=1}^{N}(\partial_{+}f_{i})(\partial_{-}f_{i})\right]\;,\label{eq:BPPRSTactLiouvapp}\eeq
with
\beq \chi\equiv \Omega-\sqrt{\frac{N}{12}}(\phi-\rho)=\sqrt{\frac{12}{N}}e^{-2\phi}+\sqrt{\frac{N}{12}}(a-1)\phi+\sqrt{\frac{N}{12}}\rho\;.\eeq
From here it is easy to see the interpolating model has restored the Kruskal gauge symmetry, $\Omega=\chi$, such that $(\Omega-\chi)$, or, equivalently, $(\rho-\phi)$, may be treated as a free field. Without the $I_{a}$ term, $(\rho-\phi)$ cannot be treated as a free field in the semi-classical CGHS model.

\section{Covariant phase space and conserved charges}\label{app:Waldformalism}

\noindent Here we apply the covariant phase space formalism \cite{Crnkovic:1986ex,Lee:1990nz,Wald:1993nt,Iyer:1994ys} to derive (quasi-local) conserved energy for CGHS gravity and its semi-classical generalizations. Our presentation and notation follows Appendix C of \cite{Pedraza:2021cvx}. For previous applications of the covariant phase formalism for specific two-dimensional dilaton-gravity models, namely, the classical CGHS and JT models, see, \emph{e.g.}, \cite{Navarro-Salas:1992bwd,Navarro-Salas:1994cyz,Iyer:1994ys,Kummer:1995qv,Harlow:2019yfa,Svesko:2022txo}.

\subsection{Covariant phase space formalism}

\noindent Covariant phase space formalism provides a  way to understand the Hamiltonian dynamics of covariant Lagrangian field theories. In this context, phase space $\mathcal{P}$ is simply defined as the set of solutions to the equations of motion. Geometrically, phase space is a described by a symplectic manifold equipped with a closed and non-degenerate 2-form $\Omega$. The symplectic 2-form $\Omega$ can be directly constructed using Lagrangian mechanics.

\vspace{2mm}

\subsubsection{Lagrangian mechanics} 

\noindent Let $\psi = (g_{\mu\nu}, \Phi)$ denote a collection of dynamical fields, where $g_{\mu\nu}$ is an arbitrary background metric of a $(1+1)$-dimensional Lorenztian spacetime $M$ and $\Phi$ represents any scalar field on~$M$. For generic 2D dilaton theories characterized in Appendix \ref{app:eoms}, the covariant Lagrangian 2-form is
\beq L=L_{0}\epsilon\left[W(\Phi)R+U(\Phi)(\nabla\Phi)^{2}-V(\Phi)\right]\;,\label{eq:genericdilatonlagr}\eeq
where $\epsilon=d^{2}x\sqrt{-g}$ is the spacetime volume 2-form. A total variation of the  Lagrangian \eqref{eq:genericdilatonlagr} yields
\begin{equation} \label{eq:applagrangianvar}
	\delta L = \epsilon E_\Phi \delta \Phi + \epsilon E_{\mu \nu} \delta g^{\mu \nu} + d \theta (  \psi, \delta \psi)\;.
\end{equation}
Here $d$ denotes the spacetime exterior derivative, $\delta\psi$  the representative of field variations, and  $E_{\mu\nu}$ and $E_{\Phi}$ are the metric and $\Phi$ equations of motion, (\ref{eq:gravEOMgenapp}) and (\ref{eq:dilaEOMgenapp}), respectively. The (pre-) symplectic potential\footnote{Technically, we work with the  configuration space or pre-phase space (the set of off-shell field configurations on $\mathcal{M}$ obeying boundary conditions when $\mathcal{M}$ has a boundary) such that $\theta$, $\omega$, and $\Omega$ are the pre-symplectic potential, current, and 2-form, respectively. Physical phase space $\mathcal{P}$ is subsequently given by the quotient of pre-phase space under the action of the group of continuous transformations whose generators are the zero modes of $\Omega$ \cite{Lee:1990nz}.} (spacetime) 1-form $\theta$ for the theory  (\ref{eq:genericdilatonlagr}) is generally given by   
\beq \theta=\epsilon_{\mu}\left[2P^{\mu\alpha\beta\nu}\nabla_{\nu}\delta g_{\alpha\beta}-2(\nabla_{\nu}P^{\mu\alpha\beta\nu})\delta g_{\alpha\beta}+2L_0 U(\Phi)(\nabla^{\mu}\Phi)\delta\Phi\right]\;,\eeq
where $P^{\mu\alpha\beta\nu}\equiv\frac{\partial L}{\partial R_{\mu\alpha\beta\nu}}$ such that for $L=\epsilon W(\Phi)R$, one has $P^{\mu\alpha\beta\nu}=\frac{W}{2}(g^{\mu\beta}g^{\alpha\nu}-g^{\alpha\beta}g^{\mu\nu})\epsilon$. 

From the potential $\theta$, the (pre-) symplectic current (spacetime) 1-form is generically defined  as the antisymmetrization
\beq \omega (\psi, \delta_1 \psi, \delta_2 \psi) \equiv \delta_1 \theta (\psi, \delta_2 \psi) - \delta_2\theta (\psi, \delta_1\psi)\;.\eeq
Its explicit expression will not be needed for our purposes (see Appendix C of \cite{Pedraza:2021cvx}). Note that the infinitesimal variation $\delta$ is the exterior derivative acting on differential forms on the configuration space such that the potential $\theta$ and current $\omega$ are, respectively, 1- and 2-forms on the configuration space. Finally, let $\Sigma$ be any Cauchy slice of $M$. The (pre-) symplectic (configuration space) 2-form is defined to be
\beq \Omega(\psi, \delta_1 \psi, \delta_2 \psi) \equiv \int_{\Sigma}\omega(\psi, \delta_1 \psi, \delta_2 \psi)\;.\label{eq:symp2formgenapp}\eeq

 Next, consider the case when $\zeta$ is an arbitrary smooth vector field on $M$; the infinitesimal generator of a diffeomorphism. The associated Noether current 1-form is defined as $j_{\zeta}\equiv \theta (\psi, \mathcal{L}_{\zeta} \psi) - \zeta\cdot L$, with $\mathcal{L}_{\zeta}$ being the Lie derivative along $\zeta$. 
For on-shell field configurations, $dj_{\zeta}=0$, such that the Noether current is equal to the exterior derivative of the
 Noether charge $0$-form $Q_{\zeta}$, defined via $j_\zeta \equiv d Q_\zeta $ (on-shell). For the theory (\ref{eq:genericdilatonlagr}) we have \cite{Pedraza:2021cvx}
\begin{equation}
	Q_\zeta = -L_{0}\epsilon_{\mu \nu} \left [W(\Phi) \nabla^\mu \zeta^\nu + 2 \zeta^\mu \nabla^\nu W(\Phi) \right]\;,
\label{eq:Noethercharge}\end{equation}
where $\epsilon_{\mu\nu}$ is the volume form of a codimension-0 surface $\partial\Sigma$ embedded in a Cauchy slice $\sigma$.  A standard exercise then gives the following on-shell fundamental variational identity  
\begin{equation}
	\omega (\psi, \delta \psi, \mathcal L_\zeta \psi) =d \left[ \delta Q_\zeta - \zeta \cdot \theta (\psi, \delta \psi) \right]\;.
\label{eq:variationalid}\end{equation}
With this form of the symplectic current one can characterize the Hamiltonian of the system. We will return to this momentarily.

\subsubsection{Spacetimes with a boundary} 

\noindent  For spacetimes $M$ with boundary $\partial M$, the action  is supplemented by the integral of (minus)~$b$ at $\partial M$,
\beq I_{\text{tot}} = \int_M L - \int_{\partial M} b\;, \label{eq:totalaction1}\eeq
a Gibbons-Hawking-York (GHY) boundary term to ensure the variational problem is well-posed.\footnote{Past and future boundaries $\Sigma_{\pm}$ are neglected. Thus, the action is stationary under arbitrary variations of the dynamical fields up to  terms at the future and past boundary of $M$. 
In order for the variational principle to be well posed  we require \emph{Dirichlet} boundary conditions on the induced metric and $\Phi$ only on $B$. For alternative boundary conditions, e.g., induced conformal boundary conditions \cite{Galante:2025tnt} (see also Appendix E of \cite{Banihashemi:2025qqi}), the 1-form $b$ differs but the general analysis remains the same.} In particular, 
\beq
b =-2L_{0}\epsilon_{\partial  M} W(\Phi)  K\;,\label{eq:bformgenapp}\eeq
where $\epsilon_{\partial M}$ is the volume form on $\partial M$. We decompose the boundary as $\partial M=B\cup\Sigma_{-}\cup\Sigma_{+}$, where $\Sigma_{-}$ and $\Sigma_{+}$ are past and future boundaries, respectively, and $B$ is the timelike boundary with induced metric $\gamma_{\mu \nu} = - n_\mu n_\nu + g_{\mu \nu}$ for (outward pointing) unit normal $n_\mu$ to $B$. Further, $K$ is the trace of the extrinsic curvature, $K_{\mu \nu} = \frac{1}{2}\mathcal L_n \gamma_{\mu \nu}$.

The pullback of the potential $\theta$ to the boundary  can be written as \cite{Pedraza:2021cvx} 
\begin{equation}
\theta\big|_{\partial M}=\big  [p(\Phi) \delta \Phi + \frac{1}{2}\tau^{\mu \nu} \delta g_{\mu \nu} \big]  \epsilon_{\partial  M}+ \delta b  + d C\;,  \label{eq:pullbackoftheta}
\end{equation}
with
\begin{equation}
p(\Phi)\equiv 2L_{0}[W'(\Phi)K+ U(\Phi) n^\mu  \nabla_\mu \Phi], \quad C = c \cdot \epsilon_{\partial M}\;,\quad  c^\mu=-L_{0}W(\Phi)  \gamma^{\mu \lambda}n^\nu \delta g_{\lambda \nu}\;,
\label{eq:Cdefn}\end{equation}
and Brown-York stress tensor \cite{Brown:1992br}
\beq \tau^{\mu \nu}\equiv\frac{2}{\sqrt{-\gamma}}\frac{\delta I}{\delta \gamma_{\mu\nu}}=2L_{0}W'(\Phi)\gamma^{\mu \nu} n^\alpha \nabla_\alpha \Phi\;.
\label{eq:BYstresstensandP}\end{equation}
Here, $\epsilon_{\mu}|_{\partial M}=n_{\mu}\epsilon_{\partial M}$, and $C$ is a local $0$-form on the  boundary $\partial M$, which is  covariant under diffeomorphisms that preserve the location of the (spatial) boundary \cite{Harlow:2019yfa}. 

From the pullback \eqref{eq:pullbackoftheta} it follows that the symplectic potential restricted to $B$ has the form \cite{Harlow:2019yfa}
\beq
\theta \big |_{B} = \delta b + d C\;. \label{eq:importantrestrictiontheta}
\eeq
Consequently, using \eqref{eq:applagrangianvar} and  \eqref{eq:importantrestrictiontheta},  the variation of the total action,
\beq \delta I_{\text{tot}}=\int_{M}E_{\psi}\delta\psi+\int_{\Sigma_{+}-\Sigma_{-}}\hspace{-2mm}(\theta-\delta b-dC)\;, \label{eq:variationtotalactionapp}\eeq
is stationary up to boundary terms at $\Sigma_\pm$. 
Since the combination $\Psi   \equiv  \theta - \delta b- dC$ appears as a boundary term in the variation of the action, it is natural to consider the modified symplectic current
\beq
\tilde \omega (\psi, \delta_1 \psi, \delta_2 \psi) \equiv \delta_1 \Psi (\psi, \delta_2 \psi) - \delta_2 \Psi (\psi, \delta_1 \psi)\;.
\eeq
 By construction, the symplectic current vanishes on the spatial boundary, $\tilde \omega |_B =0$. Further,  the   on-shell variational identity (\ref{eq:variationalid}) for the new  symplectic current becomes
\beq
  \int_\Sigma \tilde \omega (\psi, \delta \psi, \mathcal L_\zeta \psi ) =   \oint_{\partial \Sigma} \left[ \delta Q_\zeta - \zeta \cdot \theta (\psi, \delta \psi)-  \delta C(\psi,\mathcal L_\zeta \psi) + \mathcal L_\zeta C(\psi,\delta \psi) \right]\;, \label{eq:identityharlowsymplecticform}
\eeq
where  $\partial\Sigma$ is the cross-section of $\Sigma$ and $B$. From here we can construct the symplectic 2-form $\tilde{\Omega}$ as in (\ref{eq:symp2formgenapp}). Moving forward we will always be working with the modified current $\tilde{\omega}$ and will therefore drop the $\tilde{.}$ notation.

\subsubsection{Hamiltonian and conserved energy}

\noindent In covariant phase space formalism, Hamilton's equations read
\beq \delta H_{\zeta}=\Omega(\psi,\delta\psi,\mathcal{L}_{\zeta}\psi)\;,\eeq
where $H_{\zeta}$ denotes the Hamiltonian which generates evolution  along the flow of the diffeomorphism generating  vector field $\zeta$.  Imposing  $\theta|_{B}= \delta b+dC $, one can ``integrate'' to find the Hamiltonian $H_{\zeta}$ \cite{Harlow:2019yfa}
\beq H_{\zeta}=\int_{\partial\Sigma}[Q_{\zeta}-\zeta\cdot b- C(\psi,\mathcal{L}_{\zeta}\psi)]\;,\eeq
up to the standard ambiguous constant for any Hamiltonian system, which we promptly set to zero.

For the generic 2D dilaton-gravity model, the Hamiltonian is
\beq
\begin{split}
H_\zeta=\oint_{\partial \Sigma} \epsilon_{\partial \Sigma} \zeta^\mu u^\nu \tau_{\mu \nu}=\oint_{\partial\Sigma}\epsilon_{\partial\Sigma}\mathcal{N}\varepsilon\;, 
\label{eq:Hzetaexp}
\end{split}
\eeq
with $\epsilon_{\mu\nu}|_{\partial\Sigma}=n_{\mu}u_{\nu}-n_{\nu}u_{\mu}$ for (future pointing) timelike unit normal $u^{\mu}$ to $\Sigma$. Here, moreover, $\mathcal{N}=-\zeta^{\mu}u_{\mu}$ is the `lapse', $\tau_{\mu\nu}$  is the   Brown-York stress-energy tensor (\ref{eq:BYstresstensandP}), and $\varepsilon$ is the quasi-local energy density,
\beq \varepsilon \equiv u_\mu u_\nu \tau^{\mu \nu} =-2L_{0}W'(\Phi) n^\alpha \nabla_\alpha \Phi\;.\label{eq:quasilocalen}\eeq
Note that in two-dimensions the quasi-local momentum and spatial stress vanish.  

\subsection{Quasi-local and asymptotic energies with semi-classical corrections}
\label{app:energies}

\noindent Following \cite{Brown:1992br}, 
the conserved quasi-local energy is defined as in (\ref{eq:quasilocalen}), while the asymptotic energy is equal to the Hamiltonian evaluated at spatial infinity. Let us apply this formalism for the 2D models of interest.

\subsubsection{Classical CGHS}

\noindent For classical CGHS, the Gibbons-Hawking-York boundary 1-form (\ref{eq:bformgenapp}) on the timelike boundary $B$ is
\beq b_{\text{CGHS}}=-\frac{\epsilon_{B}}{\pi}e^{-2\phi}K\;.\label{eq:bGHYCGHS}\eeq
The Brown-York stress-tensor (\ref{eq:BYstresstensandP}) and energy density (\ref{eq:quasilocalen}) are 
\beq \tau^{\mu\nu}_{\text{CGHS}}=-\frac{2}{\pi}e^{-2\phi}\gamma^{\mu\nu}n^{\alpha}\partial_{\alpha}\phi\;,\quad \varepsilon_{\text{CGHS}}=\frac{2}{\pi}e^{-2\phi}n^{\alpha}\partial_{\alpha}\phi\;,\label{eq:BYstressCGHSapp}\eeq
such that the quasi-local Hamiltonian (\ref{eq:Hzetaexp}) is 
\beq H^{\text{CGHS}}_{\zeta}=-\frac{2}{\pi}\oint_{\partial\Sigma}\epsilon_{\partial\Sigma}e^{-2\phi}g_{\mu\nu}\zeta^{\mu}u^{\nu}(n^{\alpha}\partial_{\alpha}\phi)\;.\label{eq:quasiHamCGHSapp}\eeq

As noted in the main text, we introduce the following local counterterm 1-form Lagrangian
\beq b_{\text{CGHS}}^{\text{ct}}=\frac{\epsilon_{B}}{\pi}e^{-2\phi}(2\lambda)\;,\eeq
such that the Brown-York stress-tensor and energy density are modified to (\ref{eq:BYtensCGHS}), using that one simply adds $-\gamma^{\mu\nu}\frac{2\lambda}{\pi}e^{-2\phi}$ to the stress-tensor. 

\subsubsection{Semi-classical CGHS}

\noindent The  Gibbons-Hawking-York boundary 1-form (\ref{eq:bformgenapp}) associated with the Polyakov action is 
\beq b_{\text{Poly}}=-\frac{N}{24\pi}\epsilon_{B}ZK\;.\label{eq:bGHYpoly}\eeq
The contribution to the Brown-York stress tensor
(\ref{eq:BYstresstensandP}) and energy density (\ref{eq:quasilocalen}) will be
\beq \tau^{\mu\nu}_{\text{Poly}}=\frac{N}{24\pi}\gamma^{\mu\nu}n^{\alpha}\nabla_{\alpha}Z\;,\quad \varepsilon_{\text{Poly}}=-\frac{N}{24\pi}n^{\alpha}\nabla_{\alpha}Z\;,\label{eq:eq:BYstressPolyapp}
\eeq
with quasi-local Hamiltonian (\ref{eq:Hzetaexp})
\beq H_{\zeta}^{\text{Poly}}=\frac{N}{24\pi}\oint_{\partial\Sigma}\epsilon_{\partial\Sigma}g_{\mu\nu}\zeta^{\mu}u^{\nu}(n^{\alpha}\partial_{\alpha}Z)\;.\label{eq:quasiHampolyapp}\eeq
The quasi-local Hamiltonian for the semi-classical CGHS model is then simply the sum of Hamiltonians (\ref{eq:quasiHamCGHSapp}) and (\ref{eq:quasiHampolyapp}). Additionally, one needs to include the contribution due to local counterterms. We will report on the form of these counterterms below for the full solvable model, and their derivation in Appendix \ref{ap:HJ}.

\subsubsection{RST and BPP}

\noindent The GHY boundary 1-form (\ref{eq:bformgenapp}) for the RST/BPP action is 
\beq b_{a}=-\frac{N}{12}\epsilon_{B}(a-1)\phi K\;,\eeq
from which we see for $a=1$, there is no boundary term. Consequently, the associated Brown-York stress tensor
(\ref{eq:BYstresstensandP}), energy density (\ref{eq:quasilocalen}), and Hamiltonian (\ref{eq:Hzetaexp}) are
\beq 
\begin{split}
&\tau^{\mu\nu}_{a}=\frac{N}{12\pi}(a-1)\gamma^{\mu\nu}n^{\alpha}\partial_{\alpha}\phi\;,\quad \varepsilon_{a}=-\frac{N}{12\pi}(a-1)n^{\alpha}\partial_{a}\phi\;,\quad H_{\zeta}^{a}=\frac{N}{12\pi}(a-1)\oint_{\partial\Sigma}\epsilon_{\partial\Sigma}g_{\mu\nu}\xi^{\mu}u^{\nu}(n^{\alpha}\partial_{\alpha}\phi)\;.
\end{split}
\label{eq:BYstressRSTBPPapp}\eeq
Hence, the Brown-York stress-tensor and Hamiltonian for the exactly solvable model are 
\beq
\begin{split}
\tau^{\mu\nu}=\tau_{\text{CGHS}}^{\mu\nu}+\tau^{\mu\nu}_{\text{Poly}}+\tau^{\mu\nu}_{a}\;, \quad H_{\zeta}&=H^{\text{CGHS}}_{\zeta}+H^{a}_{\zeta}+H^{\text{Poly}}_{\zeta}\;,
\end{split}
\label{eq:stresstentotalunnormapp}\eeq
yielding (\ref{eq:BYstresstot}) and (\ref{eq:quasiHamfull}) in the main text.

As described in the main text, the above Hamiltonian diverges in the same way as the CGHS model. A natural initial guess to ameliorate the IR divergence is to try the same counterterm used in the CGHS model \eqref{eq:localctCGHS}. With this counterterm we find
\bea
H_{\zeta}&=&\frac{2}{\pi}\oint_{\partial\Sigma}\epsilon_{\partial\Sigma}\mathcal{N}e^{-2\phi}\left(n^{\alpha}\partial_{\alpha}\phi+\lambda\right)-\frac{N}{12\pi}(a-1)\oint_{\partial\Sigma}\epsilon_{\partial\Sigma}\mathcal{N}(n^{\alpha}\partial_{\alpha}\phi)-\frac{N}{24\pi}\oint_{\partial\Sigma}\epsilon_{\partial\Sigma}\mathcal{N}(n^{\alpha}\partial_{\alpha}Z)\nonumber\\
&=&\frac{2}{\pi}\oint_{\partial\Sigma}\epsilon_{\partial\Sigma}e^{-2\phi}\left(\partial_{1}\phi+\lambda e^{\rho}\right)-\frac{N}{12\pi}(a-1)\oint_{\partial\Sigma}\epsilon_{\partial\Sigma}\partial_{1}\phi-\frac{N}{24\pi}\oint_{\partial\Sigma}\epsilon_{\partial\Sigma}\partial_{1}Z\,.
\eea
Considering the choice of constants as before in the HH state, $Z=2\rho+2\lambda\sigma^1-2c^{\pm}$,
\bea
\label{eq:Hxi}
H_{\zeta}&=&\frac{2}{\pi}\oint_{\partial\Sigma}\epsilon_{\partial\Sigma}e^{-2\phi}\left(\partial_{1}\phi+\lambda e^{\rho}\right)-\frac{N}{12\pi}\oint_{\partial\Sigma}\epsilon_{\partial\Sigma}\left(a\partial_{1}\phi+\lambda\right)\nonumber\\
&=&\frac{1}{\pi}\oint_{\partial\Sigma}\epsilon_{\partial\Sigma}\left[\left(2e^{-2\phi}-\frac{Na}{12}\right)\partial_1\phi+2\lambda\left(e^{-\rho+2\lambda\sigma^1}-\frac{N}{12}\right)\right]\,.
\eea
Replacing the solutions for $\phi^{(\text {HH})}(\sigma^1)$ and $\rho^{(\text {HH})}(\sigma^1)$ and 
using the asymptotic expansion of the Lambert functions gives
\be
M_{\text{ADM}}=\lim_{\sigma^{1}\to+\infty}H_{\zeta}\approx\frac{M}{\pi}+\frac{N\lambda}{12\pi}\lim_{\sigma^{1}\to+\infty}\left(a\lambda\sigma^1-2\right)
\label{eq:mfindingapp}\ee
which diverges for $a\neq0$. For $a=0$, we find no divergences, such that the ADM mass in the BPP model goes like $M^{(a=0)}_{\text{ADM}}=\frac{M}{\pi}-\frac{N\lambda}{6\pi}$.

By generalizing the Hamilton-Jacobi method employed in \cite{Grumiller:2007ju}, we precisely determine the local counterterm which removes the IR divergences
(for details, see Appendix \ref{ap:HJ}),
\be
I_{\text{ct}}=-\frac{1}{\pi}\int_{B}dt\sqrt{-\gamma}e^{-2\phi}(2\lambda)+\frac{1}{\pi}\int_{B}dt\sqrt{-\gamma}\phi \left[\frac{N\lambda}{24}\left(1-2a+\mathfrak{c}\right)\right]+\frac{1}{\pi}\int_{B}dt\sqrt{-\gamma}C_0\;,
\ee
for state-dependent constant $\mathfrak{c}$ and arbitrary constant $C_{0}$. For example, for the Hartle-Hawking state, $\mathfrak{c}_{\text{HH}}=-1$, such that the middle contribution of the counterterm vanishes for $a=0$, consistent with the finding (\ref{eq:mfindingapp}). Associated with this boundary term is the boundary 1-form
\beq b_{\text{ct}}=\epsilon_{B}\left(\frac{2\lambda}{\pi}e^{-2\phi}-\phi\left[\frac{N\lambda}{24\pi}(1-2a+\mathfrak{c})\right]+\frac{C_{0}}{\pi}\right)\;.\eeq
This will give an additional contribution to the stress-tensor and quasi-local Hamiltonian (\ref{eq:stresstentotalunnormapp}), specifically (\ref{eq:ctstressandHam}).

\subsection{ADM energy: pseudo-tensor method}\label{ap:adm_ptm}

\noindent For completeness, here we review the standard, pseudo-tensor method for computing the ADM energy, extending \cite{Kim:1995jta} to the interpolating semi-classical dilaton model. 

\subsubsection{Classical CGHS}

\noindent One way to construct the ADM mass is to look at deviations of the metric $g_{\mu\nu}$ and the dilaton $\phi$ at asymptotic spatial infinity $\sigma^{1}\to\infty$ (the linear dilaton vacuum) \cite{Witten:1991yr}, in analogy with Weinberg's method for general relativity \cite{Weinberg:1972kfs}. Working in Eddington-Finkelstein coordinates, for spatial coordinate $\sigma^{1}=\frac{1}{2}(\sigma^{+}-\sigma^{-})$, 
decompose the dilaton and the metric as\footnote{More covariantly, the dilaton may be expanded as $\phi=-\lambda x^{\mu}\eta_{\mu\nu}\epsilon^{\mu}$ for vector $\epsilon^{\mu}$ obeying $\eta_{\mu\nu}\epsilon^{\mu}\epsilon^{\nu}=1$.}
\be \label{eq:decomposition01}
\phi=-\sigma^1 \lambda +\varphi \,, \qquad g_{\mu \nu}=\eta_{\mu \nu}+h_{\mu \nu} \,.
\ee
Here $-\sigma^1 \lambda$ is the value of the dilaton in the linear dilaton vacuum, and $\varphi$ and $h_{\mu\nu}$ denote small deviations away from the dilaton and metric at spatial infinity respectively; both $\varphi$ and $h_{\mu\nu}$ vanish as $\sigma^{1}\to\infty$ but need not be small in the interior spacetime.  

The idea is to then linearize the metric equations of motion \eqref{eq:CGHSeoms}, up to linear order in perturbations, where indices are raised using the Minkowski metric $\eta_{\mu\nu}$. We find\footnote{It is useful to know 
\beq
\begin{split}
&\nabla_{\mu}\nabla_{\nu}\phi\approx \partial_{\mu}\partial_{\nu}\varphi+\frac{\lambda}{2}(\partial_{\nu}h_{\mu1}+\partial_{\mu}h_{\nu1}-\partial_{1}h_{\mu\nu})\;,\\
&-g_{\mu\nu}\Box\varphi
\approx -\eta_{\mu\nu}[\partial^{2}\varphi-\frac{\lambda}{2}\eta_{\mu\nu}(2\partial_{0}h_{01}-\partial_{1}h_{00})+\frac{\lambda}{2}\partial_{1}h_{11}]\;,\\
&g_{\mu\nu}(\nabla\phi)^{2}-g_{\mu\nu}\lambda^{2}\approx -\eta_{\mu\nu}h^{11}\lambda^{2}-2\lambda\eta_{\mu\nu}\partial_{1}\varphi\;.
\end{split}
\eeq
}
\beq
\begin{split} 
T_{\mu\nu}\approx &\;4e^{2\lambda \sigma^{1}}\biggr\{\partial_{\mu}\partial_{\nu}\varphi-\eta_{\mu\nu}\partial^{2}\varphi+\frac{\lambda}{2}(\partial_{\nu}h_{\mu 1}+\partial_{\mu}h_{\nu1}-\partial_{1}h_{\mu\nu})+\frac{\lambda}{2}\eta_{\mu\nu}(2\partial_{0}h_{01}-\partial_{1}h_{00})\\
&-\frac{\lambda}{2}\eta_{\mu\nu}\partial_{1}h_{11}-2\lambda\eta_{\mu\nu}(\partial_{1}\varphi)-\eta_{\mu\nu}h^{11}\lambda^{2}\biggr\}\;,
\end{split}
\eeq
for $\partial^{2}\varphi=\eta^{\alpha\beta}\partial_{\alpha}\partial_{\beta}\varphi$. Notice the simplification for $T_{00}$,
\beq T_{00}\approx 4e^{2\lambda \sigma^{1}}\left[\partial^{2}_{1}\varphi+\frac{\lambda}{2}\partial_{1}h_{11}+2\lambda\partial_{1}\varphi+h^{11}\lambda^{2}\right]=4\partial_{1}\left[e^{2\lambda\sigma^{1}}\partial_{1}\varphi+e^{2\lambda\sigma^{1}}\frac{\lambda}{2}h_{11}\right]\;,\eeq
where here used $h^{11}=h_{11}$ ($h^{11}=\eta^{\alpha1}\eta^{\beta1}h_{\alpha\beta}=h_{11}$). 

 By the gravitational Bianchi identities, if there is a vector field $v^{\mu}$ that generates an 
asymptotic symmetry, then $T_{\mu\nu}v^{\nu}$ is an asymptotic conserved current (density). Using $v^{\mu}=\delta^{\mu}_{0}$, we construct the total momentum $P_{\mu}$, a `volume' integral of the conserved current density (the factor of $1/4\pi$ is our convention)
\beq P_{\mu}=\frac{1}{4\pi}\lim_{\sigma^{1}\to\infty}\int d\sigma^{1} T_{\mu0}\;.\eeq
Specifically, the ADM energy $M_{\text{ADM}}=P_{0}$ is 
\beq \label{eq:ADMmass}M_{\text{ADM}}\equiv \frac{1}{4\pi}\lim_{\sigma^{1}\to\infty}\int d\sigma^{1}T_{00}=\frac{1}{\pi}e^{2\sigma^1 \lambda}\left( \partial_1 \varphi+\frac{\lambda}{2}h_{11} \right)\biggr|_{\sigma^{1}\to\infty}\;,\eeq
matching Eq. (15) of \cite{Kim:1995jta}.\footnote{Up to an unimportant overall factor of two (a consequence of using a different definition of the matter stress-tensor). Further, Eq. (16) of \cite{Kim:1995jta} follows from using $h_{11}=g_{11}-\eta_{11}=e^{2\rho}-1\approx 2\rho$.}

Indeed, applying this definition to the eternal black hole (\ref{eq:eternalBH}) in Kruskal coordinates \eqref{eqn:kruskdil}, where $h_{11}=0$ and $\varphi=-\frac{1}{2} \ln \left(1+\frac{M}{\lambda} e^{-2\lambda \sigma^1}\right)$, it is easy to show $M_{\text{ADM}}=M/\pi$.

\subsubsection{Semi-classical models}

\noindent Let us now pass to the semi-classical extension of the CGHS model. Immediately, one finds that the definition of the ADM energy (\ref{eq:ADMmass}) diverges when semi-classical corrections are included, for either RST or BPP models. For example, in the Hartle-Hawking state, using the asymptotic expansion of the dilaton in \eqref{eq:largeOmega}, it is not hard to find 
\be
\label{eq:ADMHH}
M_{\text{ADM}}=\frac{M}{\pi} +\frac{aN\lambda}{12\pi}(\lambda \sigma^1-1)\Big|_{\sigma^1 \to \infty} \to \infty\,.
\ee
Thus, we find divergences, except in the BPP model, $a=0$. The reason for this divergence is that the black hole of mass $M$ is now surrounded by thermal radiation, which also contributes to the ADM energy. 

Similarly, for the Boulware vacuum, this definition of ADM energy leads to a divergence
\be
M_{\text{ADM}}=\left.\frac{M}{\pi}+\frac{N\lambda}{24\pi}(2a-1)(\lambda\sigma^1-1)\right|_{\sigma^1\rightarrow\infty}\rightarrow\infty\,,
\ee
except for the RST model $a=\frac{1}{2}$. In the Boulware vacuum, there is no thermal radiation coming from the quantum fluctuation of the matter fields, $t_{\pm}(\sigma^{\pm})=0$. However, $T^{(a)}\neq 0$ in \eqref{eq:Tmatter} and has an asymptotic expansion
\beq
T_{\sigma^\pm \sigma^\pm}^{(a)} = \lambda^2(1-2a)\frac{N}{48}
\eeq
which is non-zero except for $a=1/2$. This asymptotic flux due to $I_a$ can be considered as the source of the infinite contribution to the ADM energy in these states.

Thus, an alternative decomposition to compute a modified version of the ADM energy is necessary. As such, we find it useful to consider the following splitting of fields
\beq
\phi=-\lambda \sigma^1 +\phi^{(0,a)}+ \bar \varphi\, , \qquad g_{\mu \nu}=\eta_{\mu \nu}+h_{\mu \nu}^{(0,a)}+\bar h_{\mu \nu}\, ,
\eeq
where 
\beq \bar{\varphi}\equiv \varphi-\phi^{(0,a)}\;,\quad \bar{h}_{\mu\nu}\equiv h_{\mu\nu}-h_{\mu\nu}^{(0,a)}\;,\eeq
with
\be
\phi^{(0,a)}= r_a^{(0)}\, \left(-\lambda\sigma_1 e^{-2\lambda \sigma^1}\right)\, , \qquad h_{\mu \nu}^{(0,a)}=2\phi^{(0,a)}\eta_{\mu \nu}\, .
\ee
Here the constant $r_a^{(0)}$ is related to the asymptotic behavior of the theory, and depends both on the model (via the parameter $a$) and on the vacuum state. We will show later how to identify this constant. 

Let us briefly explain the notation. Here the superscript $(0,a)$ specifies which vacuum state the matter fields are in (namely, Hartle-Hawking or Boulware), and the model, indicated by $a$. The barred notation indicates radiation-like contributions to be removed.

Once this decomposition is done, we can split the ADM mass as
\beq
M_{\text{ADM}}=\mathcal{R}_{a}^{(0)}+\overline M_{\text{ADM}}^{(0,a)}\, ,
\eeq
where 
\beq
\mathcal{R}_a^{(0)}=\frac{1}{\pi}e^{2\lambda \sigma^1}\left( \partial_1 \phi^{(0,a)}+\frac{\lambda}{2}g^{(0,a)}_{11} \right)\biggr|_{\sigma^{1}\to\infty}=\frac{r_a^{(0)}}{\pi}(\lambda \sigma^{1}-1)\Big|_{\sigma^{(1)}\to\infty}\approx \frac{r_a^{(0)}}{\pi} \lambda L \, ,
\eeq
with $L\equiv\sigma_{\max}$ is the spatial `volume'  and where 
\beq \label{eq:subtractedBHmass}
\overline M_{\text{ADM}}^{(0,a)}= \frac{1}{\pi}e^{2\lambda \sigma^1 }\left( \partial_1 \bar \varphi+\frac{\lambda}{2}\bar h_{11} \right)\biggr|_{\sigma^{1}\to\infty}\, ,\qquad \text{with} \qquad \bar \varphi=\varphi-\phi^{(0,a)} \, , \quad \bar h_{11}= h_{11}-h^{(0,a)}_{11}\, ,
\eeq
is finite. To find the $r_a^{(0)}$ coefficient, we can directly expand the solutions $\rho$ and $\phi$ at spatial infinity $(\sigma^{+}-\sigma^{-})\to\infty$ for a given model and vacuum state and identify the term that goes as $\lambda \sigma^1 e^{-2\lambda \sigma^1}$, or to evaluate the stress-energy tensor of the semi-classical corrections in the same limit,
\beq
\langle 0| T_{\pm\pm}(\sigma^{\pm}) |0 \rangle + T_{\pm\pm}^{a}(\sigma^\pm)= \left(\frac{dx^{\pm}}{d\sigma^{\pm}}\right)^{2}\Big( \langle 0| T_{\pm\pm}(x^{\pm}) |0 \rangle + T_{\pm\pm}^{a}(x^{\pm})\Big)\, ,
\eeq
with $\langle 0| T_{\pm\pm}(x^{\pm}) |0 \rangle$ given in eq. \eqref{eq:qstressrelnorm}, and identify $r^{(0)}_a$ from the constant term. 

In what follows we will compute the ADM mass using this decomposition for Hartle-Hawking, and Boulware vacua and for general $a$. As a final remark, it is worth noticing that, in the RST model, the contribution from $T^{a}_{\pm}$ vanishes, so we can interpret $r^{(0)}_{a}$ as the amount (or absence) of thermal radiation.

\vspace{2mm}

\noindent $\bullet$ \emph{Hartle-Hawking vacuum.} To compute the ADM mass, we are only interested in the large $\sigma^1$ (i.e., large $\Omega$) behavior. Using the expansion given in  \eqref{eq:largeOmega} with $\Omega=\Omega^{(\text{HH})}$ given in \eqref{eq:omegaH}, and the relationship between Kruskal, Eddington-Finkelstein, and tortoise coordinates, we obtain 
\beq
\phi^{(\text{HH})}= -\lambda \sigma^1 - e^{-2\lambda \sigma^1}\left(\frac{M}{2\lambda}+a \frac{N}{24}\lambda \sigma^1\right)+\mathcal{O}(e^{-4\lambda \sigma^1})\, , \qquad \rho^{(\text{HH})} = - e^{-2\lambda \sigma^1}\left(\frac{M}{2\lambda}+a\frac{N}{24}\lambda \sigma^1\right)+\mathcal{O}(e^{-4\lambda \sigma^1})\, ,
\eeq
where we have used \eqref{eq:rhosigma-rhoxbis} to express $\rho$ in tortoise coordinates. Using previous results, it is not hard to get 
\beq
\begin{aligned}
    \langle \text{HH}| T_{\pm\pm}(\sigma^{\pm}) |\text{HH} \rangle + T^{a}_{\pm\pm}(\sigma^{\pm}) 
    &= \lambda^2a\frac{N}{24} + \mathcal{O}(e^{-2\lambda\sigma^{1}})
    \end{aligned}
\eeq

From any of these expansions, it is straightforward to identify 
\beq
r^{(\text{HH})}_a=a\frac{N}{24}\, .
\eeq
From this result we can easily compute $\overline M_{\text{ADM}}$:
\beq
\overline M_{\text{ADM}}^{(\text{HH},a)}=\frac{M}{\pi}\, .
\eeq

\vspace{2mm}

\noindent $\bullet$ \emph{Boulware vacuum.} In this case, we can repeat the same process using $\Omega^{(\text{B})}$ given in \eqref{eq:omegaB}. Using the large $\Omega$ expansion and further expanding it for large $\sigma^1$, we obtain
\beq
\phi^{(\text{B})}=-\lambda \sigma^1 -e^{-2\lambda \sigma^1}\left(\frac{M}{2\lambda}+\frac{N}{48}\lambda\sigma^1\left(2a-1\right)\right)+\mathcal{O}(e^{-4\lambda\sigma^1})\,, \qquad \rho^{(\text{B})}=\phi^{(\text{B})}+\lambda \sigma^1\, .
\eeq
In this case the stress-energy tensor of the semi-classical corrections reads 
\beq
\begin{aligned}
    \langle \text{B}| T_{\pm\pm}(\sigma^{\pm}) |\text{B} \rangle + T^{a}_{\pm\pm}(\sigma^{\pm}) &= 0 + (2a-1)\lambda^2 \frac{N}{48} + \mathcal{O}(e^{-2\lambda\sigma^{1}}) \;.
    \end{aligned}
\eeq
It is again straightforward to identify
\beq
r^{(\text{B})}_a=\left(2a-1\right)\frac{N}{48}\, .
\eeq
Using this definition the modified ADM mass gives
\beq
\overline M_{\text{ADM}}^{(\text{B},a)}=\frac{M}{\pi}\, .
\eeq

\subsection{Komar mass}

\noindent It is amusing to note that if we apply the Komar mass formula for general relativity (\ref{eq:komarmassGR}), we recover $M_{\text{Komar}}=\frac{M}{\pi}=M_{\text{ADM}}$ for the eternal black hole. To see this, recall that we identify the two-dimensional gravitational coupling as $1/8G_{2}=e^{-2\phi}$, such that for $d=2$, the Komar mass formula (\ref{eq:komarmassGR}) is 
\beq M_{\text{Komar}}=\frac{1}{\pi}\oint_{S_{\infty}}\epsilon_{S_{\infty}}e^{-2\phi}u_{\alpha}n_{\beta}\nabla^{\alpha}\zeta^{\beta}_{(t)}\;,\eeq
where $\epsilon_{S_{\infty}}$ is the area element for the `2-sphere' at spatial infinity. Evaluating this for the eternal CGHS black hole yields $M_{\text{Komar}}=M/\pi$. 

From this formula, moreover, we can arrive at the expression for the ADM Hamiltonian derived in Appendix C of \cite{Carrasco:2023fcj}. In particular, using that $\xi^{\alpha}_{(t)}$ obeys Killing's equation and $\zeta^{\mu}_{(t)}=\mathcal{N}u^{\mu}$, we can perform an integration by parts (dropping a total derivative) such that
\beq 
\begin{split} 
M_{\text{Komar}}&=\frac{1}{\pi}\int_{S_{\infty}}\epsilon_{S_{\infty}}\left[e^{-2\phi}u_{\alpha}\zeta^{\alpha}_{(t)}(\nabla^{\beta}n_{\beta}-2n^{\beta}\partial_{\beta}\phi)+e^{-2\phi}n_{\beta}\zeta^{\alpha}_{(t)}\nabla^{\beta}u_{\alpha}\right]\\
&=-\frac{1}{\pi}\oint_{S_{\infty}}\epsilon_{S_{\infty}}\mathcal{N}e^{-2\phi}\left(\nabla^{\beta}n_{\beta}-n_{\beta}a^{\beta}-2n^{\beta}\nabla_{\beta}\phi\right)\;,
\end{split}
\eeq
for `acceleration' $a^{\beta}\equiv u^{\alpha}\nabla^{\beta}u_{\alpha}$. Further, recall that the trace of the extrinsic curvature $K$ of $B$ embedded in $\mathcal{M}$ is related to the trace of the extrinsic curvature $k$ of $S_{\infty}$ via $K=k+n_{\alpha}a^{\beta}$ (cf. \cite{Brown:1992br}). Therefore, 
\beq M_{\text{Komar}}=-\frac{1}{\pi}\oint_{S_{\infty}}\mathcal{N}e^{-2\phi}(k-2n^{\beta}\partial_{\beta}\phi)\;,\eeq
matching the boundary contribution to the ADM Hamiltonian in Eq. (C.22) of \cite{Carrasco:2023fcj} (for vanishing shift vector).

\section{Hamilton-Jacobi counterterm} \label{ap:HJ}

\noindent Here we briefly recall the Hamilton-Jacobi method for determining local counterterm Lagrangians specific to two-dimensional dilaton gravity \cite{Davis:2004xi,Grumiller:2007ju}. We generalize this approach to semi-classical dilaton theories, deriving the local counterterm action used to render the analytically solvable model finite in the IR. 
    
To this end,  consider the following generic Euclidean action for two-dimensional dilaton gravity,
\be
\label{eq:Idilaton}
I_{E}=-L_{0}\int_{\mathcal{M}}d^2x\sqrt{g}\left[\Phi R-U(\Phi)(\nabla \Phi)^2-2V(\Phi)\right]-2L_{0}\int_{\partial\mathcal{M}}dy\sqrt{\gamma} \Phi K\,,
\ee
where $\Phi$ is a scalar field. In what follows we set $2L_{0}=1$.It is always possible to fix a gauge such that the solution to the (Euclidean) equations of motion have the form
\beq \Phi=\Phi(r)\;,\quad ds^{2}=\Xi(r)d\tau^{2}+\Xi^{-1}(r)dr^{2}\;,\label{eq:gensoleuc}\eeq
where
\beq \partial_{r}\Phi=e^{-Q(\Phi)}\;,\quad \Xi(\Phi)=w(\Phi)e^{Q(\Phi)}\left(1-\frac{2M}{w(\Phi)}\right)\;.\label{eq:metfuncsan}\eeq
Here 
\bea 
\label{eq:Q}Q(\Phi)&\equiv&Q_0+\int^\Phi d\tilde{\Phi} U(\tilde{\Phi})\;,\\
\label{eq:omega} w(\Phi)&\equiv& w_0-2\int^\Phi d\tilde{\Phi}V(\tilde{\Phi})e^{Q(\tilde{\Phi})}\,,
\eea
are model-dependent functions, with constants $Q_0$, $\omega_0$, and $M$. When $M=0$, the norm of the Killing vector $\partial_{\tau}$ has the `ground state' value, $\Xi(\Phi)|_{M=0}\equiv\Xi_{0}(\Phi)=w(\Phi)e^{Q(\Phi)}$. 

Consider a small variation of the action (\ref{eq:Idilaton}). In addition to a term which will vanish upon imposing the equations of motion,  there is a boundary term that depends on the momenta conjugate to the boundary metric and $\Phi$, 
\beq \delta I_{E}=\int_{\mathcal{M}}d^{2}x\sqrt{g}(\text{EOM})+\int_{\partial\mathcal{M}}dy\sqrt{\gamma}(\pi^{ab}\delta\gamma_{ab}+\pi_{\Phi}\delta\Phi)\;,\eeq
with
\beq \pi^{ab}=-\frac{1}{2}\gamma^{ab}n^{\mu}\nabla_{\mu}\Phi\;,\quad \pi_{\Phi}=U(\Phi)n^{\mu}\nabla_{\mu}\Phi-K\;.\label{eq:conjmom}\eeq
In general, the boundary term does not evaluate to zero. The GHY boundary term, further, does not guarantee the elimination of this boundary term for \emph{arbitrary} field variations $\delta \gamma_{ab}$ and $\delta\Phi$ that preserves (Dirichlet) boundary conditions. Indeed, evaluating the action variation for the solution (\ref{eq:gensoleuc}) yields the boundary term
\beq \delta I_{E}=\int_{\partial \mathcal{M}}d\tau\left[-\frac{1}{2}(\partial_{r}\Phi)\delta\Xi+\left(U\Xi\partial_{r}\Phi-\frac{1}{2}\partial_{r}\Xi\right)\delta\Phi\right]\;,\eeq
which is to be evaluated at some IR regulator surface. Assuming boundary conditions that are preserved under variations with an asymptotic behavior like the subleading term in (\ref{eq:metfuncsan}), i.e., $\delta\Xi=\delta Me^{Q}$, it follows $\delta I_{E}=\int_{\partial\mathcal{M}}d\tau\delta M\neq0$. 

The goal then is to modify the original action that is finite and stationary for arbitrary solutions to the (semi-)classical equations of motion. This is done by introducing a counterterm action, which has the generic form\footnote{Our terminology for $I_{\text{ct}}$ differs from \cite{Grumiller:2007ju} by an overall minus sign. Here we add $I_{\text{ct}}$ to the original action, generating their ``improved action''.}
\beq I_{\text{ct}}=-\int_{\partial\mathcal{M}}dyL_{\text{ct}}(\gamma,\Phi)\;.\eeq
The boundary counterterm is identified with a solution
of the Hamilton-Jacobi equation for the on-shell action. In words, one starts by rewriting the Hamiltonian constraint of the theory in terms of the momenta (\ref{eq:conjmom}), cast as functional derivatives of the on-shell action. The simplicity of working in 2D allows one to realize $L_{\text{ct}}(\gamma,\Phi)=\sqrt{\gamma}\mathcal{L}_{\text{ct}}(\Phi)$, such that conjugate momenta simplify and the Hamiltonian constraint becomes a linear differential equation for $\mathcal{L}^{2}_{\text{ct}}$ \cite{Davis:2004xi,Grumiller:2007ju}:
\beq \frac{1}{2}\partial_{\Phi}(\mathcal{L}_{\text{ct}}^{2})+\frac{1}{2}U(\Phi)\mathcal{L}_{\text{ct}}^{2}+V(\Phi)=0\;,\eeq
with general solution $\mathcal{L}_{\text{ct}}=-\sqrt{e^{-Q(\Phi)}w(\Phi)}$.
Thus, the counterterm action to be added to the original action (\ref{eq:Idilaton}) is
\be
\label{eq:Ict_gen}
I_{\text{ct}}=\int_{\partial \mathcal{M}}dy\sqrt{\gamma}\sqrt{\omega(\Phi)e^{-Q(\Phi)}}\,.
\ee
In principle, the $w$ function may be shifted by a constant.

\subsubsection{CGHS counterterm} 

\noindent As a warm-up, let us apply the above formalism to determine the counterterm action for classical CGHS. To this end, we notice we can bring the CGHS action to the form (\ref{eq:Idilaton}) upon the field redefinition $e^{-2\phi}=\Phi$, such that $U(\Phi)=-\Phi^{-1}$ and $-2V(\Phi)=4\lambda^{2}\Phi$. Consequently, $Q(\Phi)=-\ln(\Phi)+Q_{0}$ and $w(\Phi)=w_{0}+4\lambda^{2}e^{Q_{0}}\Phi$. We set $w_{0}=0$ and $Q_{0}=0$ such that (restoring factors of $2L_{0}=\pi^{-1}$)
\be
I^{\text{CGHS}}_{\text{ct}}=\frac{1}{\pi}\int_{\partial \mathcal{M}}dy\sqrt{\gamma}\sqrt{\omega(\Phi)e^{-Q(\Phi)}}=\frac{1}{\pi}\int_{\partial\mathcal{M}}dy\sqrt{\gamma}\sqrt{4\lambda^{2}\Phi^{2}}=\frac{1}{\pi}\int_{\partial\mathcal{M}}dy\sqrt{\gamma}e^{-2\phi}(2\lambda)\;.
\ee
This is precisely the counterterm action (in Euclidean signature) used in Section \ref{sec:CGHSbhs}.

\subsubsection{Counterterm for semi-classical corrections} 

\noindent We are interested in the scenario where we incorporate semi-classical corrections, where, at least, one appends the classical action (\ref{eq:Idilaton}) with the semi-classical Polyakov term (restated here in Euclidean signature)
\beq I_{E}^{\text{Poly}}=\frac{N}{48\pi}\int_{\mathcal{M}}d^{2}x\sqrt{g}\left(-ZR+\frac{1}{2}(\nabla Z)^{2}\right)\;.\eeq
The authors of \cite{Bergamin:2007sm} argue there are two ways to proceed to complement the Hamilton-Jacobi procedure outlined above: (i) to treat $Z$ as a function of the classical field $\Phi$, or (ii) to treat $Z$ as an independent field. Either viewpoint requires a modification. Focusing on the former,\footnote{An example of the latter is semi-classical AdS JT gravity, where the counterterm for the Polyakov action is explicitly found by examining IR behavior of the on-shell action \cite{Almheiri:2014cka}.} one considers a field redefinition $\hat{\Phi}=\Phi+\frac{N}{24\pi}Z(\Phi)$, such that the semi-classical generalization of (\ref{eq:Idilaton}) is 
\beq I_{E}^{\text{semi}}=-L_{0}\int_{\mathcal{M}}d^{2}x\sqrt{g}\left[\hat{\Phi}R-\left(U(\Phi)+\frac{N}{48\pi}(\partial_{\Phi}Z)^{2}\right)(\nabla\Phi)^{2}-2V(\Phi)\right]\;.\eeq
Presently, the situation of interest has that $Z=2\rho-2\xi$ and $\phi$ are related upon imposed Kruskal symmetry, $\rho=\phi$. Notably, however, $Z$ is not in general solely a function of $\phi$, complicating the matter. 

As it happens, we will be able to eliminate all divergences just by accounting for the counterterm associated with the RST/BPP contribution.

\subsubsection{RST/BPP counterterm}

Let us proceed by considering only the explicit gravi-dilaton sector of the interpolating model. To write the action $I=I_{\text{CGHS}}+I_a$ in the form \eqref{eq:Idilaton}, we introduce the field redefinition $\Phi=e^{-2\phi}$. In Lorentzian signature, 
\bea
I&=&\frac{1}{2\pi}\int_{\mathcal{M}}d^2x\sqrt{-g}\left[R\Phi+\frac{1}{\Phi}(\nabla \Phi)^2+4\lambda^2 \Phi\right]
+C_a\int_{\mathcal{M}}d^2x\sqrt{-g}\left[R\frac{1}{2}\left(1-a\right)\ln(\Phi)+\frac{1-2a}{4\Phi^2}(\nabla \Phi)^2\right]\,,
\eea
for $C_a=\frac{N}{24\pi}$. We can further bring the above to the form
\be
I=\frac{1}{2\pi}\int_{\mathcal{M}}d^2x\sqrt{-g}\left(Y-U(Y)(\nabla Y)^2-\frac{1}{2}V(Y)\right)\,,
\ee
upon another field redefinition
\beq
\Psi=\Phi+\pi C_a(1-a)\ln \Phi\;\;\Longrightarrow \;\;\Phi= (1-a) \pi C_a W_0\left(\tfrac{e^{\frac{\Psi}{C_a\pi(1-a)}}}{C_a\pi(1-a)}\right)\;,\eeq
for Lambert function $W_{0}$, and where
\beq
\begin{split}
&U(\Psi)=-\left(\frac{1}{\Phi}+\frac{\pi}{\Phi^2}\frac{C_a(1-2a)}{2}\right)\left(\frac{d\Phi}{d\Psi}\right)^2=-\frac{(1-2a)+2(1-a) W_0\left(\tfrac{e^{\frac{\Psi}{C_a\pi(1-a)}}}{C_a\pi(1-a)}\right)}{2(1-a)^2C_a\pi\left(1+ W_0\left(\tfrac{e^{\frac{\Psi}{C_a\pi(1-a)}}}{C_a\pi(1-a)}\right)\right)^2}\;,\\
&V(\Psi)=-2\lambda^2 \Phi(\Psi)=-2(1-a)C_a\pi\lambda^2 W_0\left(\tfrac{e^{\frac{\Psi}{C_a\pi(1-a)}}}{C_a\pi(1-a)}\right)\;.
\end{split}
\eeq
Then, \eqref{eq:Q}, \eqref{eq:omega} and \eqref{eq:Ict_gen} exactly evaluate to
\bea
Q(\Psi)&=&Q_0-s(1-2a)\ln\left( W \right)-s\ln\left(1+ W\right)\;,\\
\omega(\Psi)&=&\omega_0+ {}_2F_1\!\left(s-1,s;s+1;-W \right) c\, W^s\;,\\
\sqrt{\omega(\Phi)e^{-Q(\Phi)}}&=&\sqrt{e^{-Q_0}(1+W)^sW^{s(1-2a)}\left(\omega_0+{}_2F_1\!\left(s-1,s;s+1;-W \right)c\,W^s\right)}\;,
\eea
where we introduced notation
\beq W\equiv W_0\left(\tfrac{e^{\frac{\Psi}{C_a\pi(1-a)}}}{C_a\pi(1-a)}\right)\;,\quad  s\equiv\frac{1}{2(1-a)}\;,\quad  c\equiv 8(1-a)^3C_a^2\pi^2\lambda^2e^{Q_0}\;.\eeq

It is enough to consider these equations in the asymptotic limit $\sigma^1\rightarrow\infty$, $\phi\rightarrow-\infty$, such that $W\rightarrow\infty$. Taking $W\rightarrow\infty$  gives
\be
\label{eq:ct_1}
\sqrt{\omega(\Phi)e^{-Q(\Phi)}}=e^{-\frac{Q_0}{2}}\sqrt{c\, s}\, W-\frac{e^{-\frac{Q_0}{2}}}{2}\sqrt{c\, s}\left((s-1)\ln(W)-\frac{\omega_0}{c\, s}-(s-1)H_{s-1}-1 \right)+ \mathcal{O}\!\left(\frac{1}{W}\right)\,,
\ee
where $H_s$ are harmonic numbers. Further, note the asymptotic expansion of the Lambert function $W_{0}$
\be
W=\frac{\Psi}{k}-\ln \Psi+\frac{k \ln \Psi}{\Psi}+\mathcal{O}\!\left(\frac{1}{\Psi^2}\right)\;,
\ee
with $k=\pi C_a(1-a)$. Substituting this in \eqref{eq:ct_1} and expressing $\Phi$ in terms of $\phi$,
we have
\be
\sqrt{\omega(\Phi)e^{-Q(\Phi)}}=2\lambda e^{-2\phi}+(-1+2a)C_a\pi\lambda\phi+C_0+\mathcal{O}(\phi^{2}e^{4\phi})\;,
\ee
where 
\beq C_0=\frac{(1-a) C_a \pi}{2} \lambda\left(2+\frac{e^{-Q_0} \omega_0}{2(1-a)^2 C_a^2 \pi^2 \lambda^2}-\frac{(1-2 a)}{1-a} \left(\ln \left[(1-a) C_a \pi\right]+\gamma_E+\psi^{(0)}\!\left(\tfrac{1}{2-2a}\right)
\right)\right)\;,\eeq
with Euler gamma constant $\gamma_E$ and $\psi^{(n)}$ is the polygamma function. Note that for $a=1/2$ (RST) and $\omega_0=0$, we have $C^{\text{RST}}_0=\frac{N\lambda}{48}$. 

In summary, the dominant contributions to the local counterterm action for the gravi-dilaton sector of the semi-classical interpolating model is
\be
I^{\text{CGHS}+a}_{\text{ct}}=-\frac{1}{\pi}\int_{\mathcal{M}}dx\sqrt{\gamma}e^{-2\phi}(2\lambda)+\frac{1}{\pi}\int_{\mathcal{M}}dx\sqrt{\gamma}\phi \left(1-2a\right)\frac{N\lambda}{24}-\frac{1}{\pi}\int_{\mathcal{M}}dx\sqrt{\gamma}C_0\;,
\ee
as reported in the main text \eqref{eq:Ict_semi}.

\newpage

\bibliographystyle{utphys}
\bibliography{biblio}

\end{document}